\documentclass[usenatbib,lscape]{emulateapj}
\usepackage{graphicx,ifthen,url,float,lscape}
\newcommand {\kms}{km s$^{-1}$}

\newcommand {\lya}{Ly$\alpha$}
\def\ltsima{$\; \buildrel < \over \sim \;$}
\def\simlt{\lower.5ex\hbox{\ltsima}}
\def\gtsima{$\; \buildrel > \over \sim \;$}
\def\simgt{\lower.5ex\hbox{\gtsima}}
\newcommand {\uJy}{$\mu$Jy}
\newcommand {\um}{$\mu$m}

\def\um     {$\mu$m}
\def\ts     {\thinspace}
\def\kms    {\ifmmode{{\rm \ts km\ts s}^{-1}}\else{\ts km\ts s$^{-1}$}\fi}
\def\msol   {\ifmmode{{\rm M}_{\odot}}\else{M$_{\odot}$}\fi}
\def\lsol   {\ifmmode{{\rm L}_{\odot}}\else{L$_{\odot}$}\fi}
\def\zsol   {\ifmmode{{\rm Z}_{\odot}}\else{Z$_{\odot}$}\fi}
\def\etal   {{\rm et\ts al.}}
\def\ci     {\ifmmode{{\rm C}{\rm \small I}}\else{C\ts {\scriptsize I}}\fi}
\def\hi     {\ifmmode{{\rm H}{\rm \small I}}\else{H\ts {\scriptsize I}}\fi}
\def\hh     {\ifmmode{{\rm H}_2}\else{H$_2$}\fi}
\def\cone {\ifmmode{{\rm C}{\rm \small I}(^3\!P_1\!\to^3\!P_0)}
     \else{C\ts {\scriptsize I}{\small$(^3\!P_1\!\to\,^3\!P_0)$}}\fi}
\def\ctwo {\ifmmode{{\rm C}{\rm \small I}(^3\!P_2\!\to\,^3\!P_1)}
     \else{C\ts {\scriptsize I}{\small$(^3\!P_2\!\to\,^3\!P_1)$}}\fi}
\def\cij {\ifmmode{{\rm C}{\rm \small I}\,(^3P_i\to^3P_j)}\else{C\ts {\scriptsize I}\,{\small$(^3P_i\to^3P_j)$}}\fi}
\def\cii    {\ifmmode{{\rm C}{\rm \small II}}\else{C\ts {\scriptsize II}}\fi}
\def\tex {\ifmmode{{T}_{\rm ex}}\else{$T_{\rm ex}$}\fi}
\def\tmb {\ifmmode{{T}_{\rm mb}}\else{$T_{\rm mb}$}\fi}
\def\tkin {\ifmmode{{T}_{\rm kin}}\else{$T_{\rm kin}$}\fi}
\def\microns {\ifmmode{\mu{\rm m}}\else{$\mu$m}\fi}
\def\nhh   {\ifmmode{n({\rm H}_2)}\else{$n$(H$_2$)}\fi}

\newcommand{\sfr}{{\rm\,M$_\odot$\,yr$^{-1}$}}
\newcommand{\lsun}{{\rm\,L$_\odot$}}

\newcommand{\ha}{{\rm\,H$\alpha$}}
\newcommand{\hb}{{\rm\,H$\beta$}}
\newcommand{\hg}{{\rm\,H$\gamma$}}
\newcommand{\siiv}{{\rm\,Si{\sc IV}}}
\newcommand{\heii}{{\rm\,He{\sc II}}}
\newcommand{\nii}{{\rm\,[N{\sc II}]}}

\newcommand{\oii}{{\rm\,[O{\sc II}]}}
\newcommand{\mgii}{{\rm\,Mg{\sc II}}}
\newcommand{\oiii}{{\rm\,[O{\sc III}]}}

\newcommand{\ciii}{{\rm\,C{\sc III}]}}
\newcommand{\civ}{{\rm\,C{\sc IV}}}

\shorttitle{High-$z$ {\it Herschel}-{\sc Spire} Galaxies}
\shortauthors{C.~M. Casey et al.}
\begin{document}

\title{A population of $\lowercase{z}>2$ Far-infrared {\it
    Herschel}-{\sc Spire} selected Starbursts}
\author{C.M.~Casey\altaffilmark{1},
S.~Berta\altaffilmark{2},
M.~B{\'e}thermin\altaffilmark{3,4},
J.~Bock\altaffilmark{5,6},
C.~Bridge\altaffilmark{5},
D.~Burgarella\altaffilmark{7},
E.~Chapin\altaffilmark{8,9},
S.C.~Chapman\altaffilmark{10,11},
D.L.~Clements\altaffilmark{12},
A.~Conley\altaffilmark{13},
C.J.~Conselice\altaffilmark{14},
A.~Cooray\altaffilmark{15,5},
D.~Farrah\altaffilmark{16},
E.~Hatziminaoglou\altaffilmark{17},
R.J.~Ivison\altaffilmark{18,19},
E.~le Floc'h\altaffilmark{3},
D.~Lutz\altaffilmark{2},
G.~Magdis\altaffilmark{3,20},
B.~Magnelli\altaffilmark{2},
S.J.~Oliver\altaffilmark{21},
M.J.~Page\altaffilmark{22},
F.~Pozzi\altaffilmark{23},
D.~Rigopoulou\altaffilmark{20,24},
L.~Riguccini\altaffilmark{3,25},
I.G.~Roseboom\altaffilmark{21,19},
D.B.~Sanders\altaffilmark{1},
Douglas~Scott\altaffilmark{8},
N.~Seymour\altaffilmark{22,26},
I.~Valtchanov\altaffilmark{9},
J.D.~Vieira\altaffilmark{5},
M.~Viero\altaffilmark{5},
J.~Wardlow\altaffilmark{15}}
\altaffiltext{1}{Institute for Astronomy, University of Hawaii, 2680 Woodlawn Drive, Honolulu, HI 96822}
\altaffiltext{2}{Max-Planck-Institut f{\"u}r Extraterrestrische Physik, Giessenbachstrasse, 85748 Garching, Germany}
\altaffiltext{3}{Laboratoire AIM-Paris-Saclay, CEA/DSM/Irfu - CNRS - Universit\'e Paris Diderot, CE-Saclay, pt courrier 131, F-91191 Gif-sur-Yvette, France}
\altaffiltext{4}{Institut d'Astrophysique Spatiale (IAS), b\^atiment 121, Universit\'e Paris-Sud 11 and CNRS (UMR 8617), 91405 Orsay, France}
\altaffiltext{5}{California Institute of Technology, 1200 E. California Blvd., Pasadena, CA 91125}
\altaffiltext{6}{Jet Propulsion Laboratory, 4800 Oak Grove Drive, Pasadena, CA 91109}
\altaffiltext{7}{Laboratoire d'Astrophysique de Marseille - LAM, Universit\'e d'Aix-Marseille \& CNRS, UMR7326, 38 rue F. Joliot-Curie, 13388 Marseille Cedex 13, France}
\altaffiltext{8}{Department of Physics \& Astronomy, University of British Columbia, 6224 Agricultural Road, Vancouver, BC V6T~1Z1, Canada}\
\altaffiltext{9}{European Space Astronomy Centre, Villanueva de la Ca\~nada, 28691 Madrid, Spain}
\altaffiltext{10}{Institute of Astronomy, University of Cambridge, Madingley Road, Cambridge CB3 0HA, UK}
\altaffiltext{11}{Department of Physics and Atmospheric Science, Dalhousie University, 6310 Coburg Rd, Halifax, NS B3H~4R2, Canada}
\altaffiltext{12}{Astrophysics Group, Imperial College London, Blackett Laboratory, Prince Consort Road, London SW7 2AZ, UK}
\altaffiltext{13}{Center for Astrophysics and Space Astronomy 389-UCB, University of Colorado, Boulder, CO 80309}
\altaffiltext{14}{School of Physics and Astronomy, University of Nottingham, NG7 2RD, UK}
\altaffiltext{15}{Dept. of Physics \& Astronomy, University of California, Irvine, CA 92697}
\altaffiltext{16}{Department of Physics, Virginia Tech, Blacksburg, VA 24061}
\altaffiltext{17}{ESO, Karl-Schwarzschild-Str. 2, 85748 Garching bei M\"unchen, Germany}
\altaffiltext{18}{UK Astronomy Technology Centre, Royal Observatory, Blackford Hill, Edinburgh EH9 3HJ, UK}
\altaffiltext{19}{Institute for Astronomy, University of Edinburgh, Royal Observatory, Blackford Hill, Edinburgh EH9 3HJ, UK}
\altaffiltext{20}{Department of Astrophysics, Denys Wilkinson Building, University of Oxford, Keble Road, Oxford OX1 3RH, UK}
\altaffiltext{21}{Astronomy Centre, Dept. of Physics \& Astronomy, University of Sussex, Brighton BN1 9QH, UK}
\altaffiltext{22}{Mullard Space Science Laboratory, University College London, Holmbury St. Mary, Dorking, Surrey RH5 6NT, UK}
\altaffiltext{23}{Dipartimento di Fisica e Astronomia, Viale Berti Pichat, 6/2, 40127 Bologna, Italy}
\altaffiltext{24}{RAL Space, Rutherford Appleton Laboratory, Chilton, Didcot, Oxfordshire OX11 0QX, UK}
\altaffiltext{25}{NASA Ames, Moffett Field, CA 94035}
\altaffiltext{26}{CSIRO Astronomy \& Space Science, PO Box 76, Epping, NSW 1710, Australia}
\label{firstpage}

\begin{abstract}
We present spectroscopic observations for a sample of 36 {\it
  Herschel}\footnote{{\it Herschel} is an ESA space observatory with
  science instruments provided by European-led Principal Investigator
  consortia and with important participation from NASA.}  -{\sc Spire}
250--500\um\ selected galaxies (HSGs) at $2<z<5$ from the {\it
  Herschel} Multi-tiered Extragalactic Survey (HerMES).  Redshifts are
confirmed as part of a large redshift survey of {\it Herschel}-{\sc
  Spire}-selected sources covering $\sim$0.93\,deg$^2$ in six
extragalactic legacy fields.  Observations were taken with the Keck~I
Low Resolution Imaging Spectrometer (LRIS) and the Keck~II DEep
Imaging Multi-Object Spectrograph (DEIMOS). Precise astrometry, needed
for spectroscopic follow-up, is determined by identification of
counterparts at 24\,\um\ or 1.4\,GHz using a cross-identification
likelihood matching method.
Individual source luminosities range from $\log(L_{\rm
  IR}/L_{\odot})$=12.5--13.6 (corresponding to star formation rates
500--9000\sfr, assuming a Salpeter IMF), constituting some of the most
intrinsically luminous, distant infrared galaxies yet discovered.  We
present both individual and composite rest-frame ultraviolet spectra
and infrared spectral energy distributions (SEDs). The selection of
these HSGs is reproducible and well characterized across large areas
of sky in contrast to most $z>2$ HyLIRGs in the literature which are
detected serendipitously or via tailored surveys searching only for
high-$z$ HyLIRGs; therefore, we can place {\it lower limits} on the
contribution of HSGs to the cosmic star formation rate density at
$(7\pm2)\times$10$^{-3}$\,\sfr\,h$^{3}$\,Mpc$^{-3}$ at $z\sim2.5$,
which is $>$10\%\ of the estimated total star formation rate density
(SFRD) of the Universe from optical surveys.  The contribution at
$z\sim4$ has a lower limit of
3$\times$10$^{-3}$\,\sfr\,h$^{3}$\,Mpc$^{-3}$, \simgt20\%\ of the
estimated total SFRD.  This highlights the importance of extremely
infrared-luminous galaxies with high star formation rates to the
build-up of stellar mass, even at the earliest epochs.
\end{abstract}


\section{Introduction}

Submillimeter galaxies \citep[SMGs, often selected by 850\,\um--1mm
  flux densities
  \simgt2\,mJy;][]{smail97a,hughes98a,barger98a,eales99a} are the most
intrinsically luminous starburst galaxies that have been identified to
date.
SMGs are thought to evolve much like local ultraluminous infrared
galaxies \citep[ULIRGs;][]{sanders88b,sanders96a} via major mergers.
The ``merger'' evolutionary scenario starts with the collision of
gas-rich disk galaxies igniting an intense, short-lived
($\tau\sim100$\,Myr) phase of gas consumption and dust production via
a starburst, followed by the formation of a quasar and eventually
(1--2\,Gyr later) a massive, elliptical galaxy.
In contrast to local ULIRGs, SMGs at $z\sim2$--3 are much more
luminous and more massive (in M$_{\star}$ and M$_{H_{2}}$), and
sometimes much larger \citep{chapman04b,biggs08a}, thus they have been
dubbed ``scaled-up'' \citep{tacconi08a}, providing evidence for cosmic
downsizing \citep{cowie96a}.

The observation that the most luminous infrared sources are at the
highest redshifts \citep[e.g. GN20;][]{daddi09a} poses a unique problem
for galaxy evolution studies.  How can these distant ULIRGs be formed
so quickly after the Big Bang with such high star formation rates?
Their extreme infrared luminosities might stem from different
evolutionary histories than the local ULIRG mergers, i.e. secular gas
accretion \citep{dekel09a,dave10a}, but solving the origin of
infrared-luminous galaxies requires large, uniformly selected samples
of ULIRGs across many epochs.

Unfortunately, most $z>2$ infrared-luminous galaxy samples number
$\sim$30 galaxies selected in non-uniform, biased ways.  It is well
known that the selection of SMGs is severely biased, first against
galaxies with warmer dust temperatures
\citep{blain04a,chapman04a,casey09b,casey11a,chapman10a,magdis10a} and
second, against galaxies at higher redshifts since they are unlikely
to have bright radio counterparts \citep{chapman05a} or
24\,\um\ counterparts \citep{ivison07a,clements08a} due to the radio
$K$-correction and surface brightness dimming.  Third, SMGs at $z>2$
are often detected in inhomogeneous, serendipitous studies with a
range of detection thresholds at different wavelengths in the
far-infrared.  Fourth, the spectroscopic follow-up and redshift
confirmation of these sources is non-uniform; their success rate could
relate to their FIR properties, e.g. color or single-band flux
density.
The lack of SMGs with confirmed redshifts $z$\simgt3.5 has been
alleviated in recent years with the discovery of several systems at
$4<z<5.3$
\citep{wang07a,wang09a,daddi09a,coppin09a,capak11a,smolcic11a,walter12a},
however this work has been severely limited by the rarity of $z>3.5$
sources and the small area, non-uniform coverage of existing
ground-based submillimeter surveys with SCUBA, MAMBO, LABOCA, and
AzTEC.

The {\it Herschel Space Observatory} \citep{pilbratt10a} has surveyed
$\sim$200\,deg$^2$ down to the $\approx5$\,mJy confusion limit of {\sc
  Spire} \citep{griffin10a,nguyen10a} as part of the {\it Herschel}
Multi-tiered Extragalactic Survey \citep[HerMES;][]{oliver12a} at 250,
350, and 500\,\um. 
Although high-$z$ infrared galaxies are spatially rare, SPIRE has
mapped much larger sky areas than previous submillimeter surveys and
thus can detect a statistically significant population of $z>2$
starbursts with a well-characterized selection.

This paper presents redshifts and spectra for 36 $2<z<5$ {\it
  Herschel} {\sc Spire}-selected galaxies identified within a large
sample of $\approx$\,1600 {\sc Spire}-selected galaxies
spectroscopically surveyed over $\sim$1\,deg$^2$.  In
section~\ref{sec:selection}, we describe the source selection, biases
in 24\um\ and radio samples, and spectroscopic observations.  In
section~\ref{sec:identification}, we present redshift identifications.
In section~\ref{sec:results}, we present our results, from derived
luminosities, dust temperatures, the FIR/radio correlation, to
composite rest-frame ultraviolet and infrared spectra.
In section~\ref{sec:discussion}, we discuss the context of our results by
calculating the {\it Herschel} {\sc Spire}-selected galaxy (HSG)
contribution to the cosmic star formation rate density (SFRD), and the
implications for infrared-luminous galaxy evolution in the early
Universe.  In section~\ref{sec:conclusions}, we conclude.  Throughout we use
a flat $\Lambda$CDM cosmology \citep{hinshaw09a} with $H_{\rm
  0}$=71\kms\,Mpc$^{-1}$ and $\Omega_{\rm M}$=0.27.  When
possible, we discuss distance and volume using the general unit,
$h^{-1}$\,Mpc.

\section{Sample and Observations}\label{sec:selection}

\subsection{The {\it Herschel}-SPIRE selected galaxy (HSG) sample}

The sources described in this paper were detected by the {\sc Spire}
instrument \citep{griffin10a} onboard the {\it Herschel Space
  Observatory\/} as part of the {\it Herschel\/} Multi-tiered
Extragalactic Survey \citep[HerMES;][]{oliver12a}.  Sources were
spectroscopically observed in a large redshift survey follow-up
program described in detail in a parallel paper, \citet{casey12c},
hereafter C12.  The results of the redshift survey have been split
between two papers due to the significant differences in the
comprehensive 731 source $z<2$ sample, identified through \oii, \oiii,
\hb, and \ha\ emission, than the 36 galaxies at $z>2$ identified
primarily through rest-frame ultraviolet features.  We refer the
reader to C12 for a detailed discussion of our source selection and
completeness and only briefly summarize those results here.

Due to the large beamsize of {\sc Spire} observations (18\arcsec,
25\arcsec, and 36\arcsec, respectively, at 250, 350, and 500\um),
counterpart identification and point source photometry is performed by
extracting flux from {\sc Spire} maps \citep{levenson10a} at known
positions of {\it Spitzer}-{MIPS} 24\,\um\ and VLA 1.4\,GHz sources (see
C12 for more details on data).  This cross-identification prior source
extraction method (called ``XID'') is described in detail in
\citet{roseboom10a} and \citet{roseboom12a}.  The disadvantage of the
XID method is that it relies on {\sc Spire}-bright sources being
detectable at 24\,\um\ and/or 1.4\,GHz, an assumption which is known to
sometimes fail at $z$\simgt3, depending on the depth of 24\,\um\ or
1.4\,GHz coverage.  For this reason, our spectroscopic survey was
conducted in the HerMES coverage areas with the deepest available
ancillary data in six different legacy fields: Lockman Hole North
(LHN; $\alpha\sim$10h\,46\arcmin, $\delta\sim$59$^\circ$), Cosmic Evolution
Survey field (COSMOS; $\delta\sim$10h\,0\arcmin, $\delta\sim$2$^\circ$),
Great Observatories Origins Deep Survey North field (GOODS-N;
$\alpha\sim$12h\,36\arcmin, $\delta\sim$62$^\circ$), Elais-N1 (EN1;
$\alpha\sim$16h\,0\arcmin, $\delta\sim$54$^\circ$), the UKIDSS Ultra-deep
field (UDS; $\alpha\sim$2h\,19\arcmin, $\delta\sim$--5$^\circ$), and the
Extended Chandra Deep Field South (CDFS; $\alpha\sim$3h\,30\arcmin,
$\delta\sim$--28$^\circ$).

Sources were selected for spectroscopic follow-up by detection at
$>$3$\sigma$ significance in at least one of the three {\sc Spire}
bands.  The absolute flux limit changes field to field depending on
XID prior source density, but averages $\sim$\,10--12\,mJy across the
three bands.  Throughout the rest of this paper we refer to this
population as {\it Herschel}-{\sc Spire} selected galaxies (HSGs) for
convenience.  Higher priority follow-up is given to sources
detected in all three bands, however the source density of all {\sc
  Spire}-detected sources is low enough such that $>$98\%\ of all HSGs
can be surveyed within one spectroscopic mask area, whether it be with
the Keck~I Low Resolution Imaging Spectrometer (LRIS; covering
5.5\arcmin$\times$7.8\arcmin) or with the Keck~II DEep Imaging
Multi-Object Spectrograph (DEIMOS; covering
5\arcmin$\times$16.7\arcmin).

\subsection{Biases in the HSG Sample}

Although the completeness of the XID source identification technique
is $>$95\%\ for sources at $z<2$
\citep{magdis10a,roseboom10a,roseboom12a,bethermin11a}, the
completeness at higher redshifts is difficult to estimate since an
increasing, non-negligible fraction of {\sc Spire}-bright sources drop
out at 24\,\um\ and/or 1.4\,GHz with increasing redshift.  This is
also a function of the 24\,\um\ and 1.4\,GHz depth, which is different
field to field.  Since this paper focuses exclusively on the $z>2$
{\sc Spire}-bright population, it is important to emphasize that the
sample here is incomplete, biased and is not representative of all
{\sc Spire}-bright galaxies at $z>2$.  Constraining the whole {\sc
  Spire} population at $z>2$ will require detailed high-resolution
submillimeter follow-up, e.g. from ALMA, of a large population of {\sc
  Spire}-bright systems, particularly those that are radio and
24\,\um-faint and for sources which fail to yield optical
spectroscopic identifications.

Note that the purity of the XID technique in counterpart
identification is not guaranteed.  In other words, XID might be
incorrect in its identification of the multiwavelength counterpart for
a {\it Herschel} source \citep[as mentioned in][]{roseboom10a}.  The
purity of this sample is impossible to gauge without direct
far-infrared interferometric observations \citep[this does away with
  confusion noise, as in][]{wang11a}.  However, we do note that radio
counterpart identifications are more robust than 24\,\um\ counterparts
due to their source rarity and radio's direct scaling with FIR
luminosity \citep{chapman03a,chapman05a}.  Half of our sample is radio
identified, with the other half showing no overall bias or trends
which would skew our results.

Another possible bias of the HSG sample is the method of spectroscopic
targeting.  Our Keck observations were centered around high-priority
`red' {\sc Spire} sources (e.g. $S_{\rm 250}<S_{\rm 350}<S_{\rm 500}$)
which are thought to be the highest-redshift {\sc Spire} sources
\citep[e.g.][]{cox11a}.  While only 1-2 `red' targets were chosen per
mask, this could skew the total redshift distribution higher than if
masks were laid down arbitrarily on the sky.  As C12 describes, and as
we discuss later in section~\ref{sec:discussion}, we measure the
impact of `red' sources on the overall redshift distribution to be
negligible since many of the `red' targets failed to yield
spectroscopic identifications.

\subsection{Spectroscopic Observations}

Spectroscopic observations were carried out at the W.M. Keck
Observatory using the Low Resolution Imaging Spectrometer (LRIS) on
Keck~I and the DEep Imaging Multi-Object Spectrograph (DEIMOS) on
Keck~II in 2011 and 2012.  LRIS observations were carried out on
2011-Feb-06, 2012-Jan-26, 2012-Jan-27, and 2012-Feb-27 with the
400/3400 grism, 560nm dichroic, and primarily the 400/8500 grating in
the red with central multi-slit wavelength 8000\AA\ for the
5.5\arcmin$\times$7.8\arcmin\ mask.  This setup yields a
1.09\AA\ dispersion in the blue (R$\sim$4000) and 0.80\AA\ dispersion
in the red (R$\sim$9000).  Integration times varied from $\sim$2700 to
5600s per mask depending on airmass and weather.  DEIMOS observations
were carried out on 2011-May-28, 2011-May-29, 2011-Nov-28,
2012-Feb-16, and 2012-Feb-17.  The 600\,lines/mm grating and
7200\AA\ blaze angle was used, resulting in a 0.65\AA\ dispersion
(R$\sim$11000).  The GG455 filter was used to block higher-order
light, and typical integration times per mask were $\sim$2700 to
4800s.  The resolution in LRIS red and DEIMOS is sufficient to
distinguish between a single emission line (e.g. \lya) and the
\oii\ doublet (rest-frame separation of 3\AA).  Data reduction for
LRIS was done using our own custom-built IDL routines, while we used
the DEEP2 DEIMOS data reduction pipeline for DEIMOS data\footnote{The
  analysis pipeline used to reduce the DEIMOS data was developed at UC
  Berkeley by Michael Cooper with support from NSF grant
  AST-0071048.}.

Twenty-five LRIS masks were observed (thirteen under photometric
conditions), and twenty-nine DEIMOS masks were observed (sixteen under
photometric conditions), surveying a total of 1594 {\sc
  Spire}-selected sources in 0.93\,deg$^{2}$ (0.43\,deg$^{2}$ observed
in photometric conditions).  Of 1594 sources surveyed, 767 have
confirmed spectroscopic redshifts identified primarily by the
\oii\ doublet, \oiii, \hb, \ha, \nii, Ca\,H and Ca\,K absorption, the
Balmer break, \hg, \lya, \civ, \ciii, \heii, and the Lyman break.  The
lower redshift sources identified through rest-frame optical
signatures are discussed in C12.  Of the 767 confirmed redshifts, 36
are above $z=2$ and thus comprise the HSG sample discussed in this
paper.  Out of the 36, we categorize 22 as secure based on the
signal-to-noise of \lya\ and/or detection of multiple spectral
features, discussed in the next section.

Note that the 36 source sample discussed in this paper was not
selected in any special way in {\sc Spire} color, photometric
redshift, or optical characteristics.  There was no special selection
imposed which would yield more high-redshift identifications, and
the selection differs in no way from the lower redshift confirmed
sources.  This paper is simply a description of all $z>2$ sources
confirmed in our large spectroscopic survey.

\begin{figure}
\centering
\includegraphics[width=0.99\columnwidth]{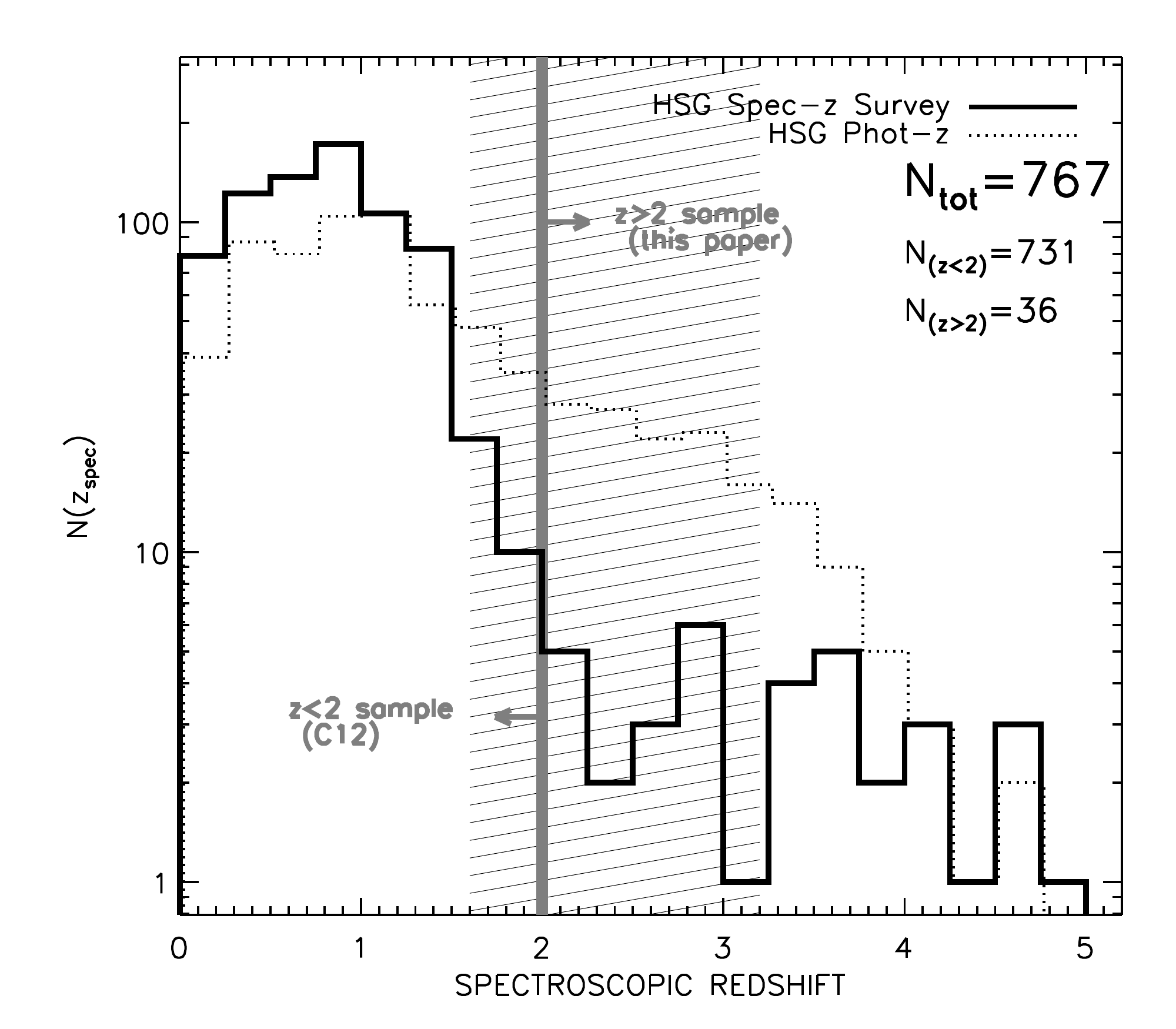}
\caption{ Redshift distribution of {\it Herschel}-{\sc Spire} selected
  galaxies from our survey.  This paper analyzes the $z>$2 sample,
  which consists of 36 sources, while the $z<$2 sample, consisting of
  731 sources, is analyzed in a parallel paper (Casey \etal\ 2012;
  C12).  The distribution of photometric redshifts comes from sources
  in the COSMOS field, where the quality of the photometric redshifts
  is high over a large area; this distribution includes sources not
  necessarily in the spectroscopic survey, but follow the same
  selection as spectroscopic targets.  The hashed area from
  $1.6<z<3.2$ highlights the DEIMOS redshift range where there is a
  deficit of sources due to spectroscopic incompleteness.}
\label{fig:nzhermes}
\end{figure}

The DEIMOS wavelength coverage roughly spans 4500--9500\AA, whereas
LRIS coverage spans 3000--10000\AA.  The limited wavelength coverage
of DEIMOS results in a gap in our redshift coverage between
$1.6<z<3.2$ which does not occur for LRIS observations.  Indeed, very
few sources observed by DEIMOS are identified in that redshift range,
with the exception of 1HERMES\,X1.4\,J104642.89+585650.0 at $z=2.841$
and 1HERMES\,X24\,J160545.99+534544.4 at $z=2.555$.  These are both
quasars with strong, broad \civ\ emission.  We take the DEIMOS
redshift desert into account when calculating the luminosity function
and contribution to the cosmic star formation rate density later in
Section~\ref{sec:discussion}.  The redshift distribution for all {\it
  Herschel}-{\sc Spire} galaxies is shown in
Figure~\ref{fig:nzhermes}.  The photometric redshift distribution
shown comes from all HSGs in the COSMOS field, not limited to those
targeted for spectroscopic observations.


\section{Redshift Identification and Spectra}\label{sec:identification}

\begin{figure*}[p]
\centering

\includegraphics[width=0.85\columnwidth]{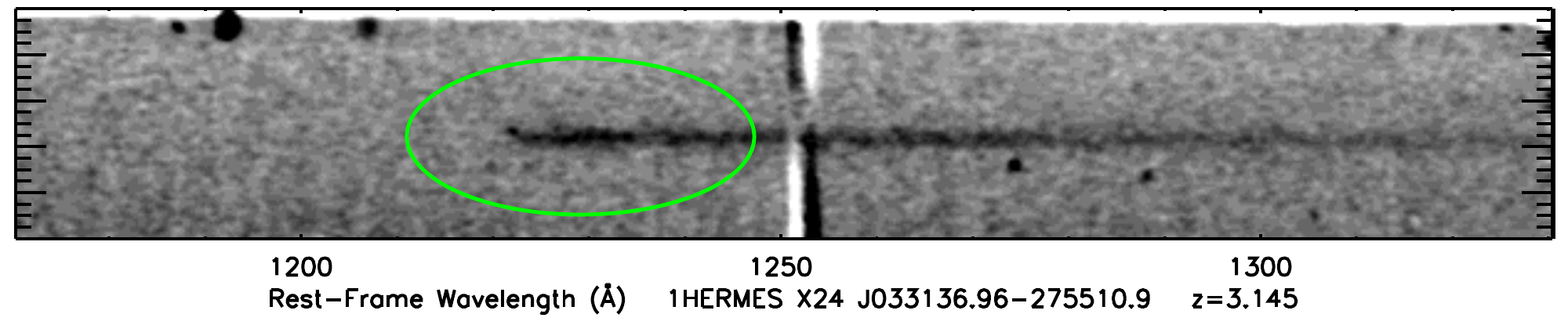}
\includegraphics[width=0.85\columnwidth]{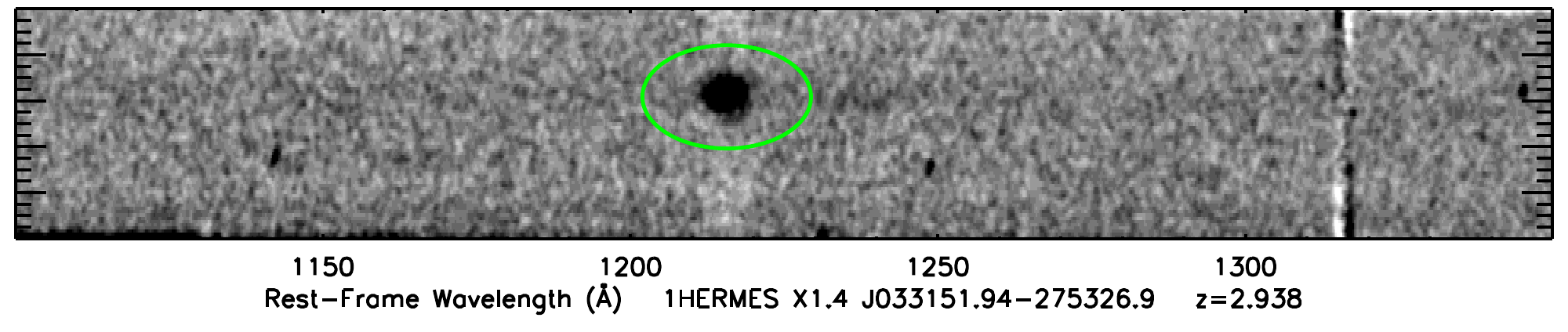}\\
\includegraphics[width=0.85\columnwidth]{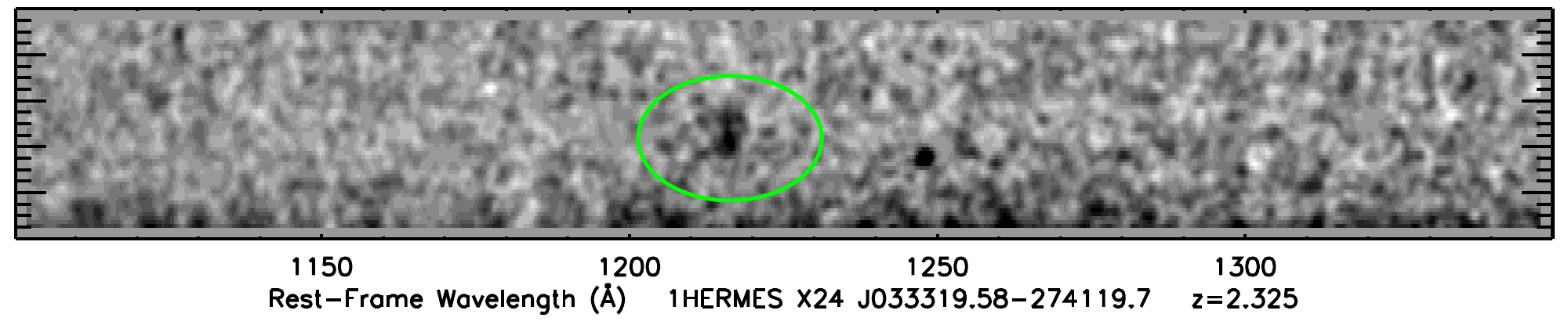}
\includegraphics[width=0.85\columnwidth]{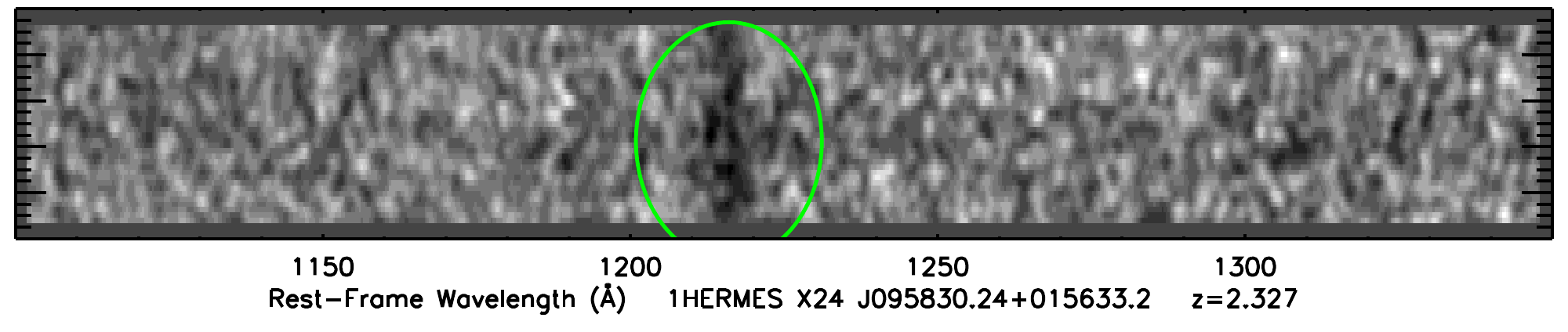}\\
\includegraphics[width=0.85\columnwidth]{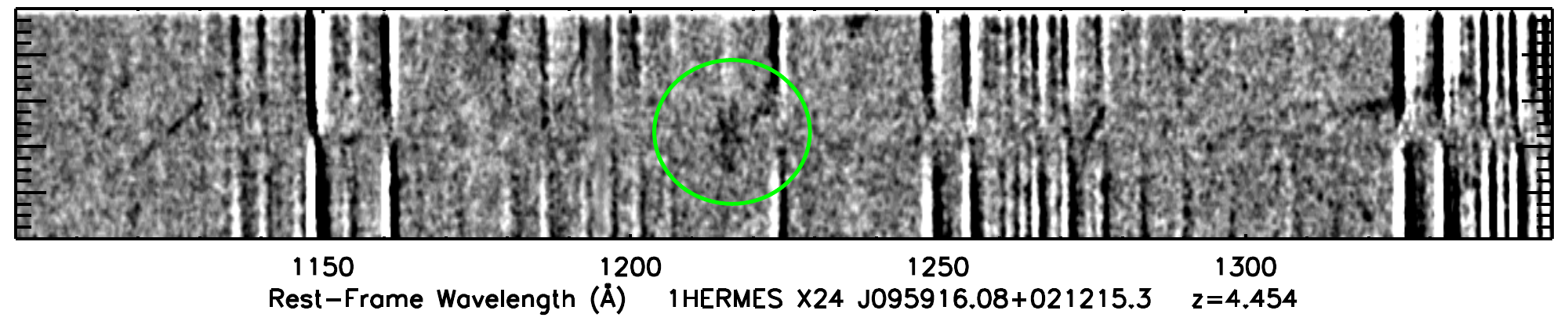}
\includegraphics[width=0.85\columnwidth]{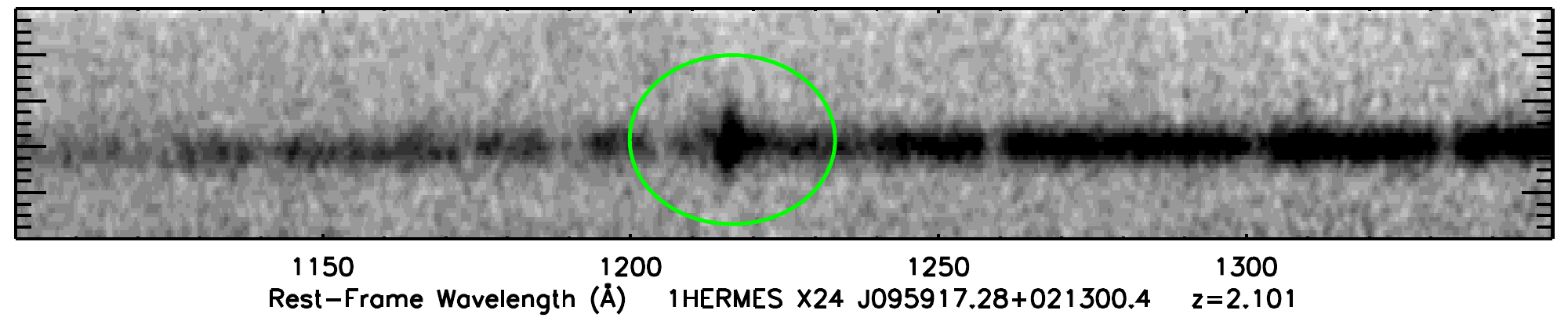}\\
\includegraphics[width=0.85\columnwidth]{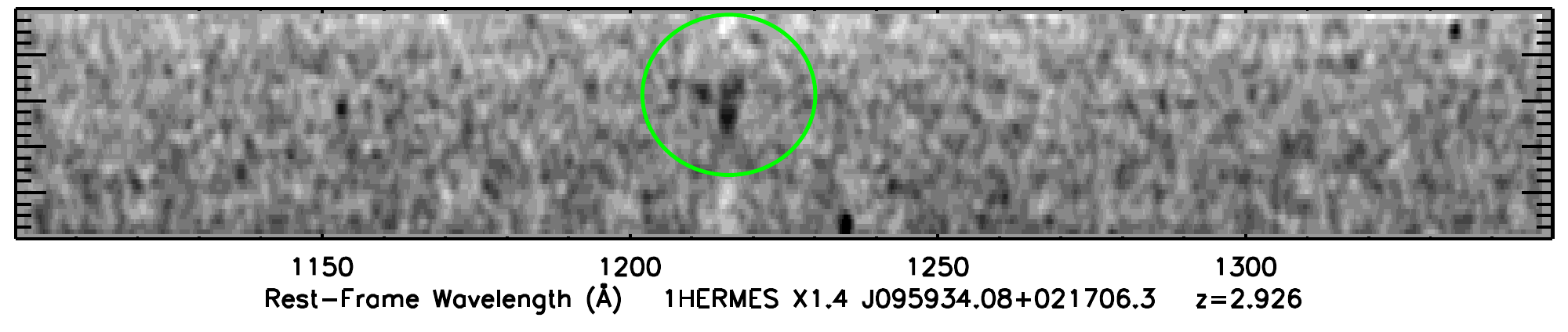}
\includegraphics[width=0.85\columnwidth]{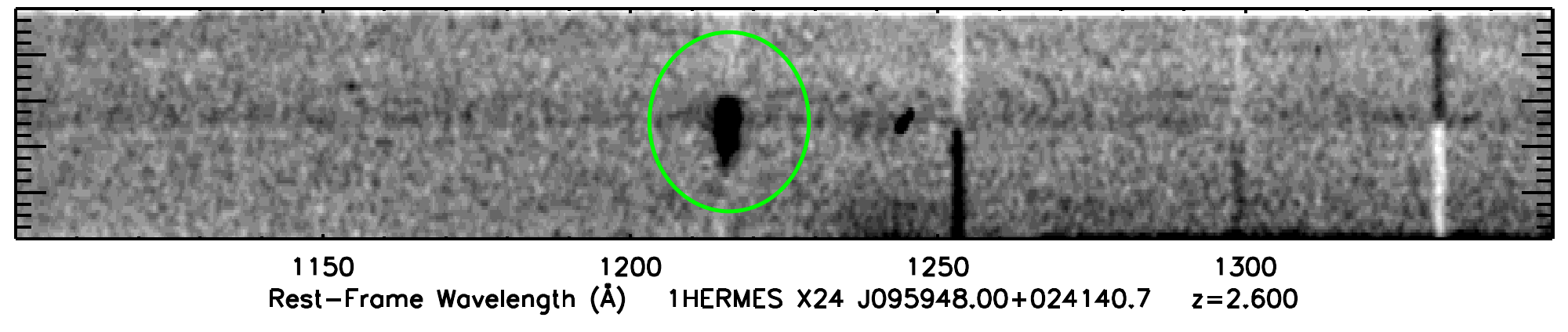}\\
\includegraphics[width=0.85\columnwidth]{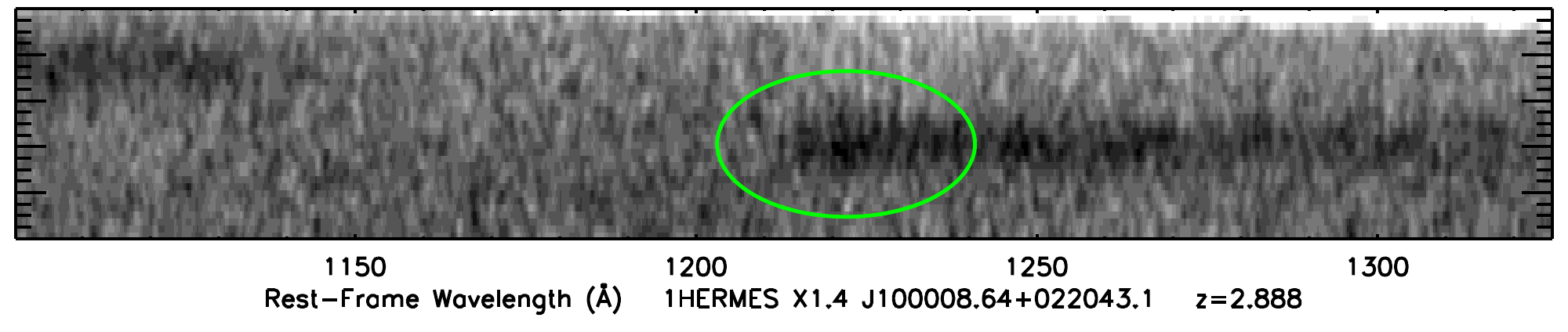}
\includegraphics[width=0.85\columnwidth]{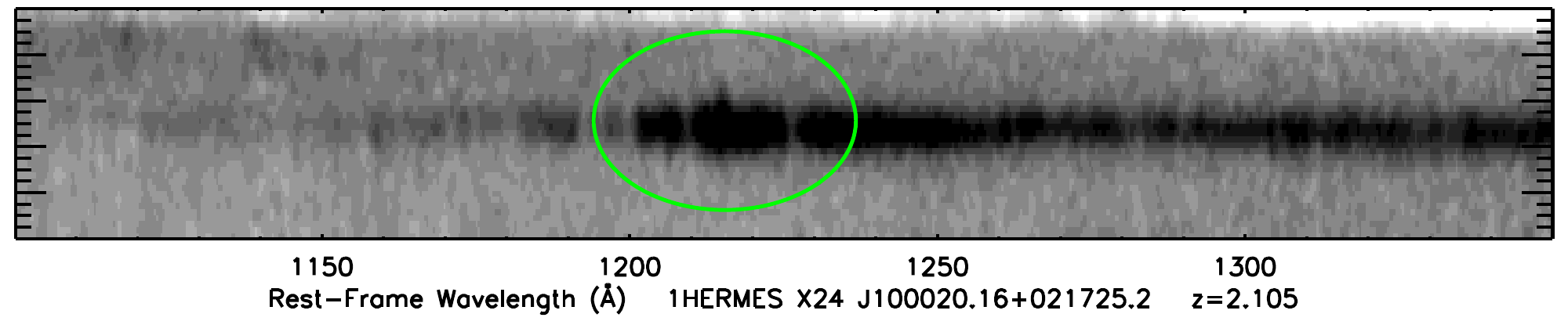}\\
\includegraphics[width=0.85\columnwidth]{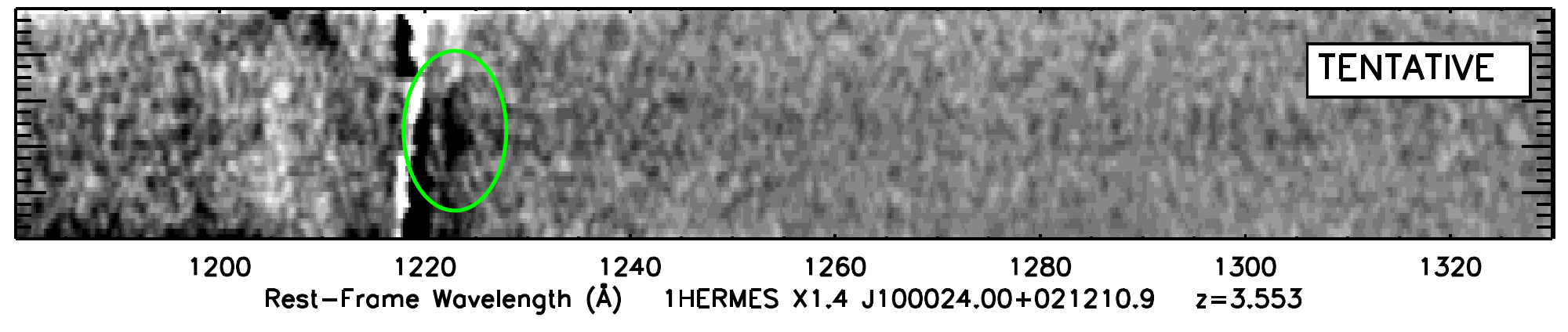}
\includegraphics[width=0.85\columnwidth]{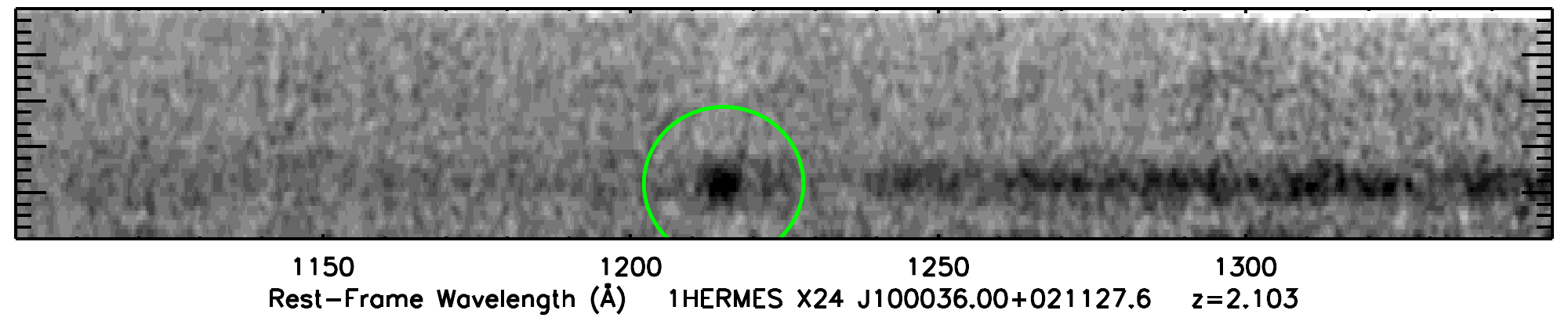}\\
\includegraphics[width=0.85\columnwidth]{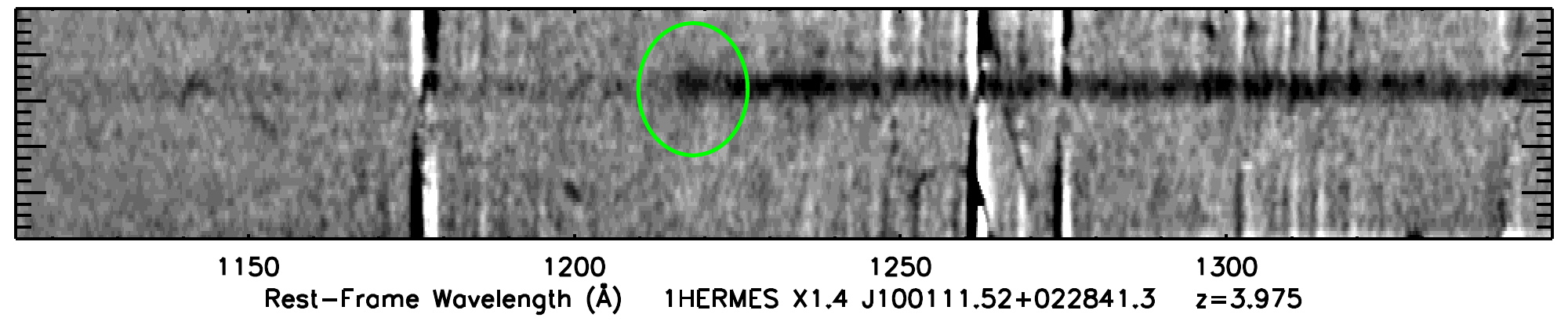}
\includegraphics[width=0.85\columnwidth]{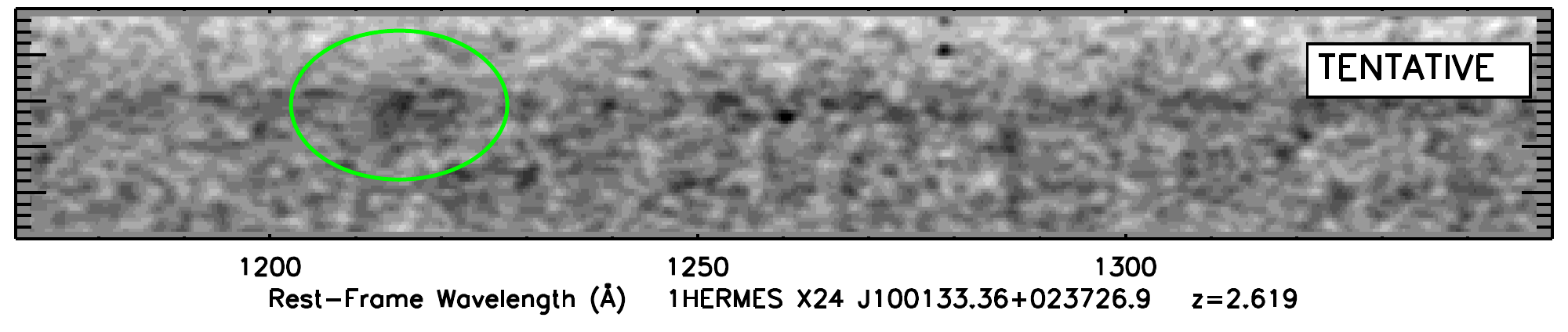}\\
\includegraphics[width=0.85\columnwidth]{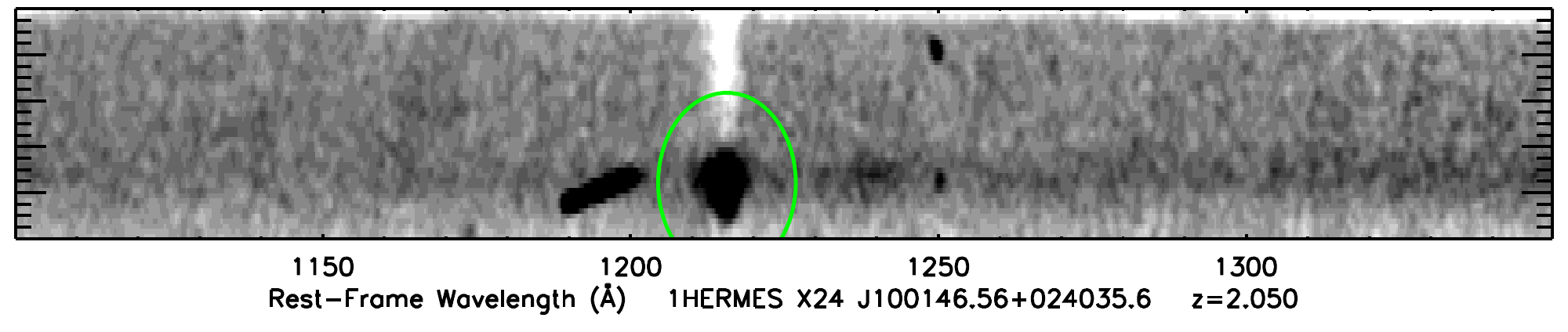}
\includegraphics[width=0.85\columnwidth]{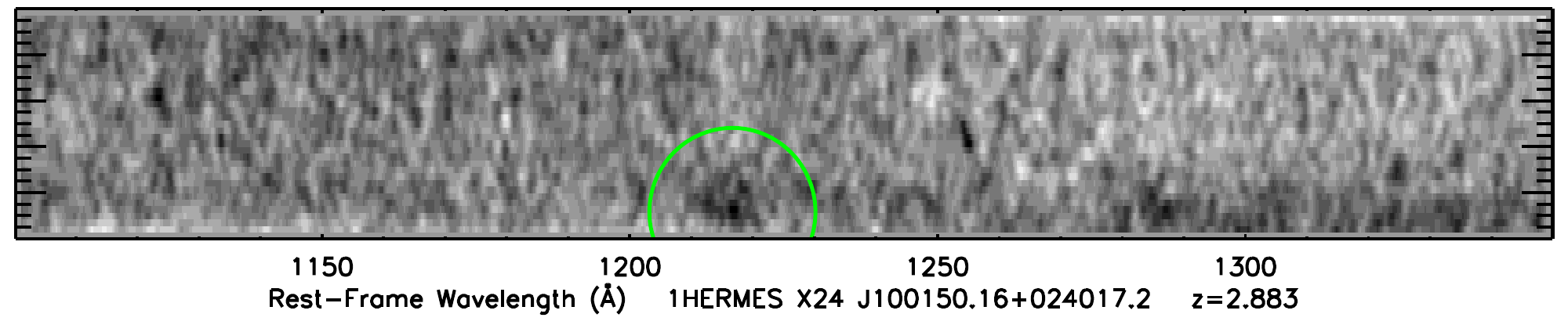}\\
\includegraphics[width=0.85\columnwidth]{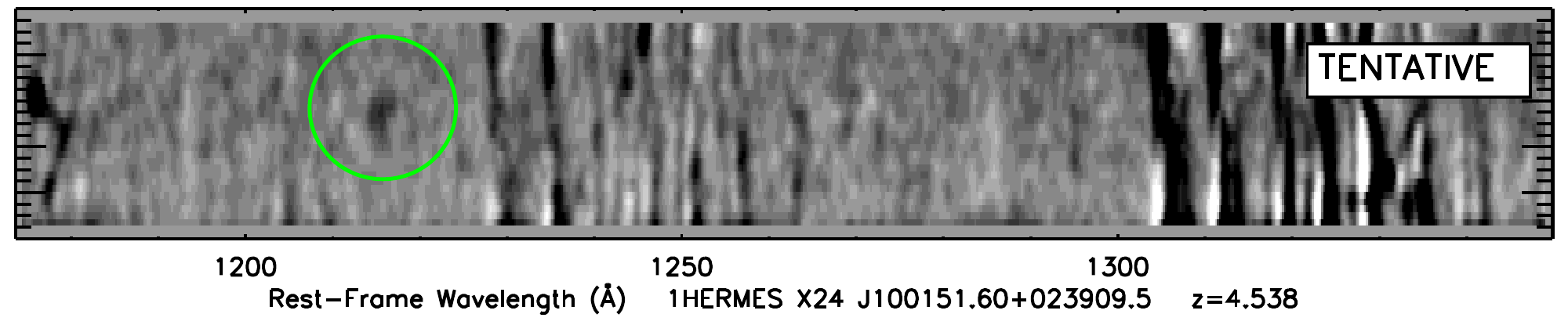}
\includegraphics[width=0.85\columnwidth]{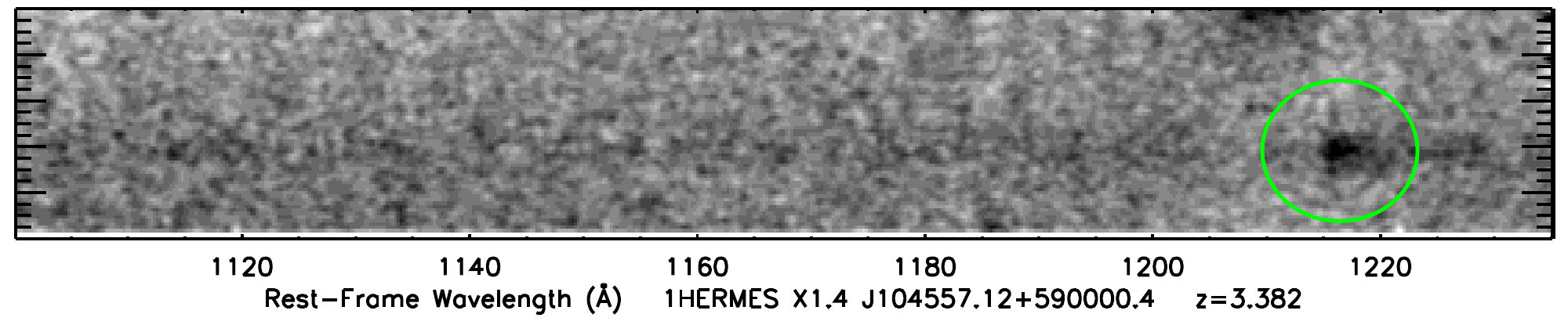}\\
\includegraphics[width=0.85\columnwidth]{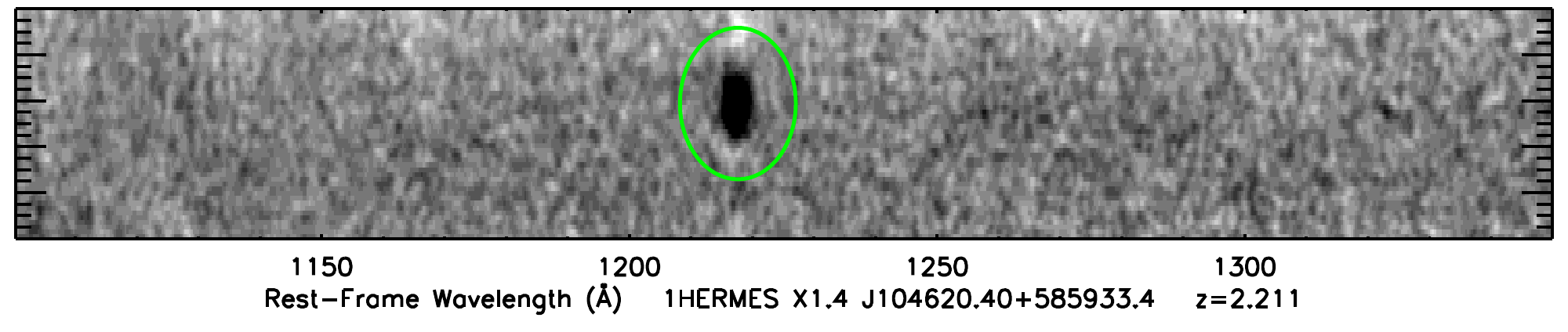}
\includegraphics[width=0.85\columnwidth]{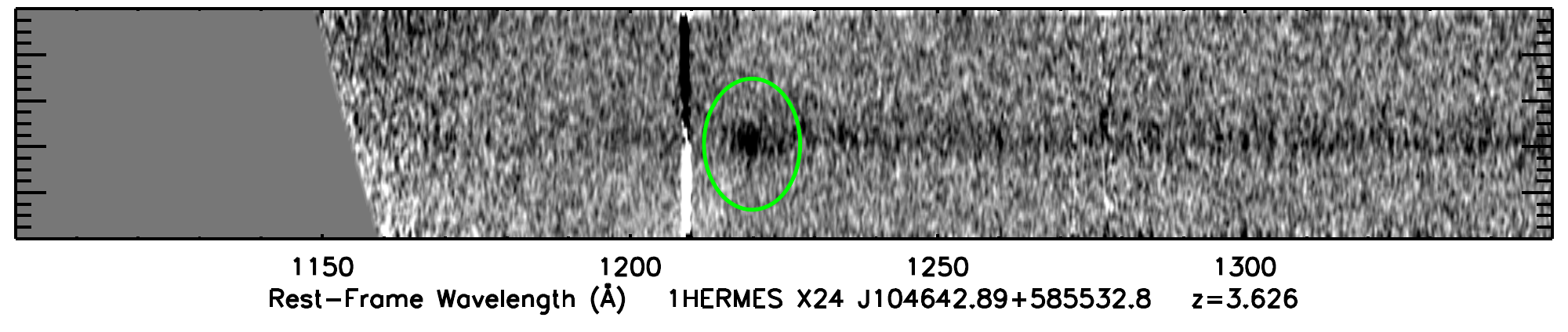}\\
\includegraphics[width=0.85\columnwidth]{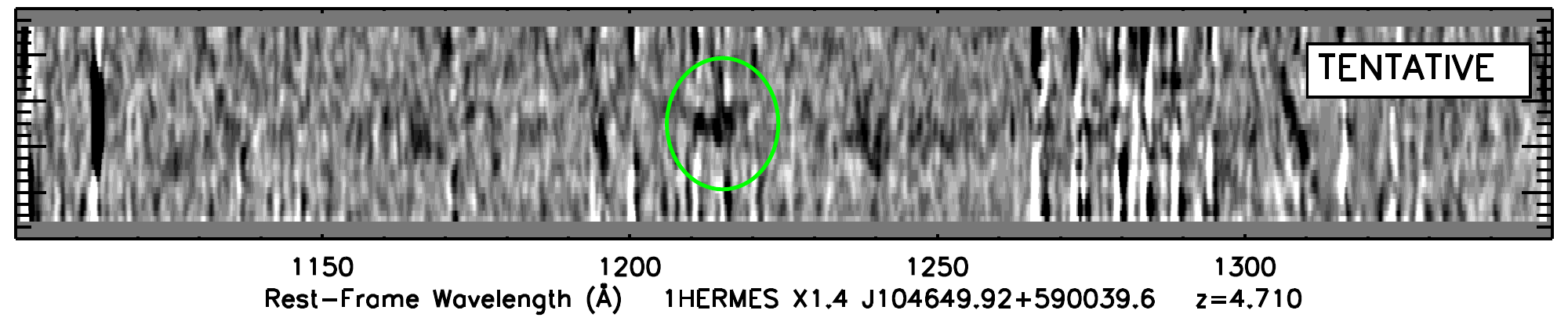}
\includegraphics[width=0.85\columnwidth]{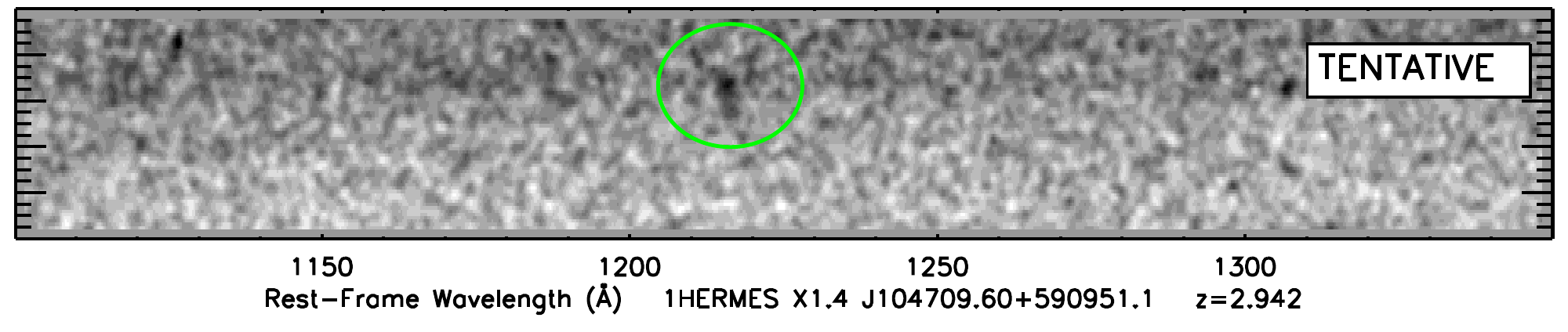}\\
\caption{Two-dimensional spectra around the identified \lya\ features
  for the sources only identified through a single emission line.
  This emission line is not thought to be \oii\ or \mgii\ in these
  cases since there is no detection of commonly bright, accompanying
  emission lines, e.g. \oii, \oiii\ and \hb, within the wavelength
  coverage of LRIS/DEIMOS observations.  Note the wide variety of
  \lya\ morphologies, from very diffuse to very compact, with and
  without redward continuum.  The sources at higher redshifts are
  observed at higher wavelengths, thus likely have one-dimensional
  extractions contaminated by the OH forest. 
}
\label{fig:spectra2d}
\end{figure*}

Table~\ref{tab_fullsample} summarizes the spectroscopic
identifications and multi-wavelength properties of the 36 {\sc
  Spire}-selected galaxies identified at $2<z<5$.  In each case, the
signal-to-noise of the identifying feature (\lya\ emission in the
majority of cases) is required to be $>$5 in the two-dimensional
spectrum.  The two-dimensional spectra for \lya\ identified sources is
shown in Figure~\ref{fig:spectra2d}.  Note that the S/N of
\lya\ changes from the two-dimensional spectra to one-dimensional
extractions of the spectra depending on the sources' compactness, the
compactness of continuum relative to \lya\ emission, and the observed
wavelength of the line.  For example, one-dimensional extractions with
\lya\ longward of $\sim$6000\AA\ are prone to contamination by the OH
forest, thus will be of significantly poorer quality in 1D than in 2D
where OH features are more easily distinguished from real lines.

Initially, we split sources into two categories: sources with multiple
spectroscopic features (e.g. \lya\ emission, \siiv\ absorption,
\civ\ emission) and single-line identifications (where \lya\ is
identified as the only emission line).  The former group of
identifications is naturally more secure than the latter.  The latter
source list is dominated by sources at the high redshift end,
3\simlt$z$\simlt5, since multiple features are naturally more
difficult to identify in higher redshift sources.  However, there are
clear cases where the identification of \lya\ at $>$5$\sigma$ is more
secure than the identification of multiple other features at low
signal-to-noise (S/N\simlt5$\sigma$).  An example of a secure
single-line source is 1\,HERMES\,X24\,J160802.63+542638.1 at $z=3.415$
where \lya\ is detected at $>$5$\sigma$.  An example of a less secure
multi-feature source is 1\,HERMES\,X1.4\,J123622.58+620340.3 at
$z=3.579$ where both \lya\ and \civ\ are detected at S/N$<$5$\sigma$.
In this paper, we choose to segregate sources with secure
spectroscopic identifications from those with tentative
identifications rather than sources with single- or multi-line
identifications.

\begin{figure*}[p]
\centering
\includegraphics[width=1.38\columnwidth]{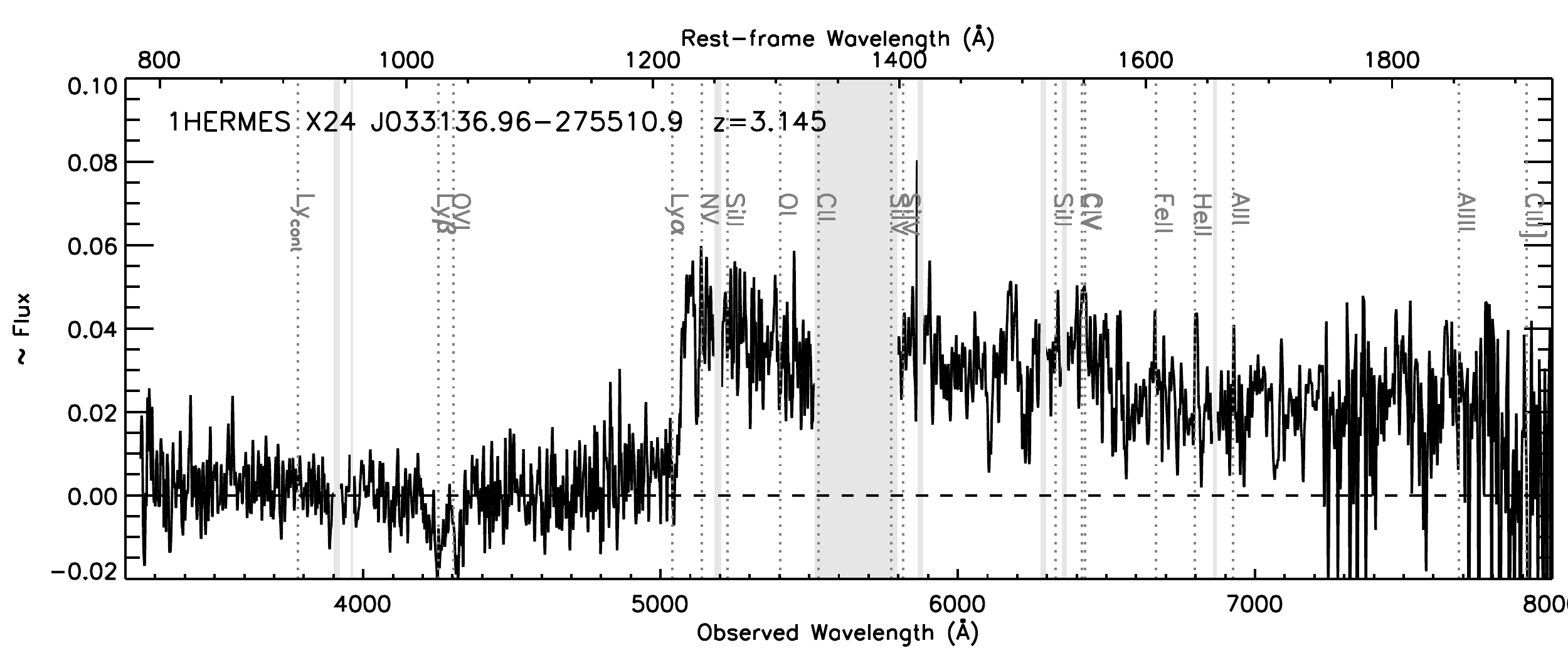}\includegraphics[width=0.58\columnwidth]{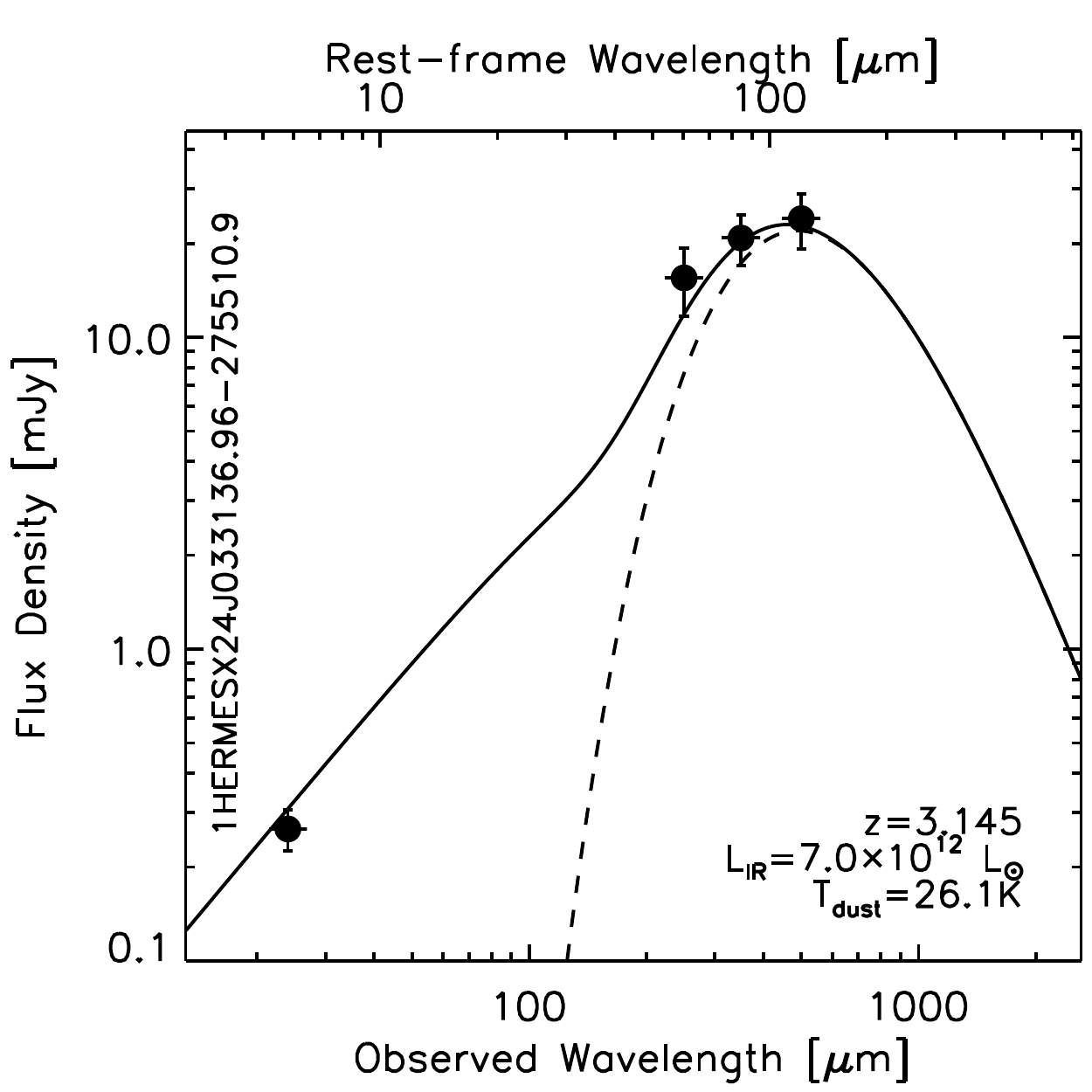}\\
\includegraphics[width=1.38\columnwidth]{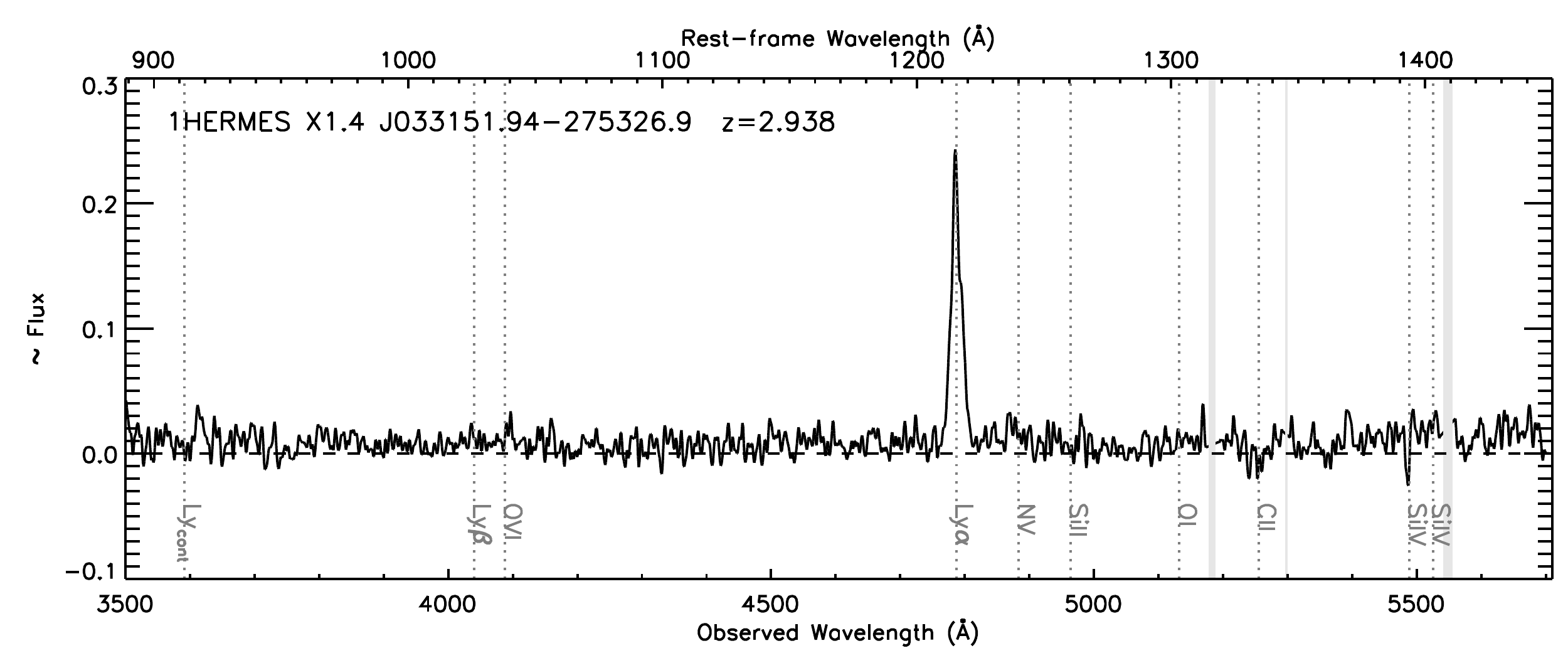}\includegraphics[width=0.58\columnwidth]{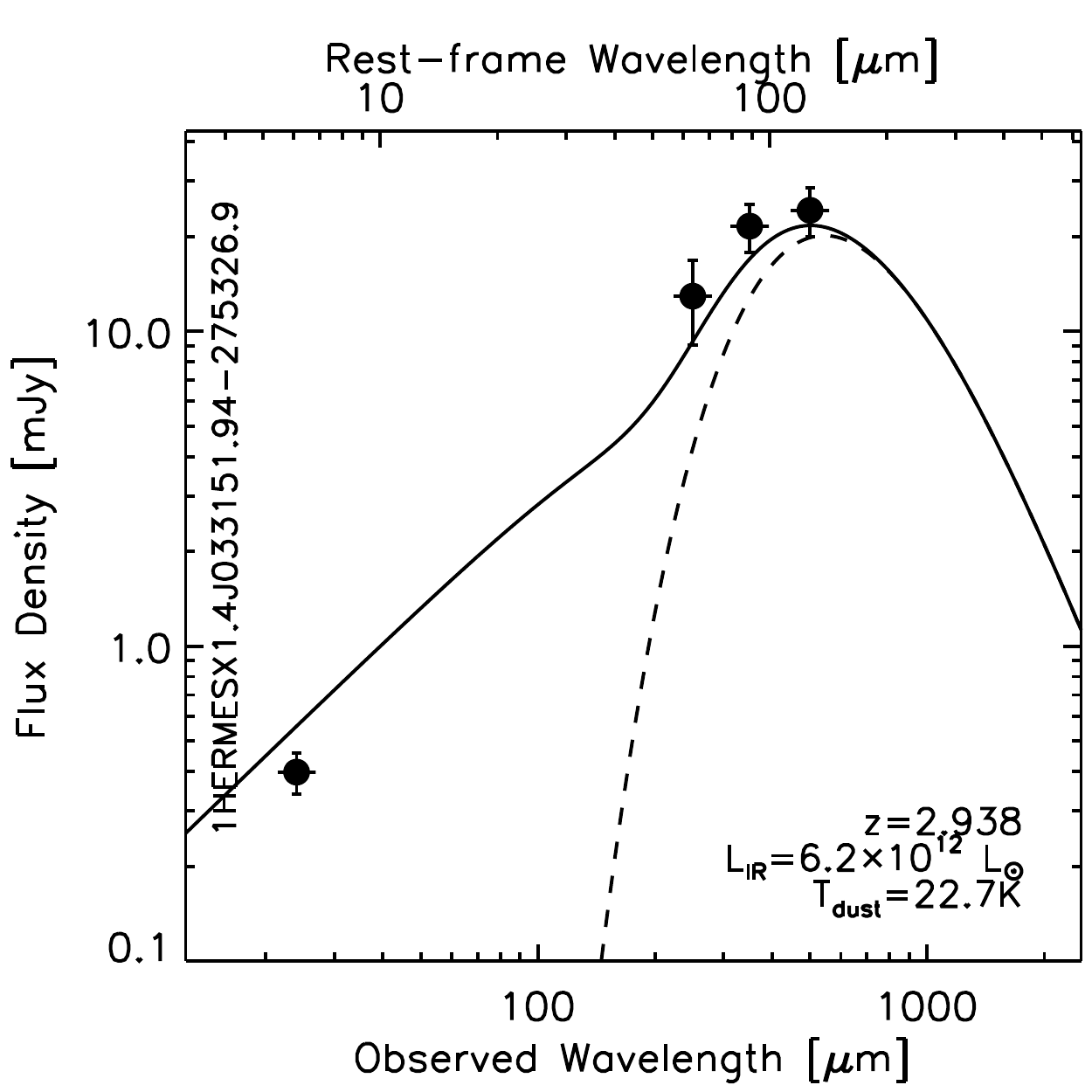}\\
\includegraphics[width=1.38\columnwidth]{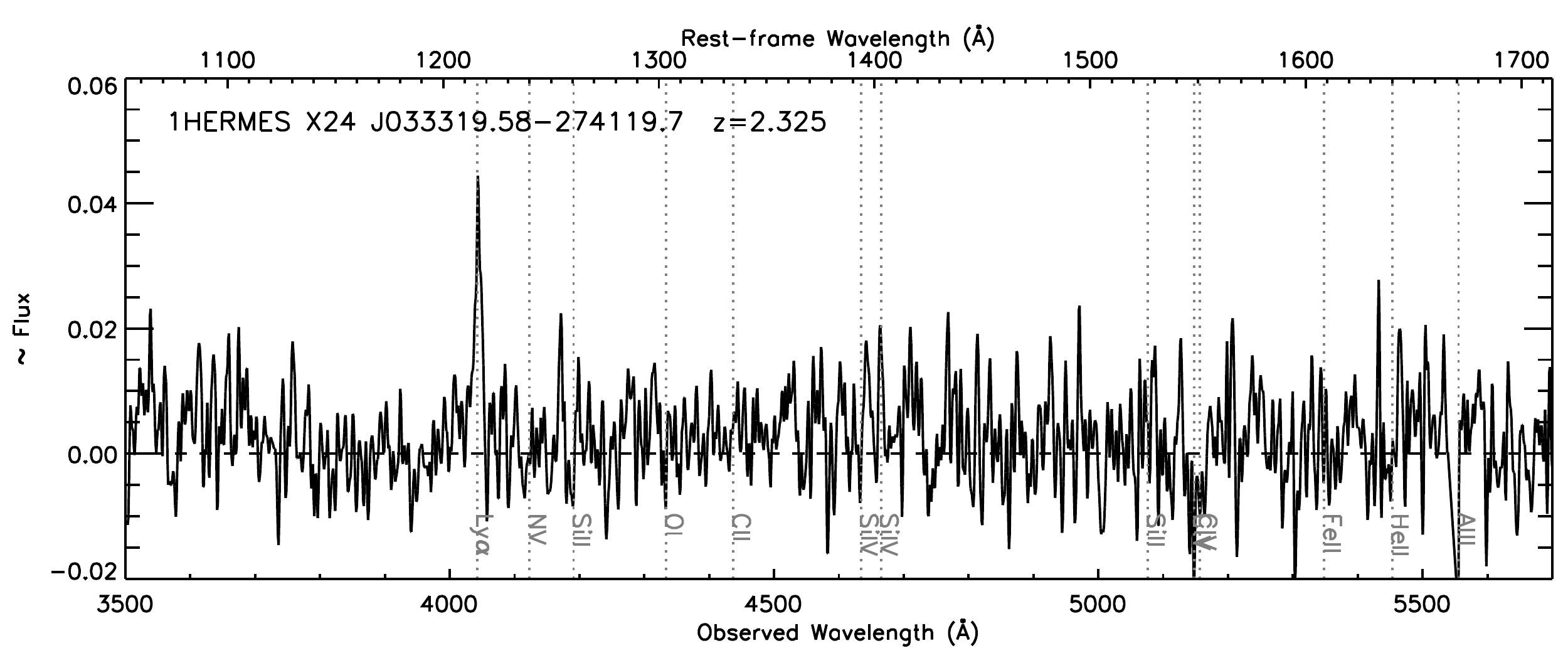}\includegraphics[width=0.58\columnwidth]{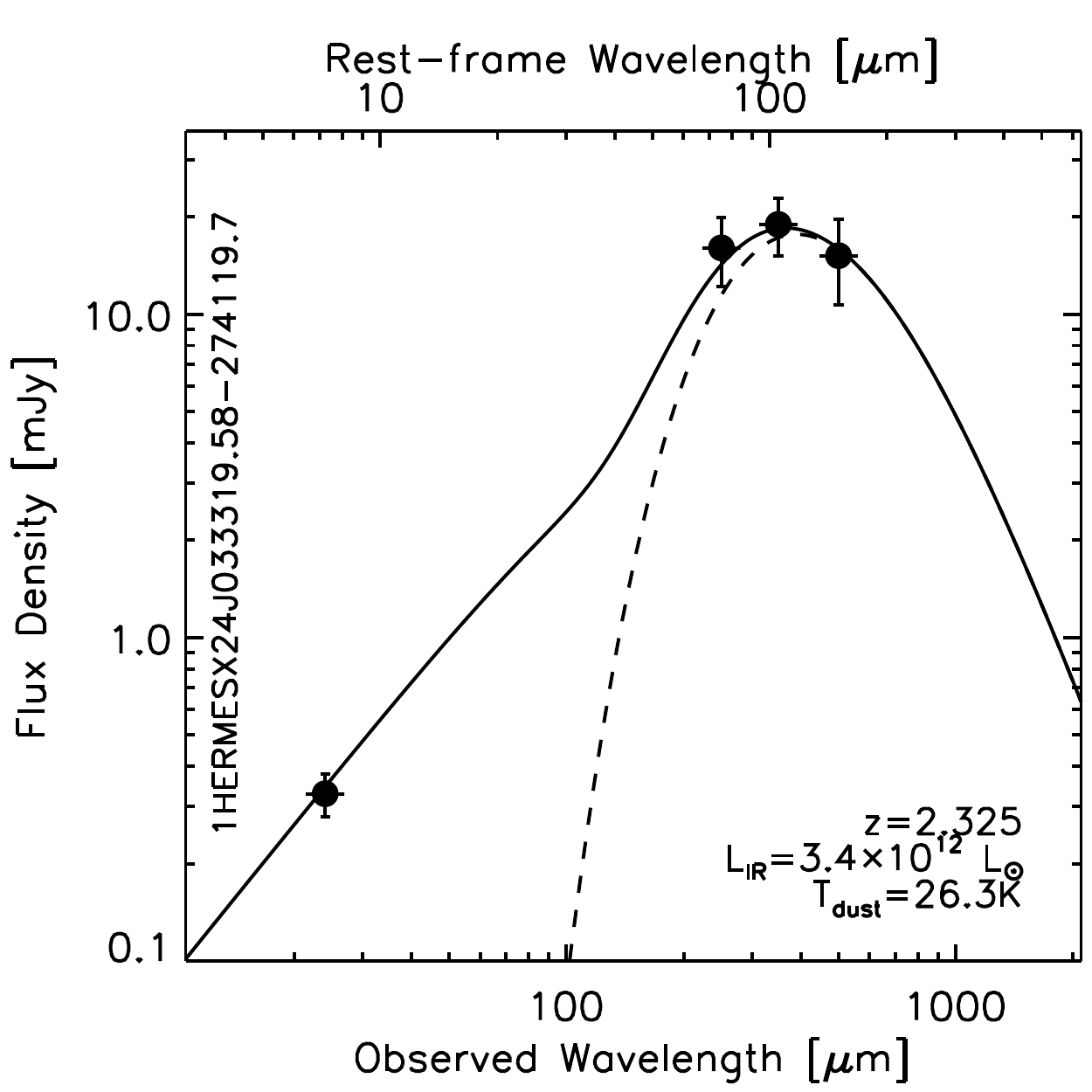}\\
\includegraphics[width=1.38\columnwidth]{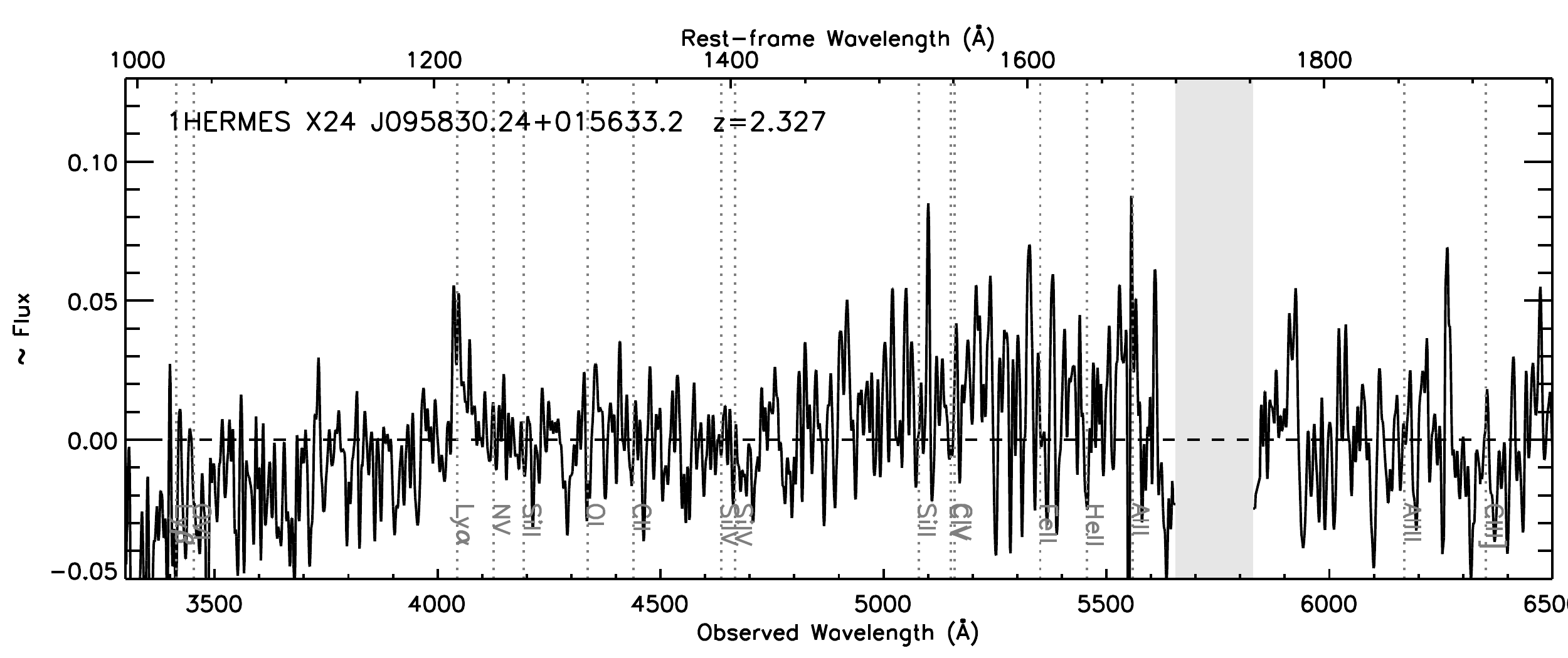}\includegraphics[width=0.58\columnwidth]{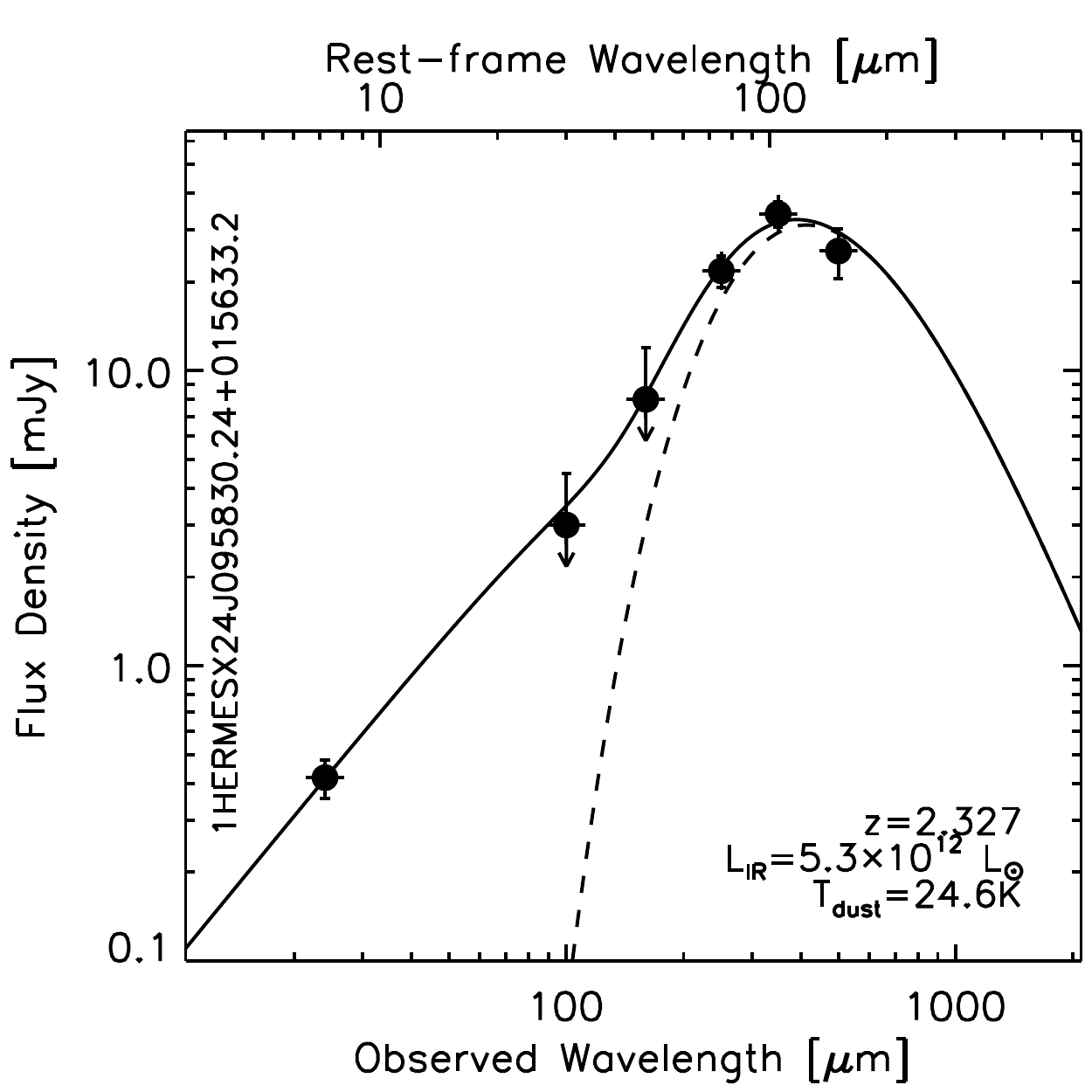}\\
\caption{One dimensional spectra of $z>$2 {\sc Spire} sources in
  observed wavelength shown in the left panels.  The flux scaling is
  arbitrary.  The wavelengths of rest-frame ultraviolet features are
  marked by vertical dotted gray lines with noted names.  All lines
  are shown irrespective of whether or not they are seen in emission,
  absorption, and whether or not they are detected.  Redshifts are
  measured off of the \lya\ redshift in all cases where \lya\ is
  detected (a minority of sources is identified by \civ\ emission).
  The \lya\ redshifts are typically redshifted with respect to other
  spectral features, evidence of stellar winds (to be discussed at
  more length in a future paper). Right panels show best-fit far-infrared
  spectral energy distributions to infrared photometry.  The total SED
  is a solid line while the underlying cold-dust modified blackbody is
  dashed.  The derived IR luminosities and dust temperatures are
  shown.}
\label{fig:spectra}
\end{figure*}

Sources are categorized as `tentative' rather than `secure' if (a) the
signal-to-noise of the \lya\ feature is $5<S/N<7$ as measured in the
two-dimensional LRIS/DEIMOS spectrum using an appropriately sized,
adjustable aperture, and if (b) out of five co-authors who did a
thorough quality ranking of each source's spectrum, at least three
ranked the source as `tentative' rather than `secure.'  
Figure~\ref{fig:spectra} show all rest-frame ultraviolet spectra for
all 36 sources in the sample.

Tentative sources are marked with a dagger ($\dagger$) in
Table~\ref{tab_fullsample} and denoted as `tentative' in
Figures~\ref{fig:spectra2d} and~\ref{fig:spectra}.  Out of 36 sources,
we classify 22 as secure and 14 as tentative.  
Note that one source, 1HERMES\,X1.4\,J100024.00+021210.9 at $z=3.553$
is classified as tentative $not$ because the \lya\ emission is of
low-S/N, but because there is peculiarly strong emission from
\siiv\ relative to \lya\ and \civ; this could be an artifact of the
noise, but since it is particularly unusual, we have categorized this
source as tentative.

Photometric redshifts are used, when available, to verify
spectroscopic redshifts to within $\Delta\,z$/$(1+z)$$\sim$\,1/2
\citep{ilbert10a,rowan-robinson08a,strazzullo10a,cardamone10a}.  This
threshold is not strict, but rather used to spot egregious
disagreements between spectroscopy and photometry. These photometric
redshift catalogs are of varying quality depending on the depth and
number of optical and near-infrared bands available, from the limited
and shallow coverage in EN1 to the extensive 30-band coverage in
COSMOS.  The loose constraint of agreement between photometric and
spectroscopic redshift is based on the fact that photometric redshifts
are notoriously unreliable for dusty starbursts. Even in the deepest
multi-band fields like COSMOS, $\sim$30\%\ of {\sc Spire} sources do
not have photometric redshifts, \citep[e.g.][]{mobasher07a,ilbert10a}.

Viewing geometry can have a very strong impact on the relative
fractions and wavelengths of escaped ultraviolet/optical light
\citep{siana07a,siana10a,scarlata09a}.  Heavy obscuration can have
dramatic impact on photometric redshift reliability in two ways.
First, by obscuring rest-frame ultraviolet and optical light so much
that the source has an artificially high photometric redshift
(i.e. the source is thought to drop out due to redshifting rather than
reddening).  Second, dusty sources often have differential absorption
of \lya\ photons and non-resonantly scattered continuum photons
\citep[e.g. the ``UV chimney'' argument, see][]{neufeld91a}, resulting
in artificially low photometric redshifts.

Even in the lower redshift samples (C12), sources with very confident
photometric redshifts from the multi-band data of COSMOS and CDFS have
photometric redshift accuracy of $\Delta\,z/(1+z)$=0.29.  This
highlights the lack of reliability in photometric redshifts of dusty
galaxies.  Therefore, the primary purpose of the photometric redshift
restriction in this paper is to verify that the detected emission line
is in fact \lya\ and not \oii\ (which would correspond to redshifts
between 0 and 1).  
\clearpage
\begin{center}
\includegraphics[width=0.85\columnwidth]{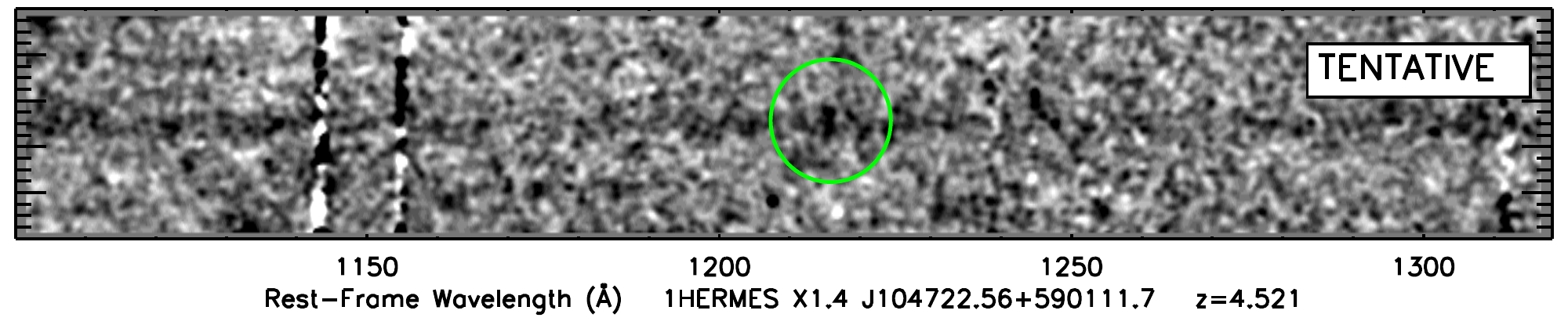}\includegraphics[width=0.85\columnwidth]{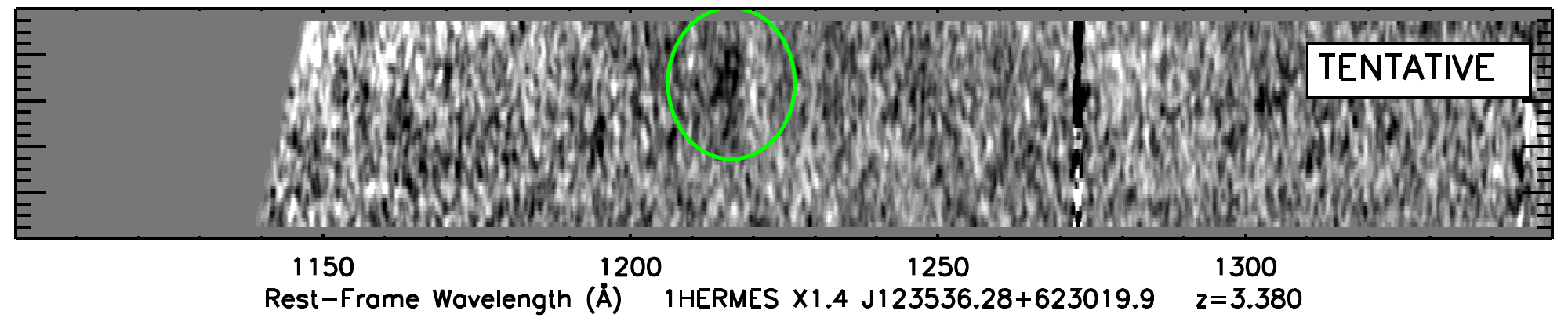}\\
\includegraphics[width=0.85\columnwidth]{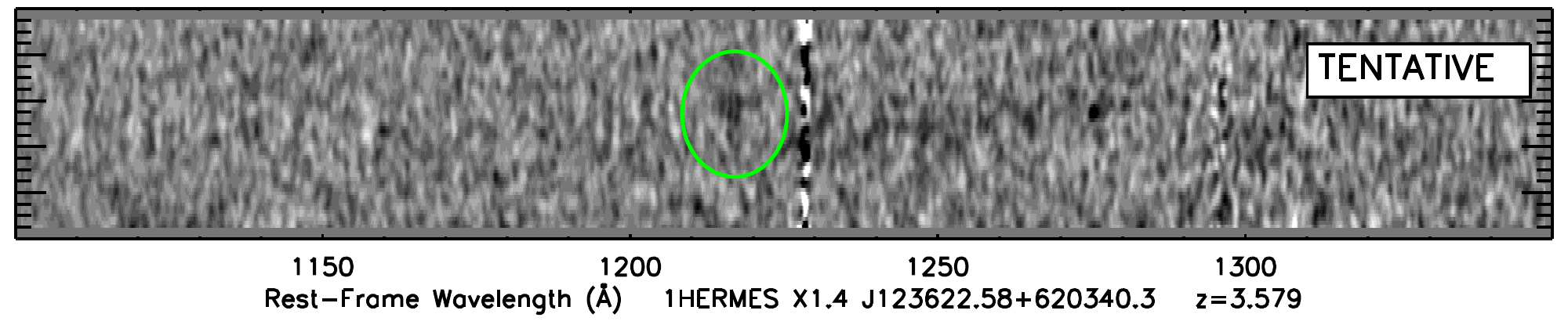}\includegraphics[width=0.85\columnwidth]{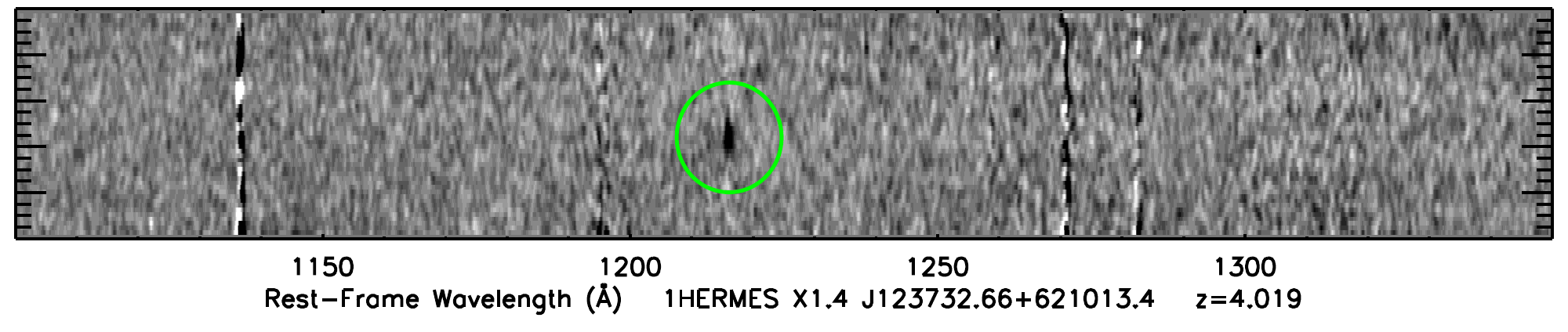}\\
\includegraphics[width=0.85\columnwidth]{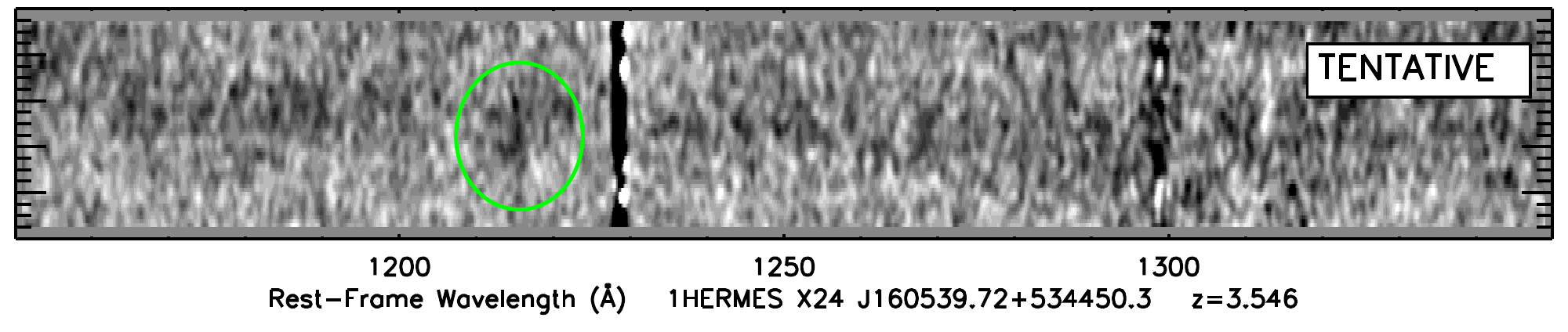}\includegraphics[width=0.85\columnwidth]{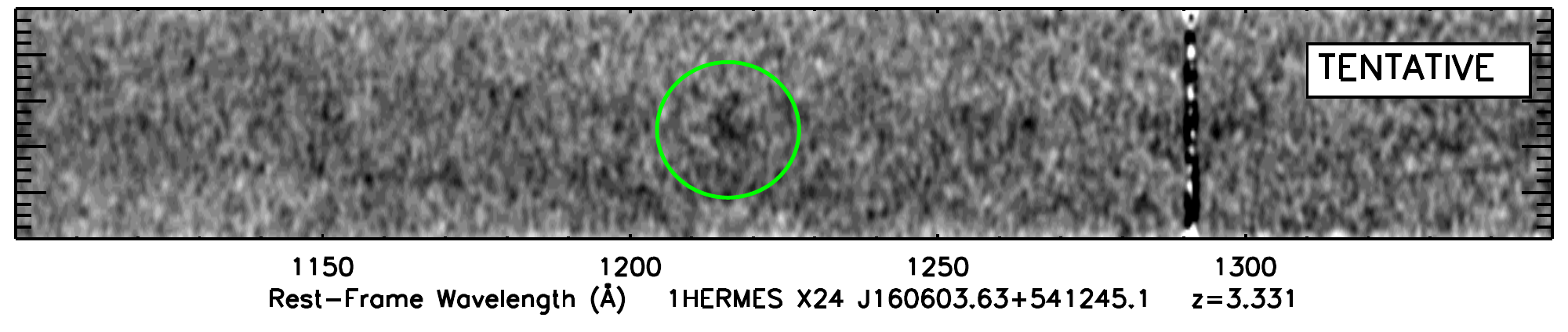}\\
\includegraphics[width=0.85\columnwidth]{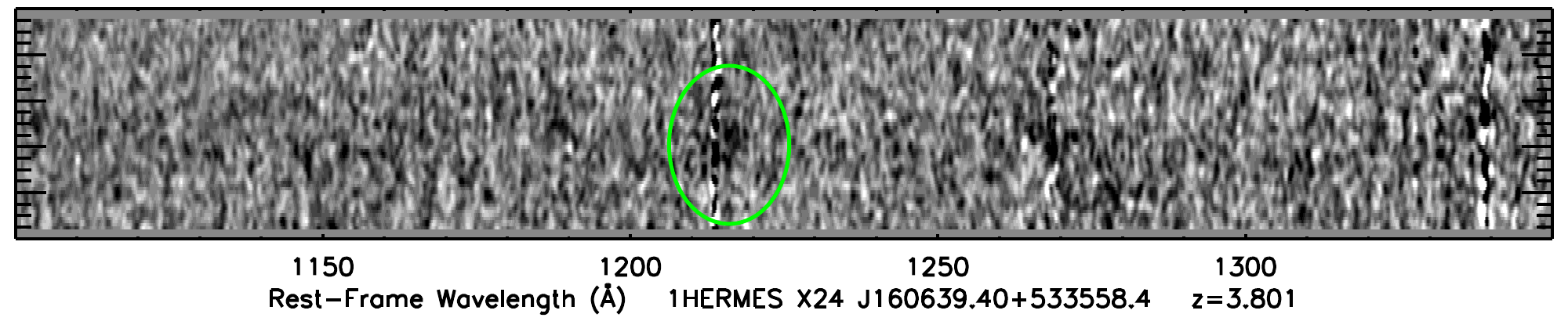}\includegraphics[width=0.85\columnwidth]{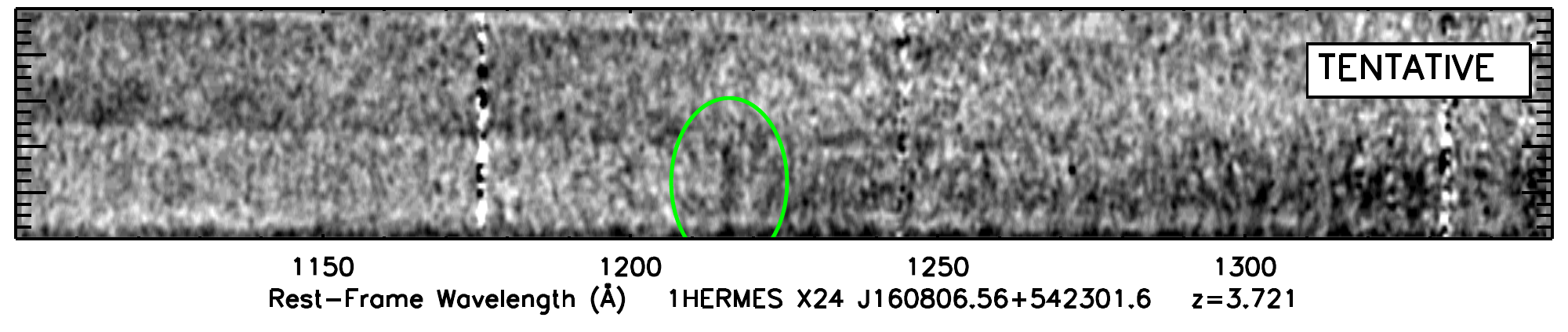}\\
\centerline{{\small Figure~\ref{fig:spectra2d} --- continued.}}
\end{center}

\clearpage
\begin{center}
\includegraphics[width=1.38\columnwidth]{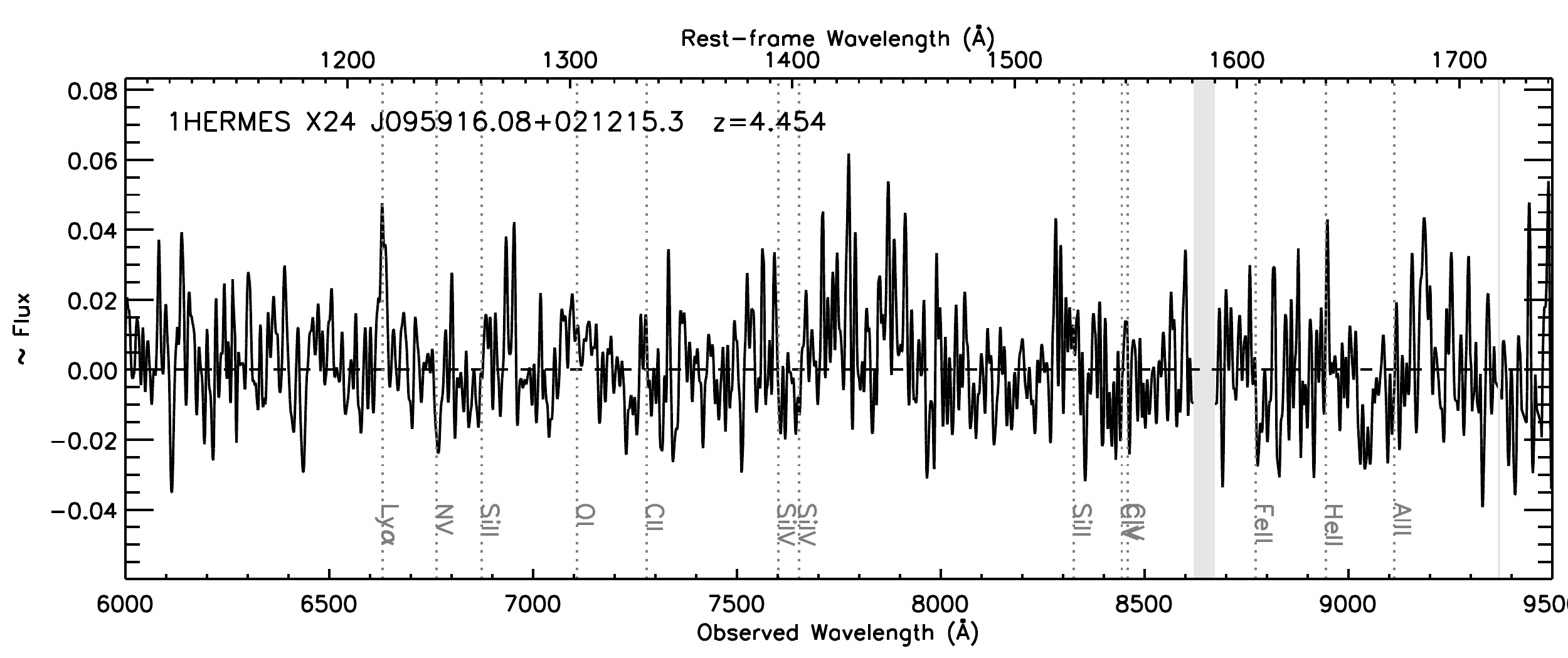}\includegraphics[width=0.58\columnwidth]{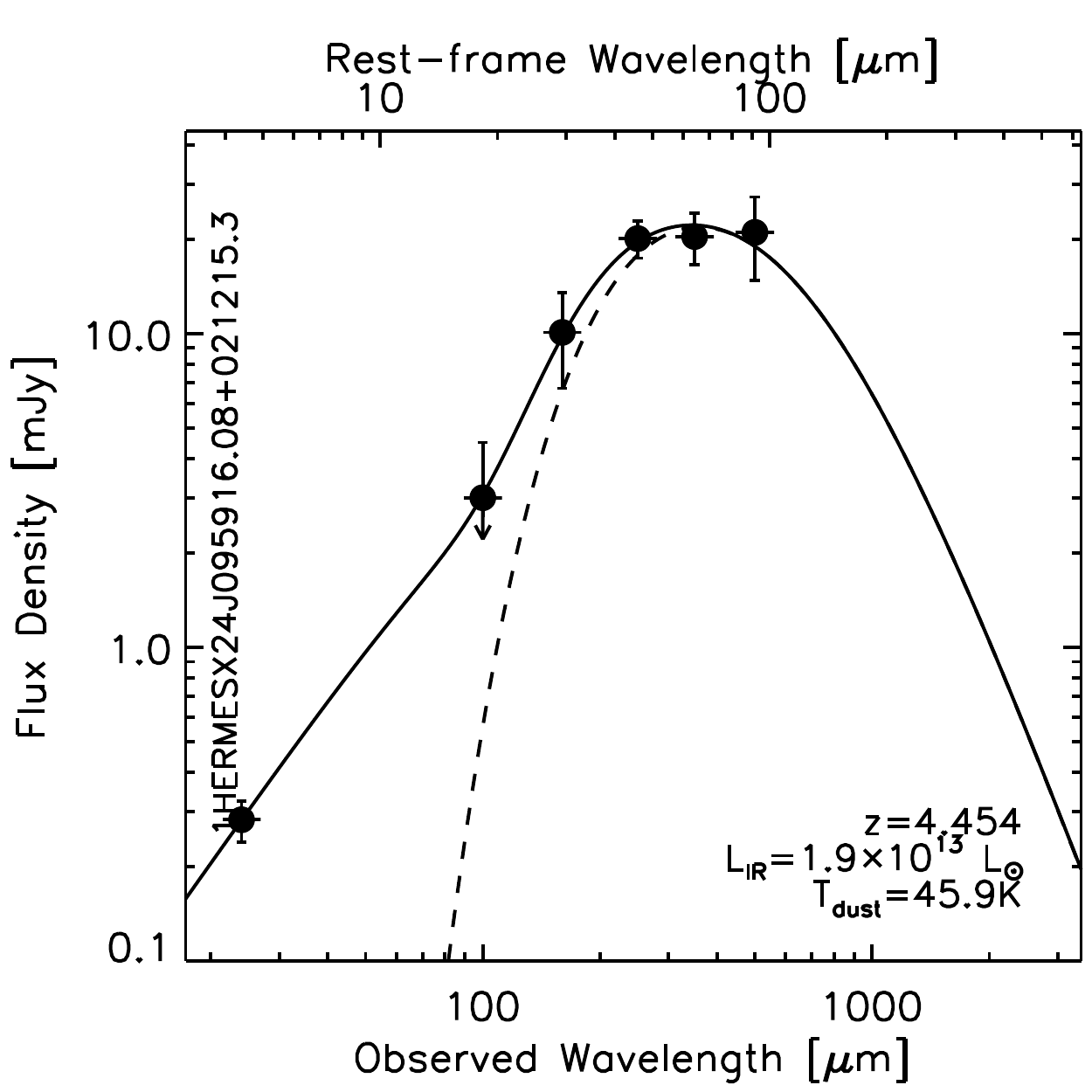}\\
\includegraphics[width=1.38\columnwidth]{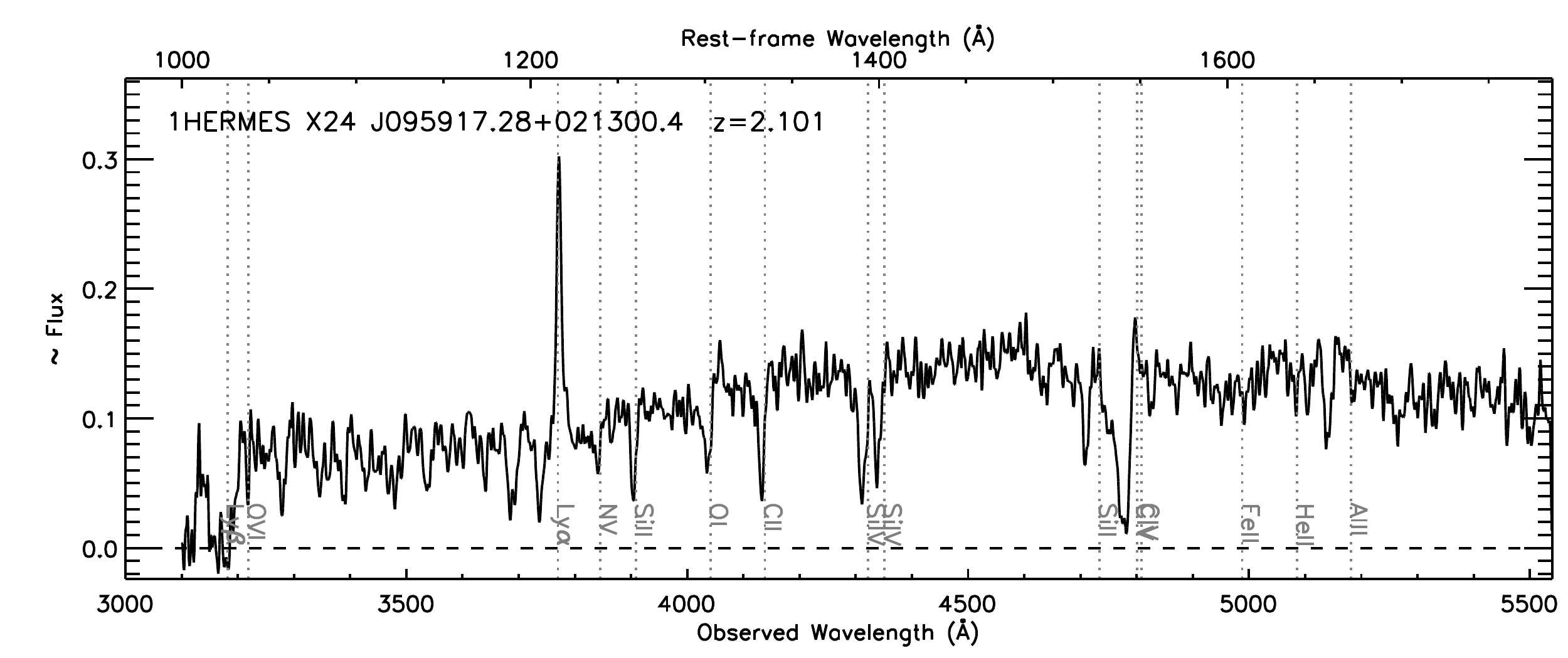}\includegraphics[width=0.58\columnwidth]{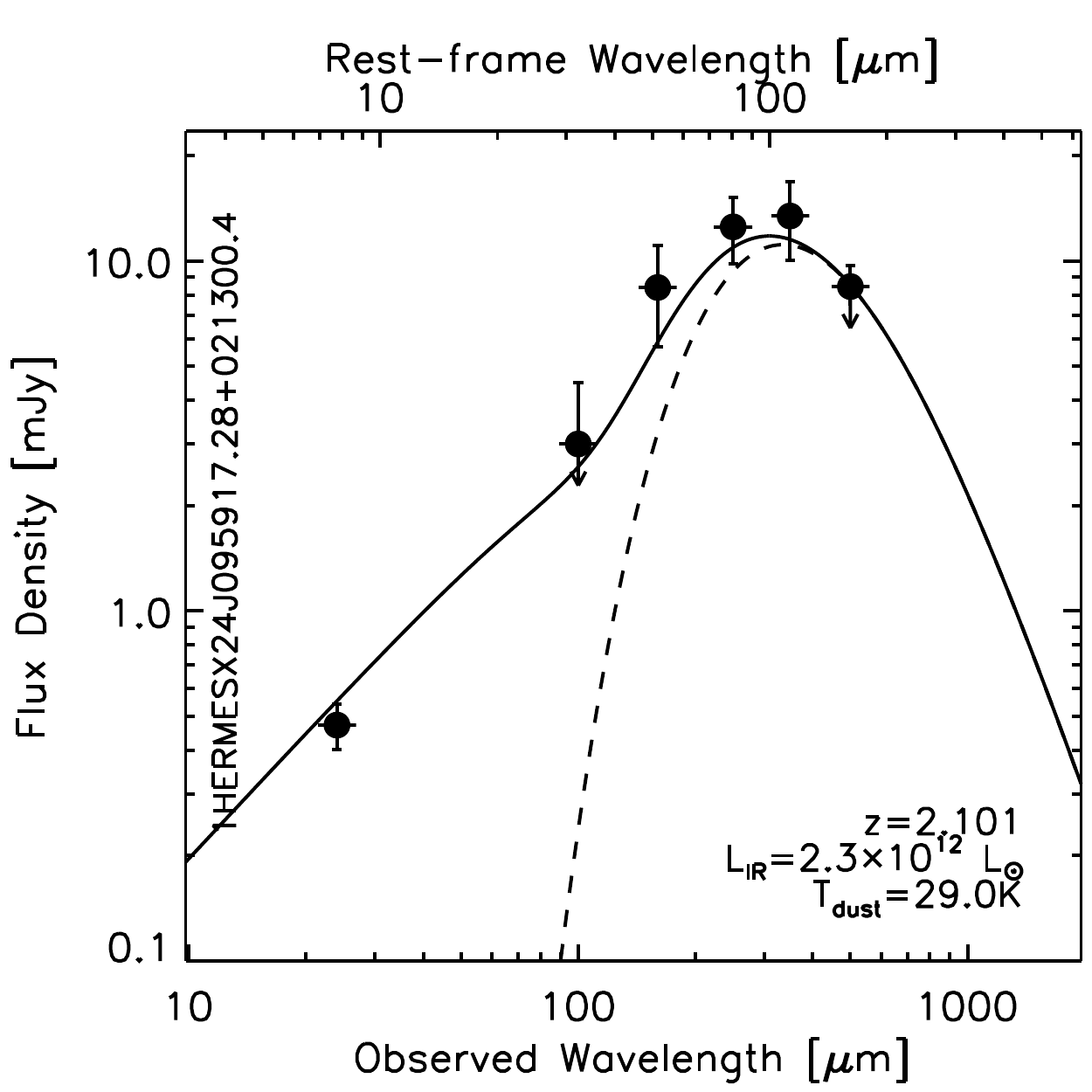}\\
\includegraphics[width=1.38\columnwidth]{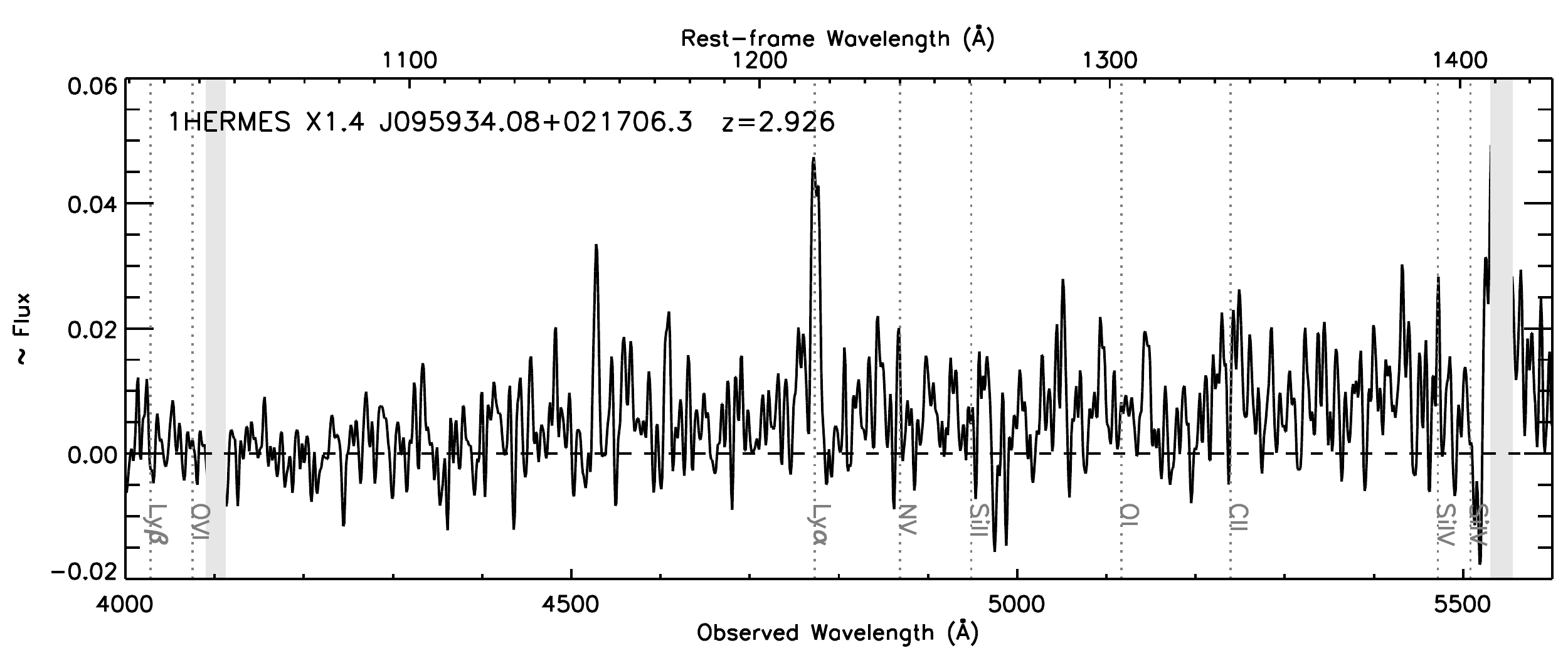}\includegraphics[width=0.58\columnwidth]{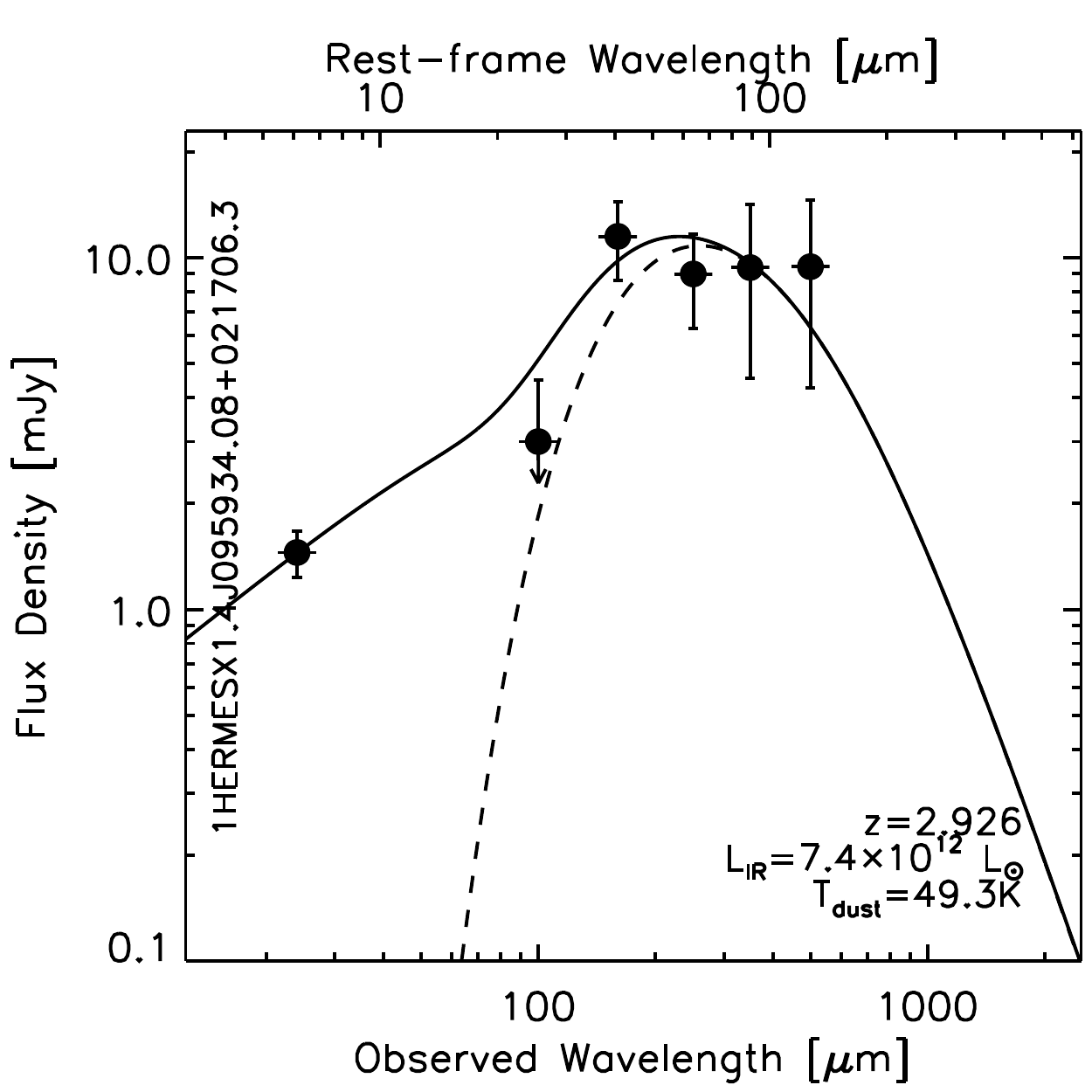}\\
\includegraphics[width=1.38\columnwidth]{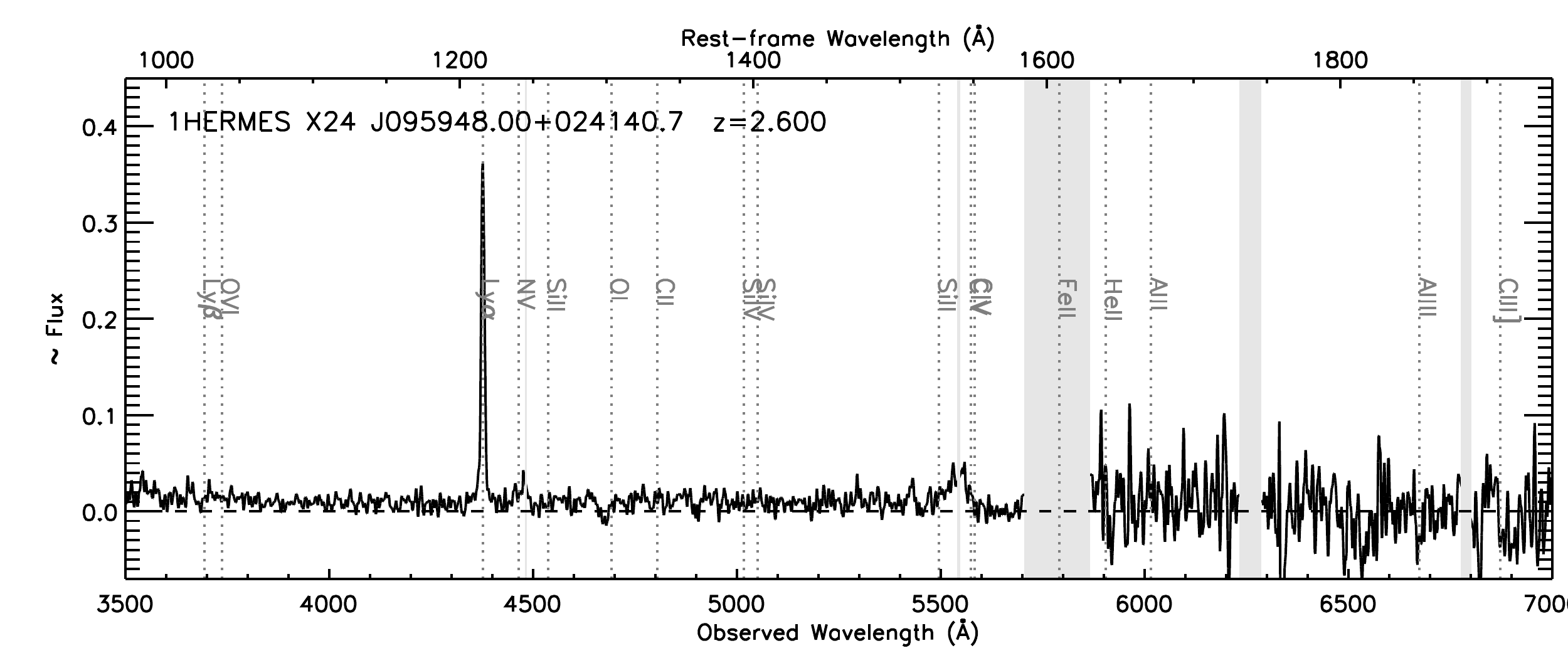}\includegraphics[width=0.58\columnwidth]{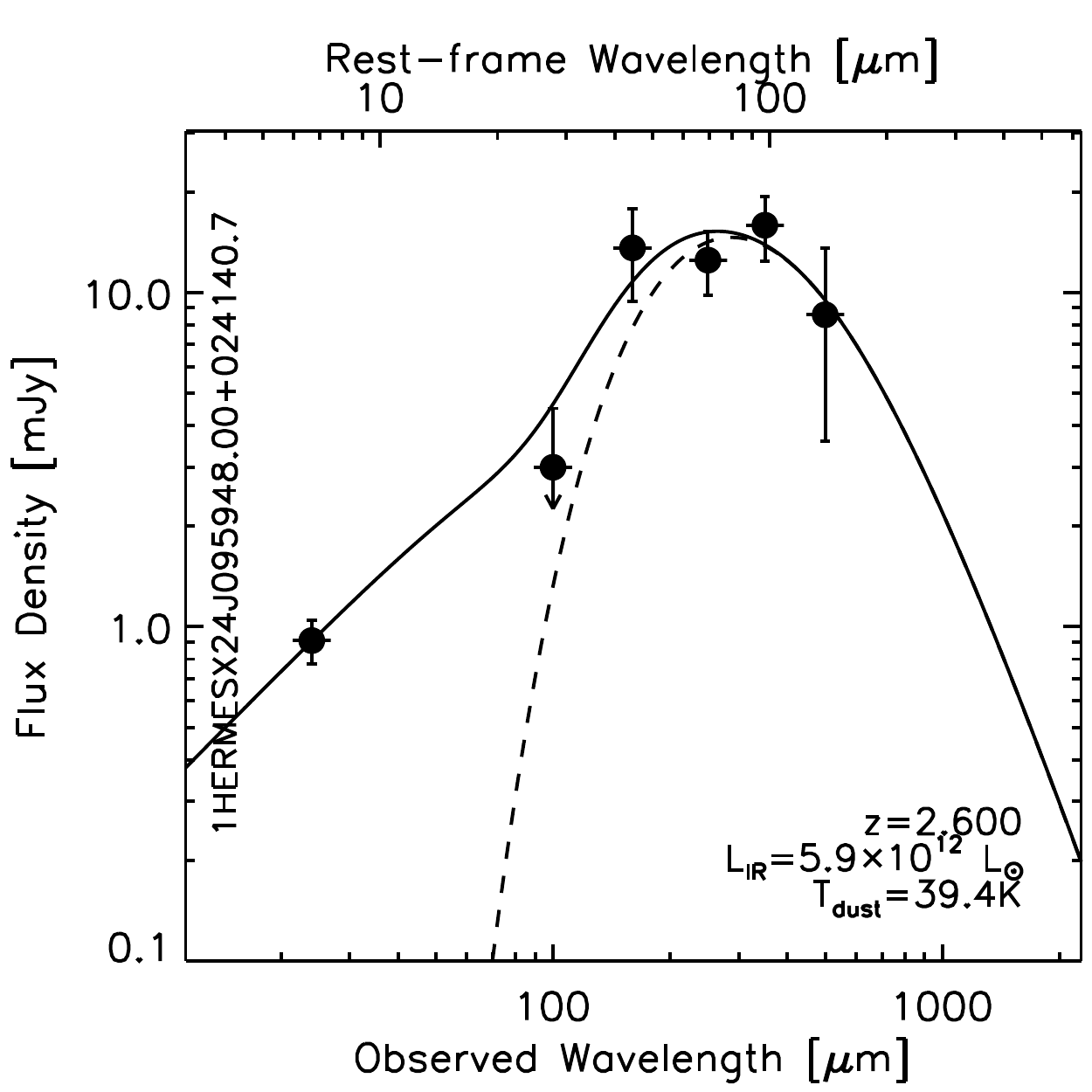}\\
\centerline{{\small Figure~\ref{fig:spectra} --- continued.}}
\end{center}
\clearpage
\begin{center}
\includegraphics[width=1.38\columnwidth]{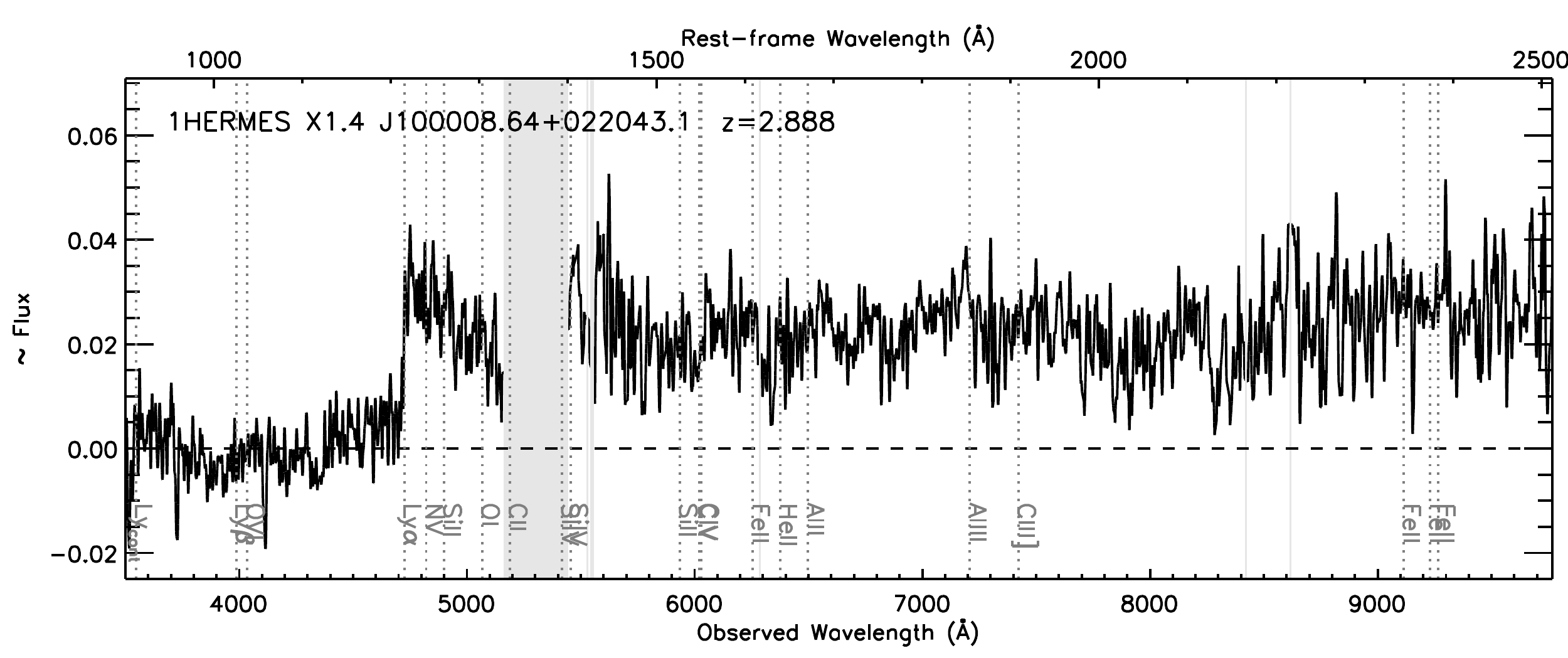}\includegraphics[width=0.58\columnwidth]{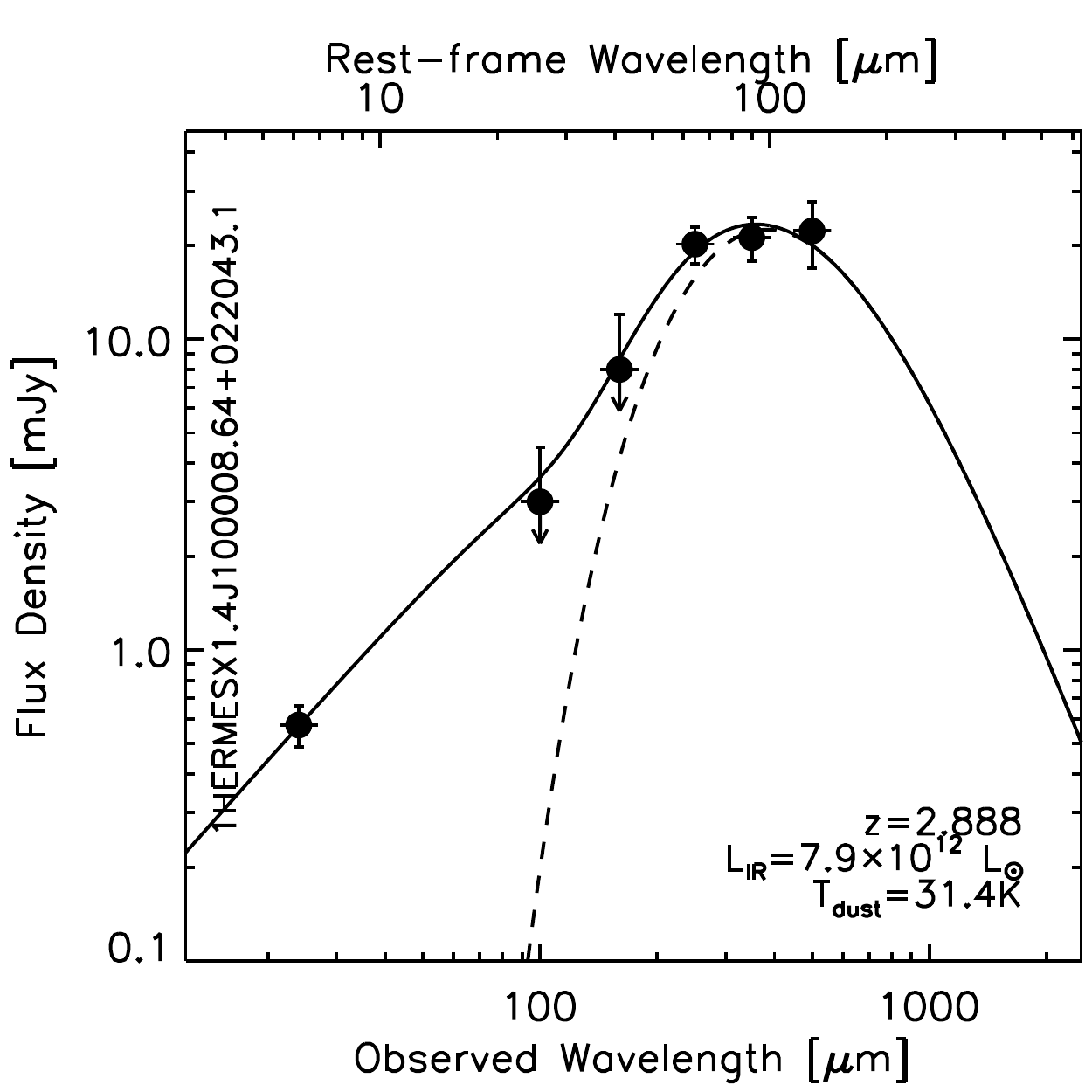}\\
\includegraphics[width=1.38\columnwidth]{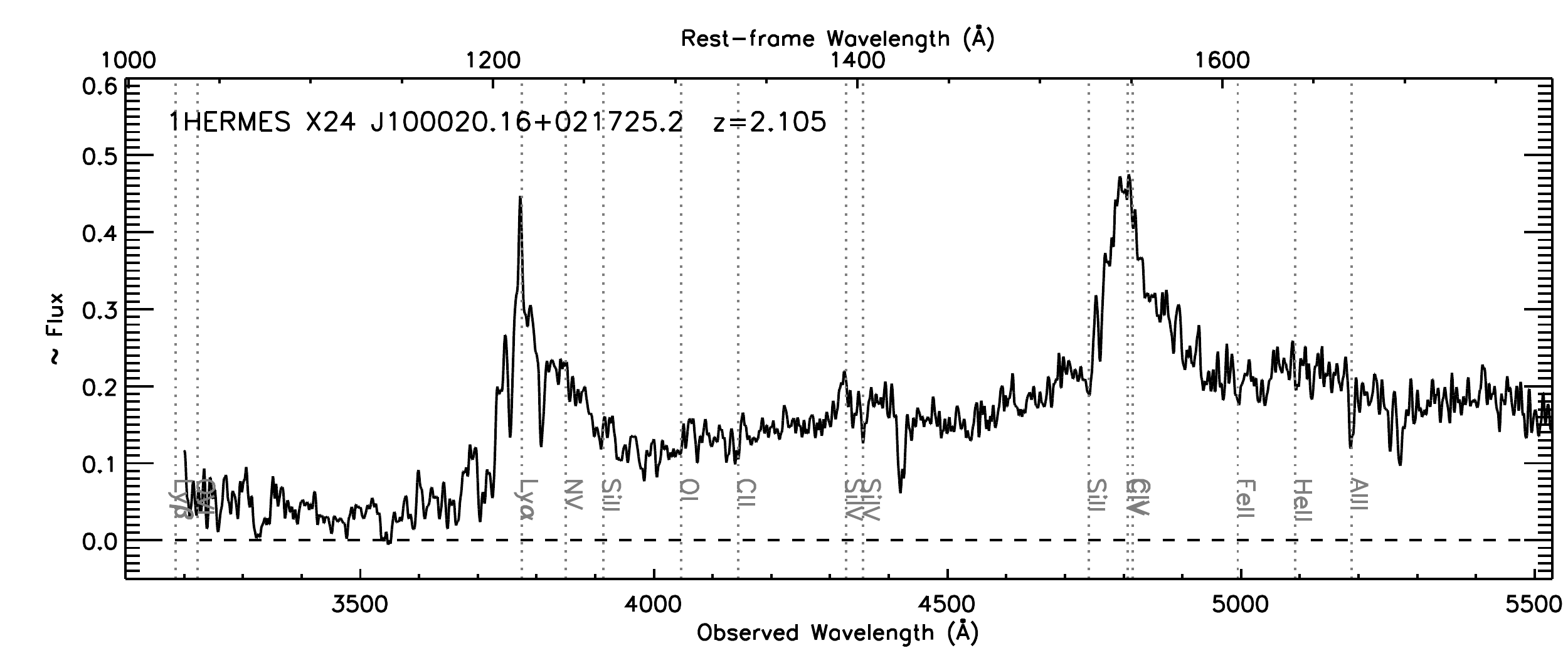}\includegraphics[width=0.58\columnwidth]{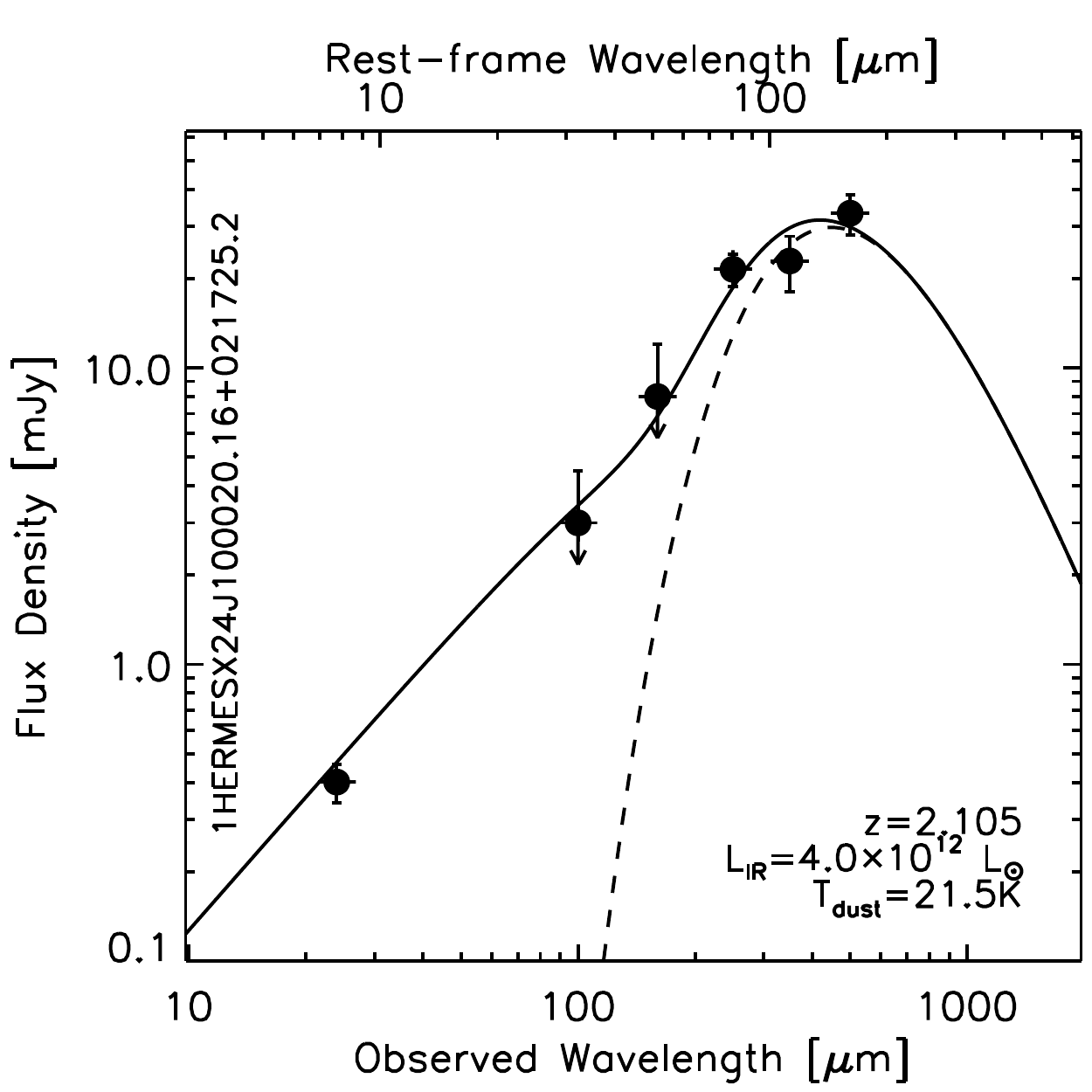}\\
\includegraphics[width=1.38\columnwidth]{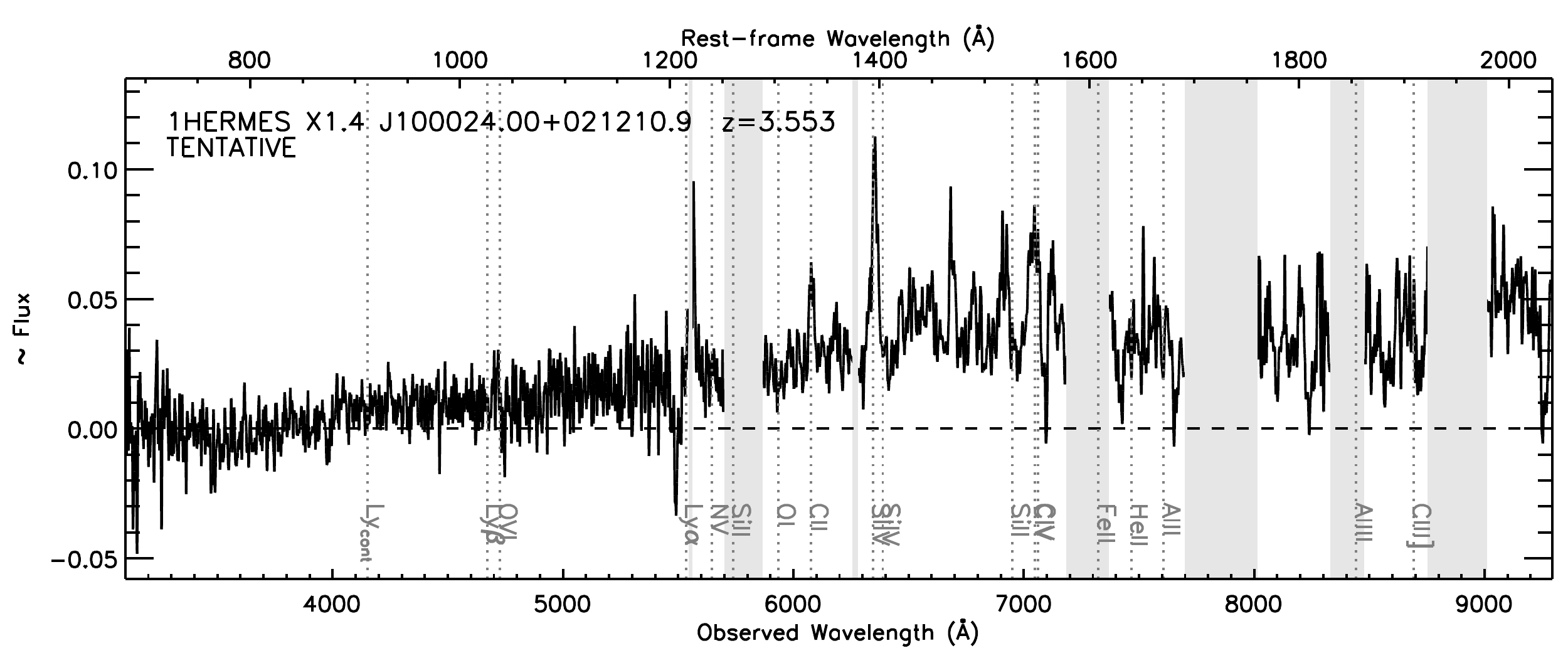}\includegraphics[width=0.58\columnwidth]{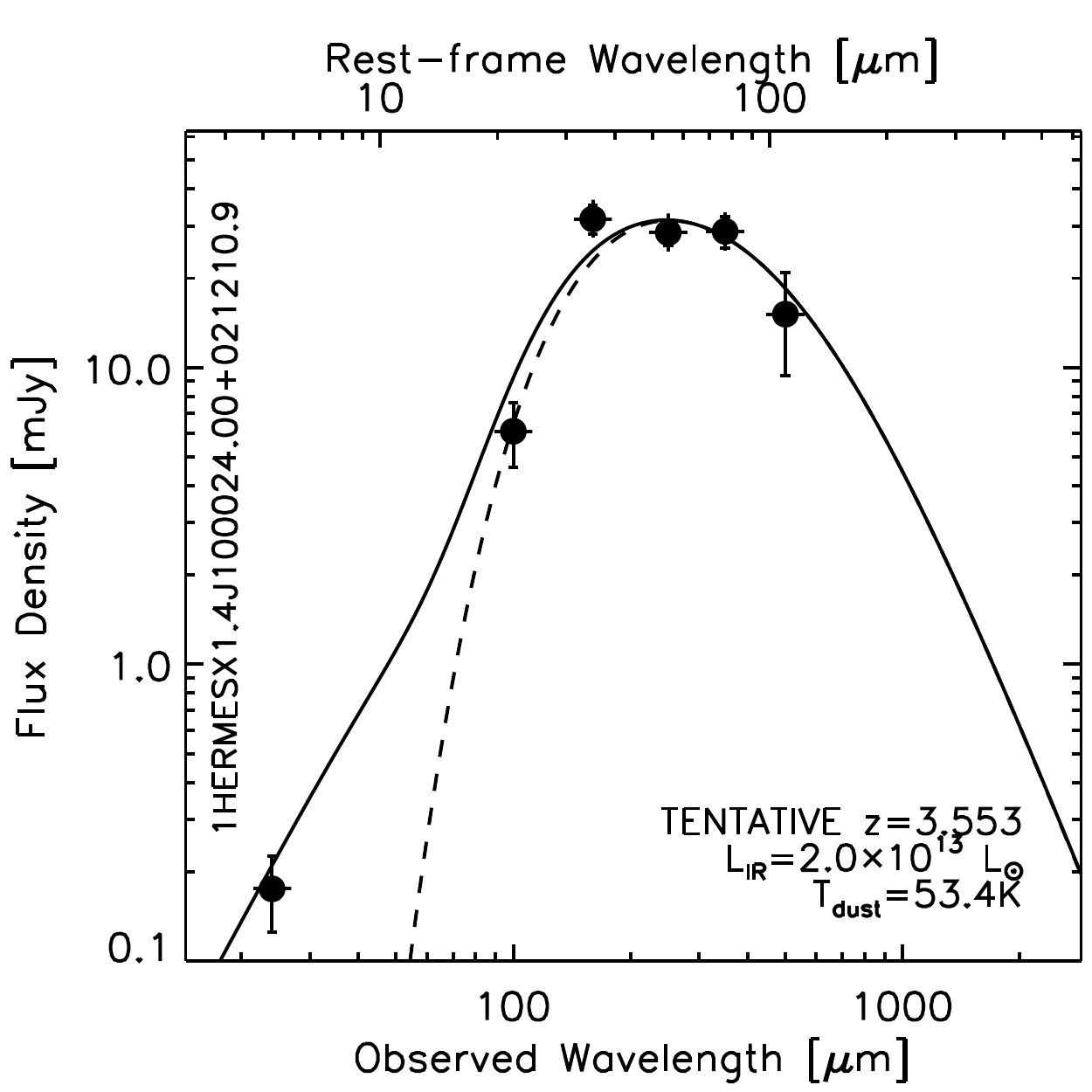}\\
\includegraphics[width=1.38\columnwidth]{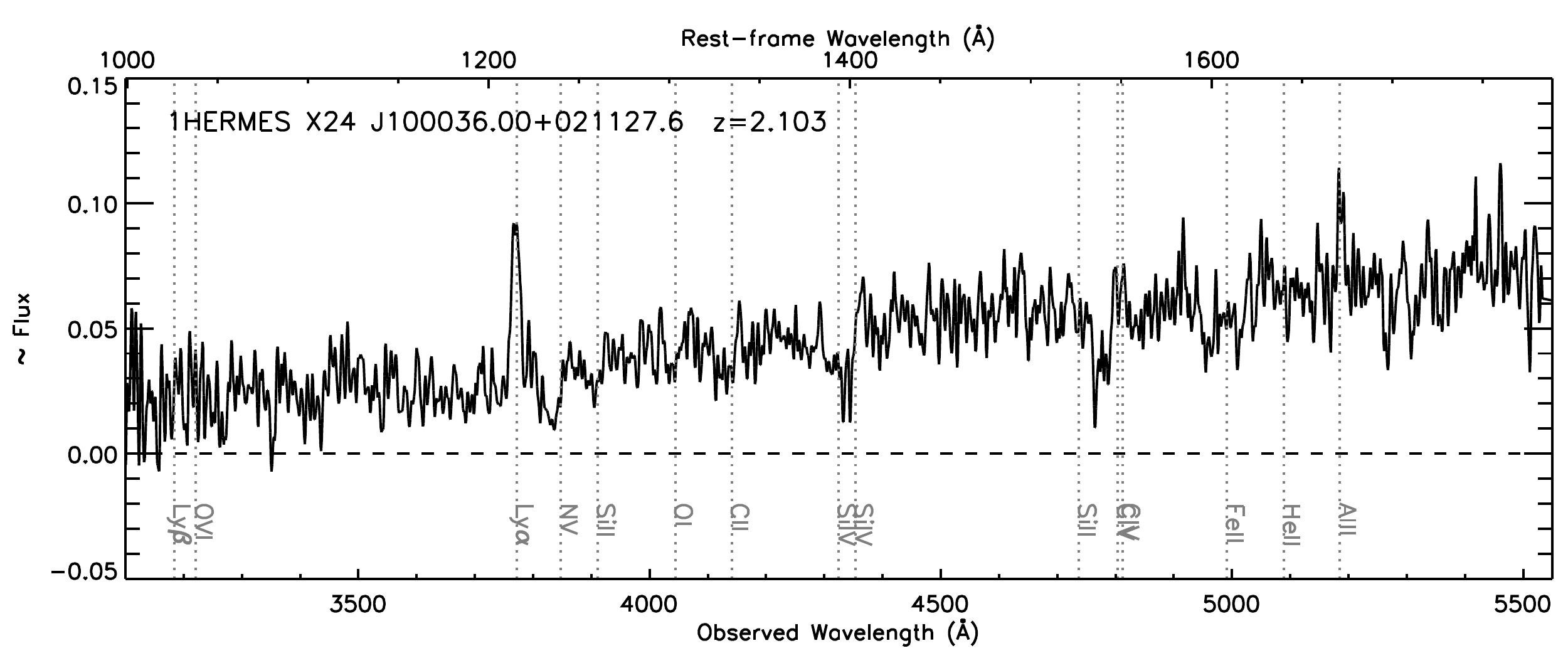}\includegraphics[width=0.58\columnwidth]{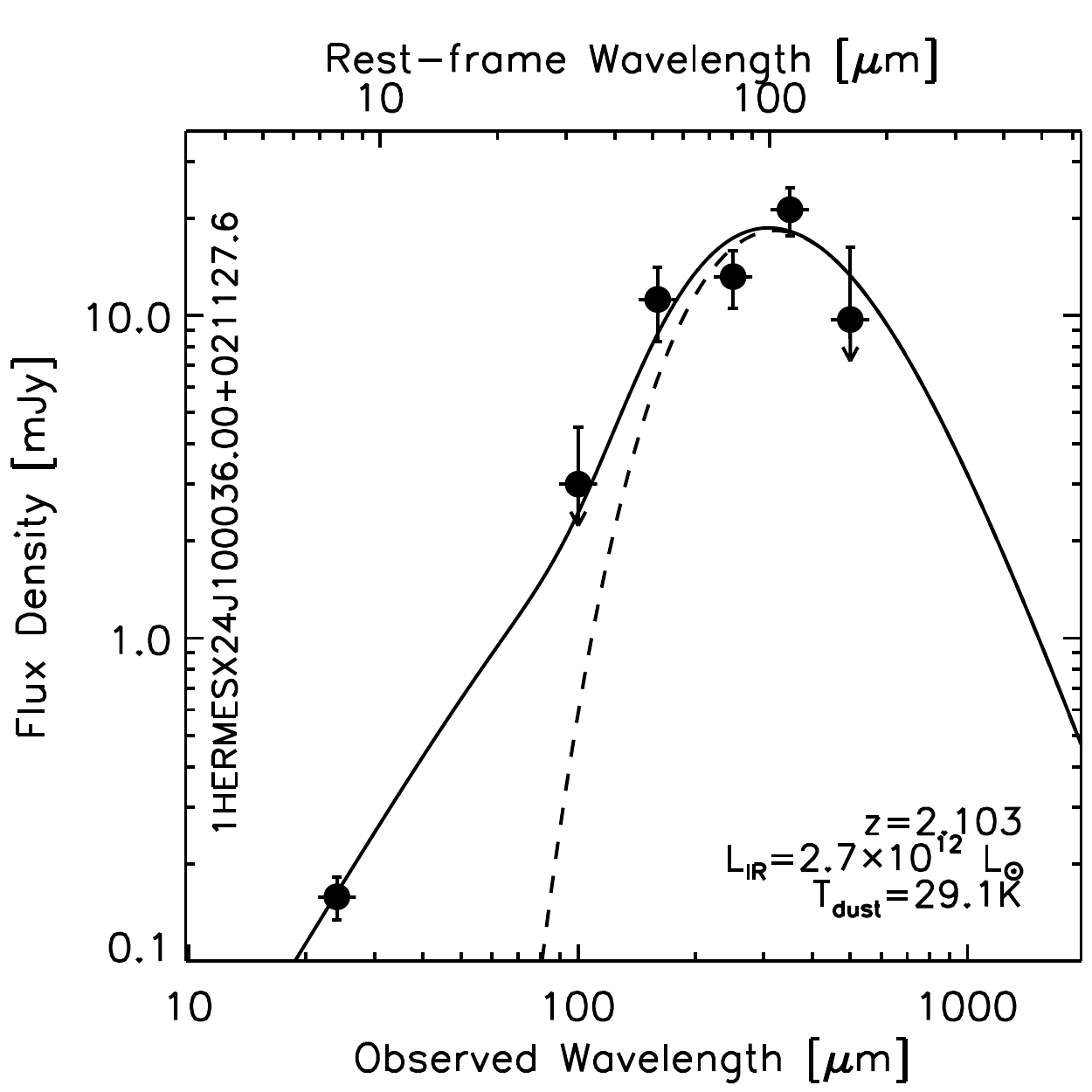}\\
\centerline{{\small Figure~\ref{fig:spectra} --- continued.}}
\end{center}
\clearpage
\begin{center}
\includegraphics[width=1.38\columnwidth]{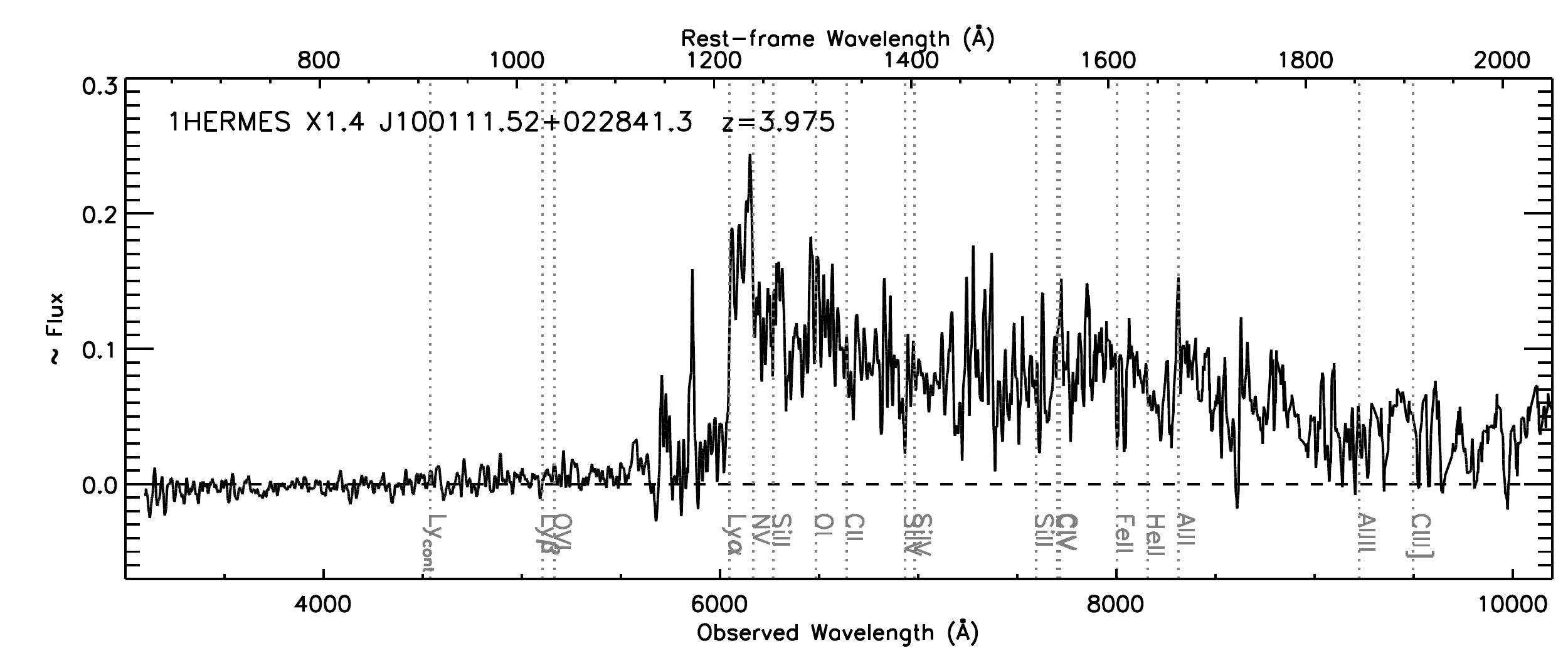}\includegraphics[width=0.58\columnwidth]{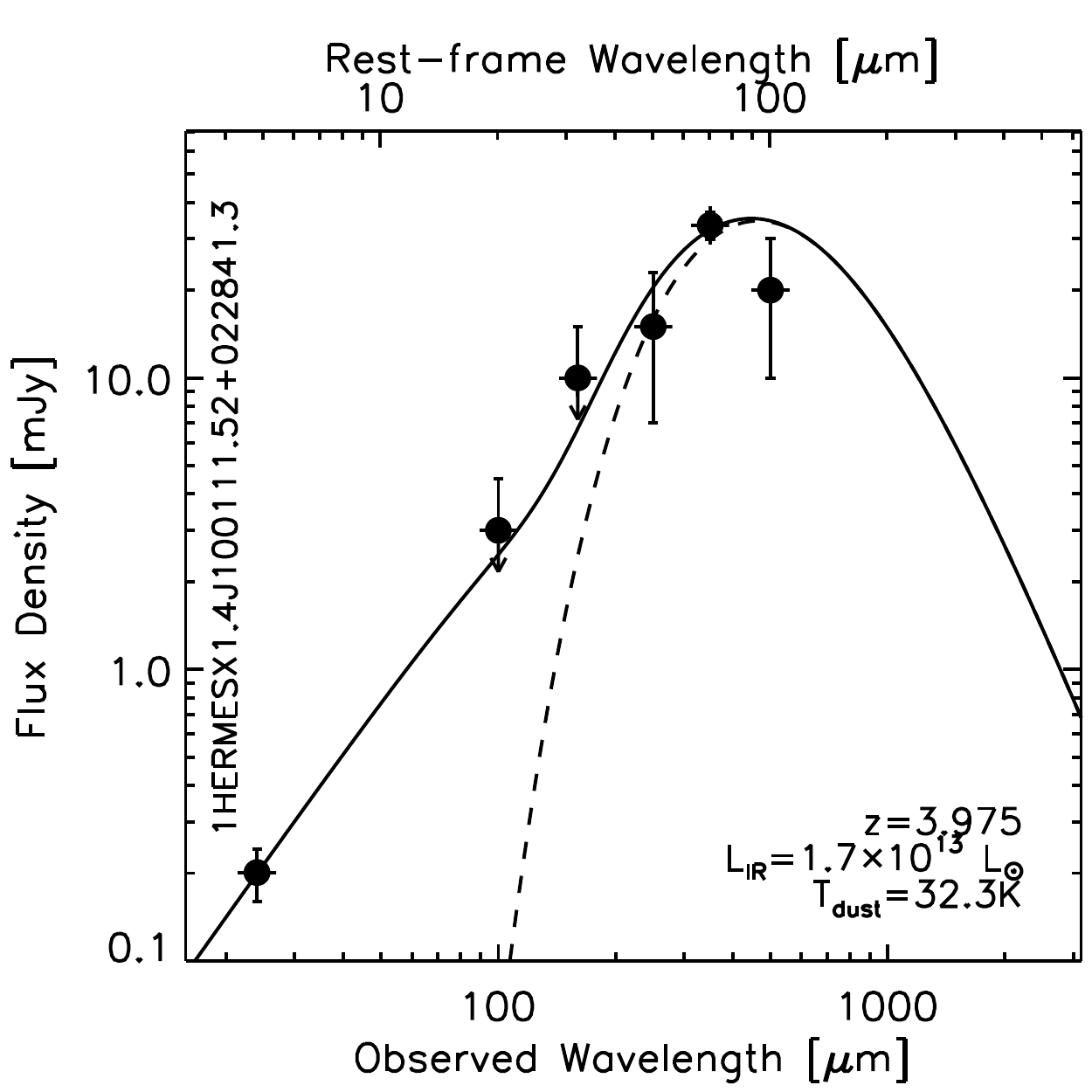}\\
\includegraphics[width=1.38\columnwidth]{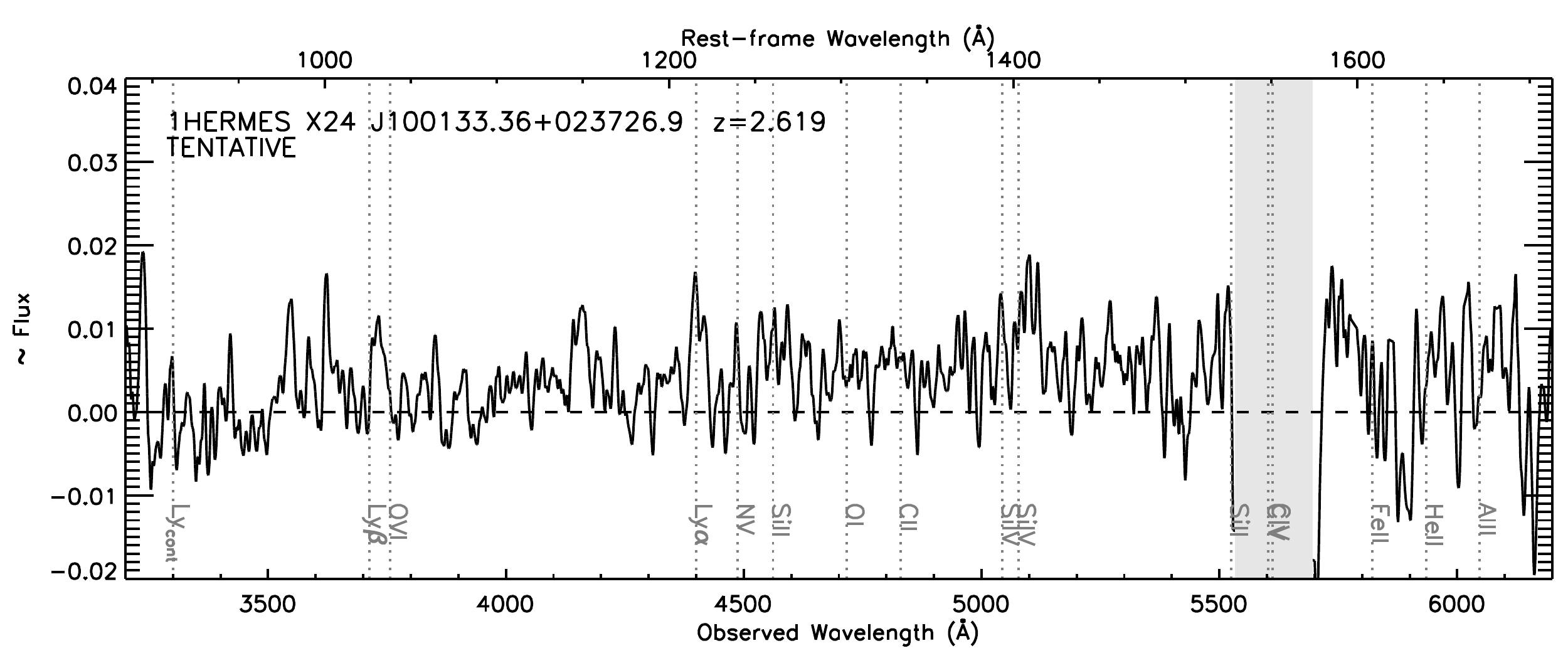}\includegraphics[width=0.58\columnwidth]{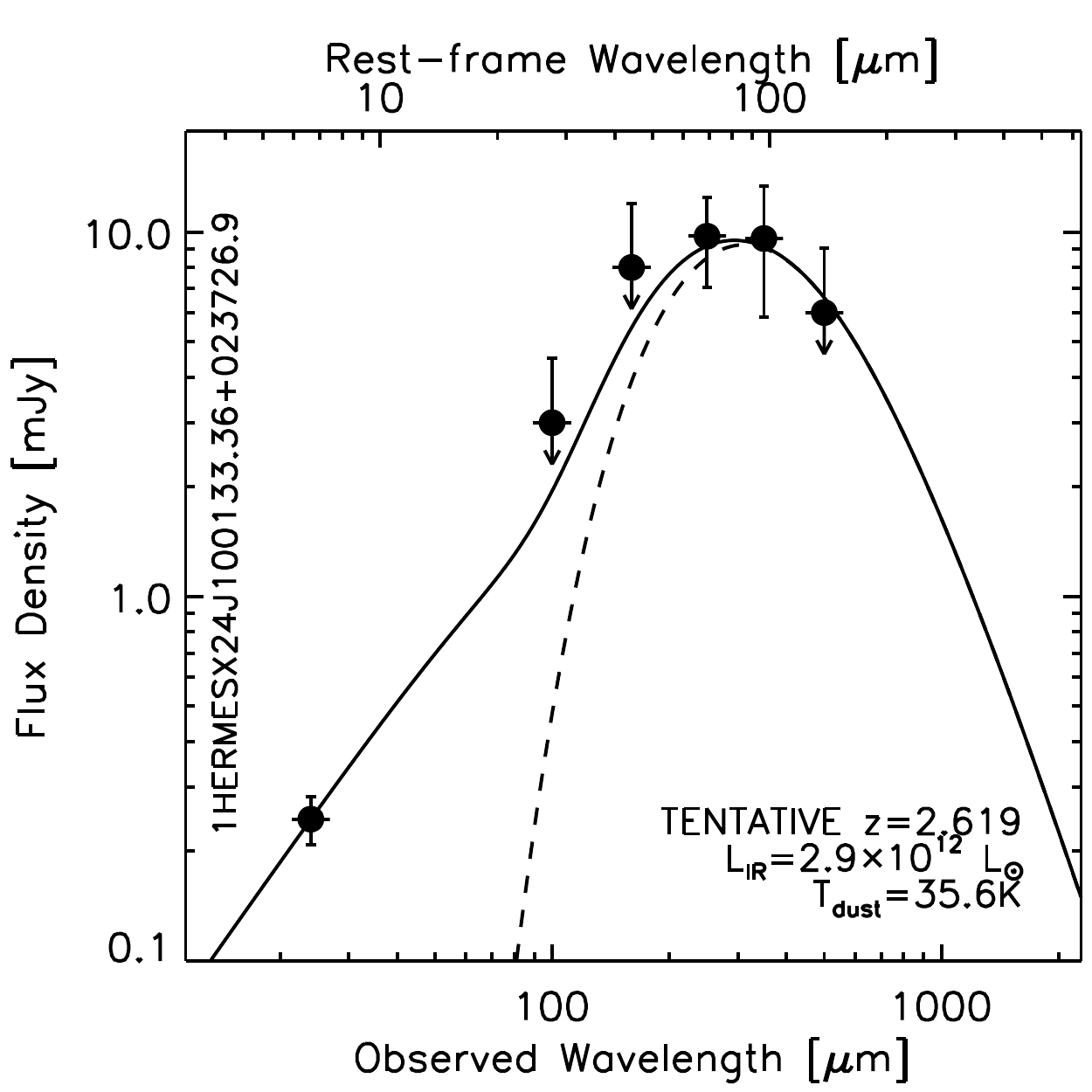}\\
\includegraphics[width=1.38\columnwidth]{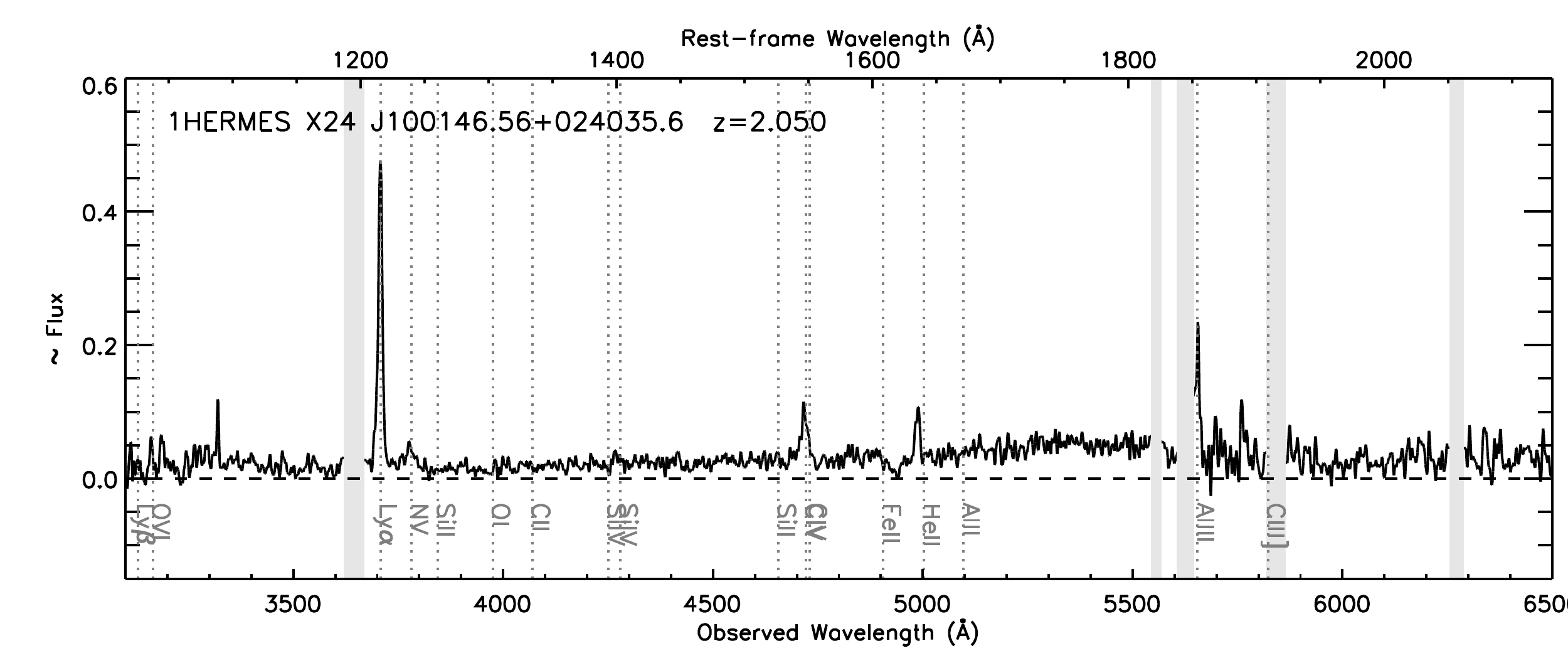}\includegraphics[width=0.58\columnwidth]{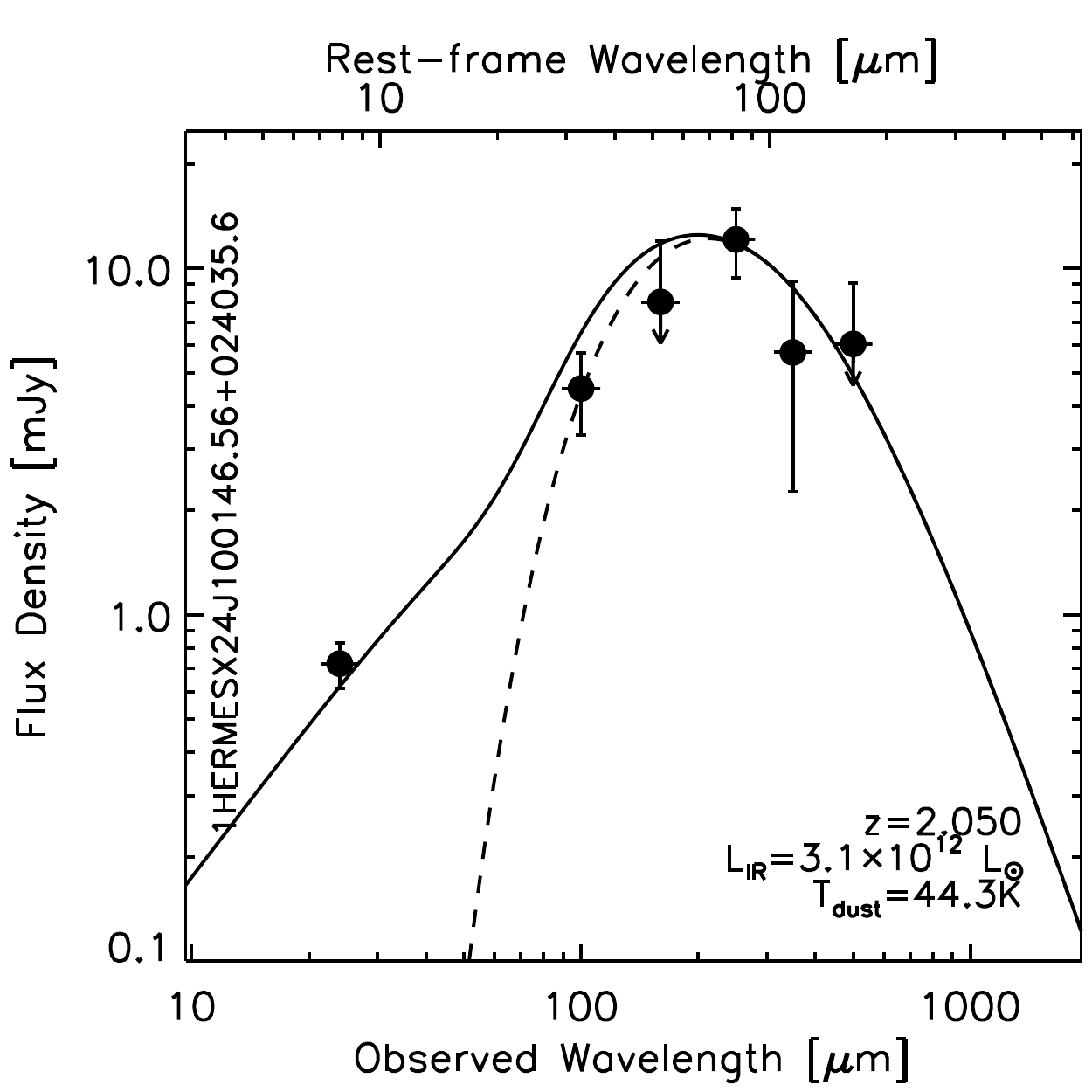}\\
\includegraphics[width=1.38\columnwidth]{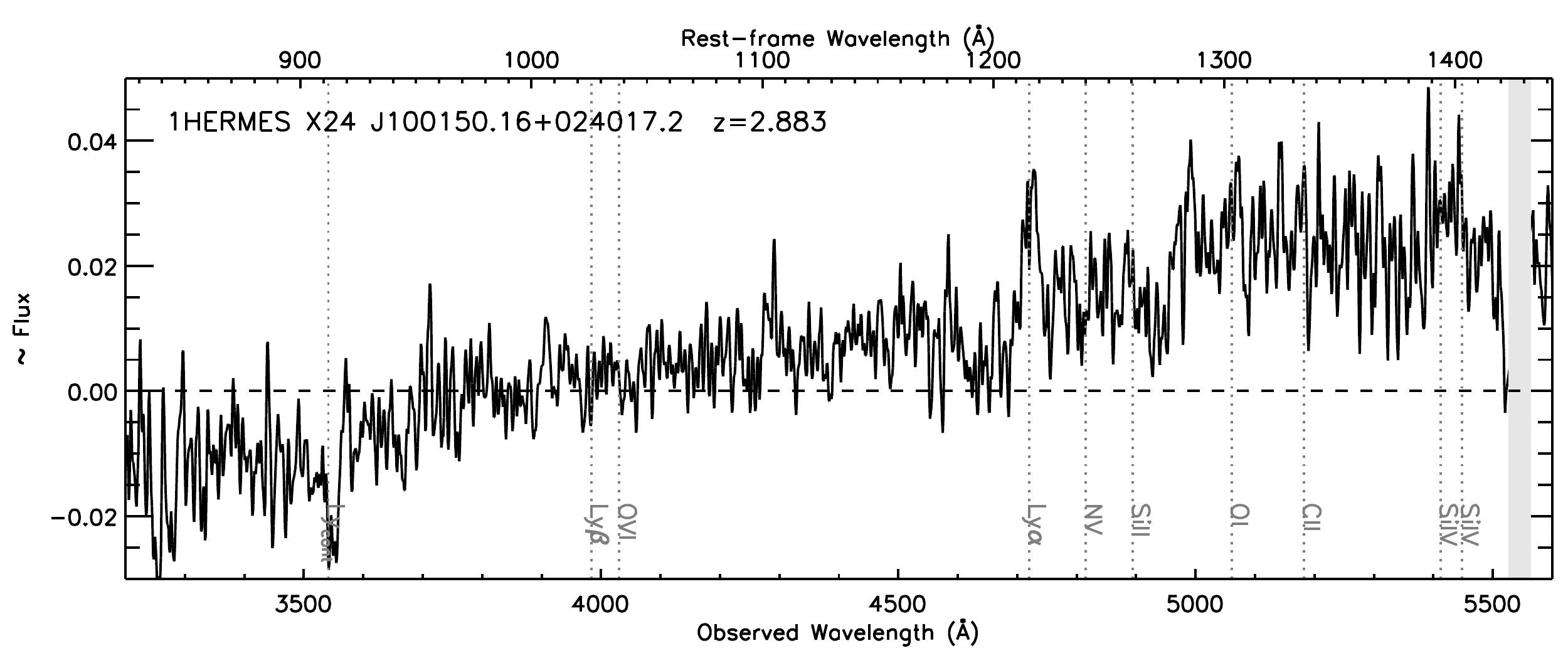}\includegraphics[width=0.58\columnwidth]{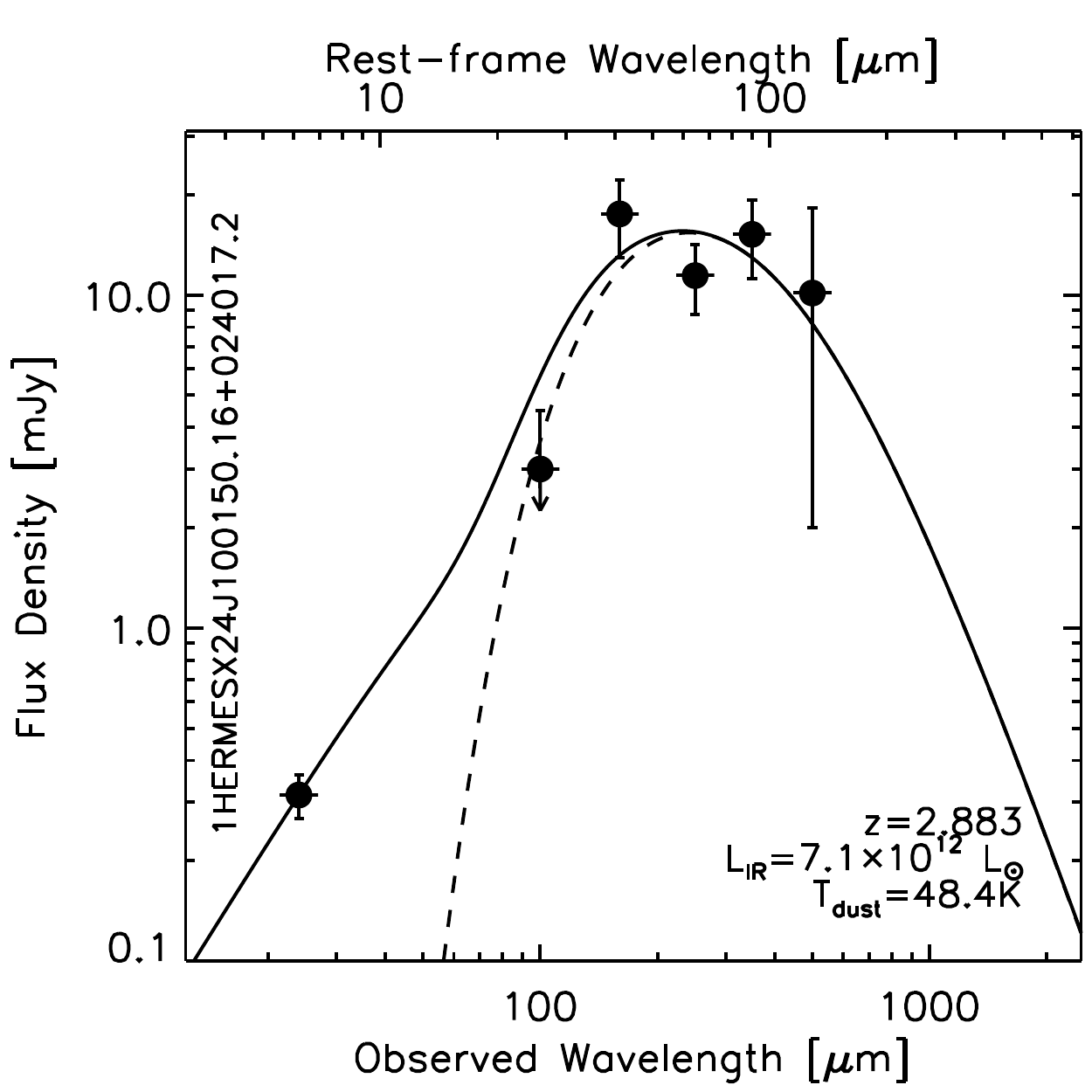}\\
\centerline{{\small Figure~\ref{fig:spectra} --- continued.}}
\end{center}
\clearpage
\begin{center}
\includegraphics[width=1.38\columnwidth]{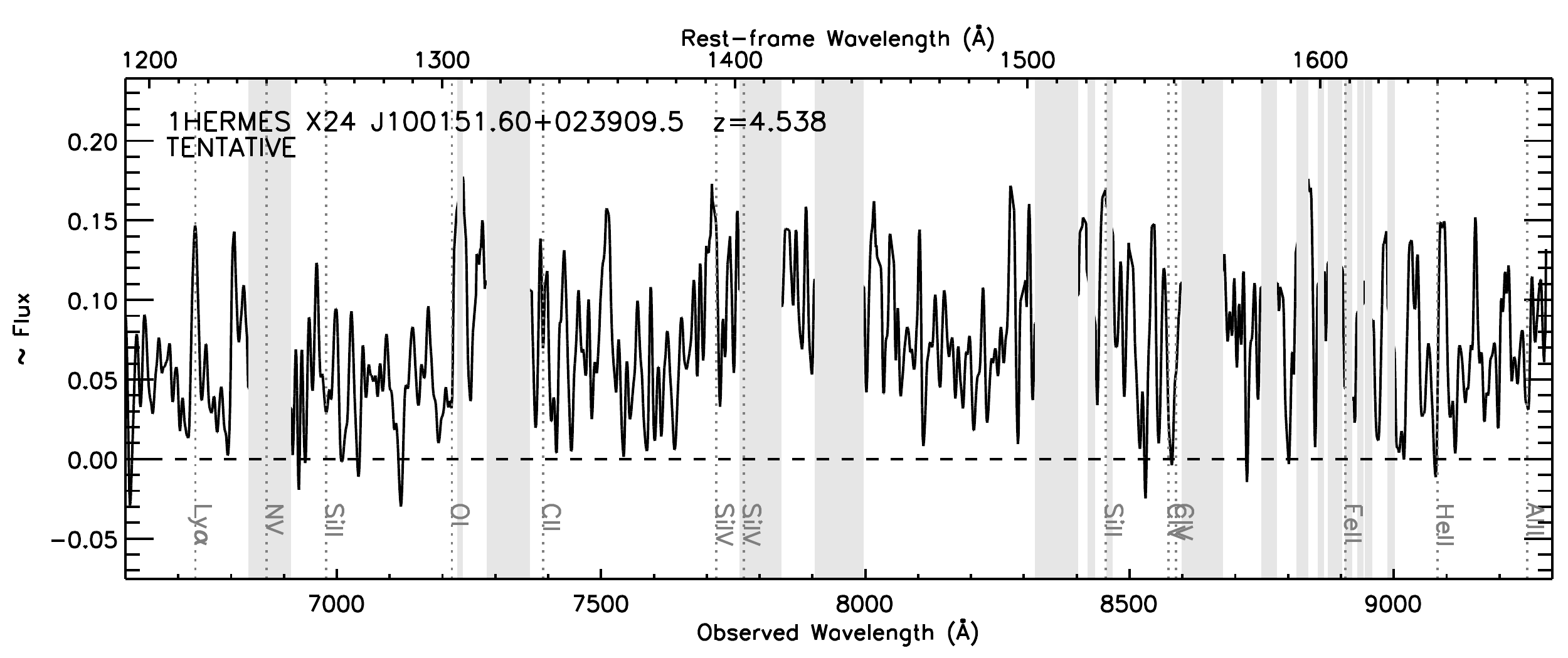}\includegraphics[width=0.58\columnwidth]{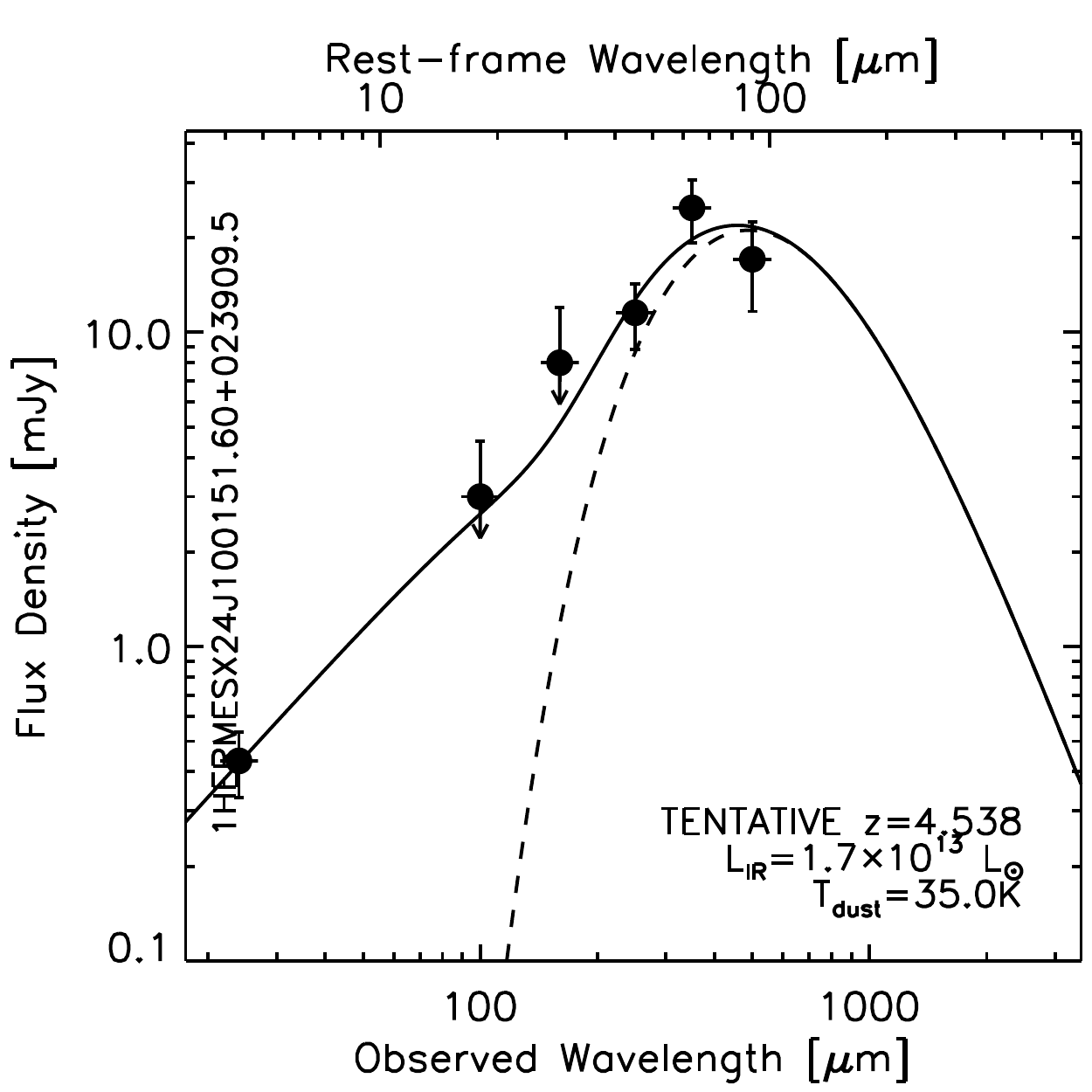}\\
\includegraphics[width=1.38\columnwidth]{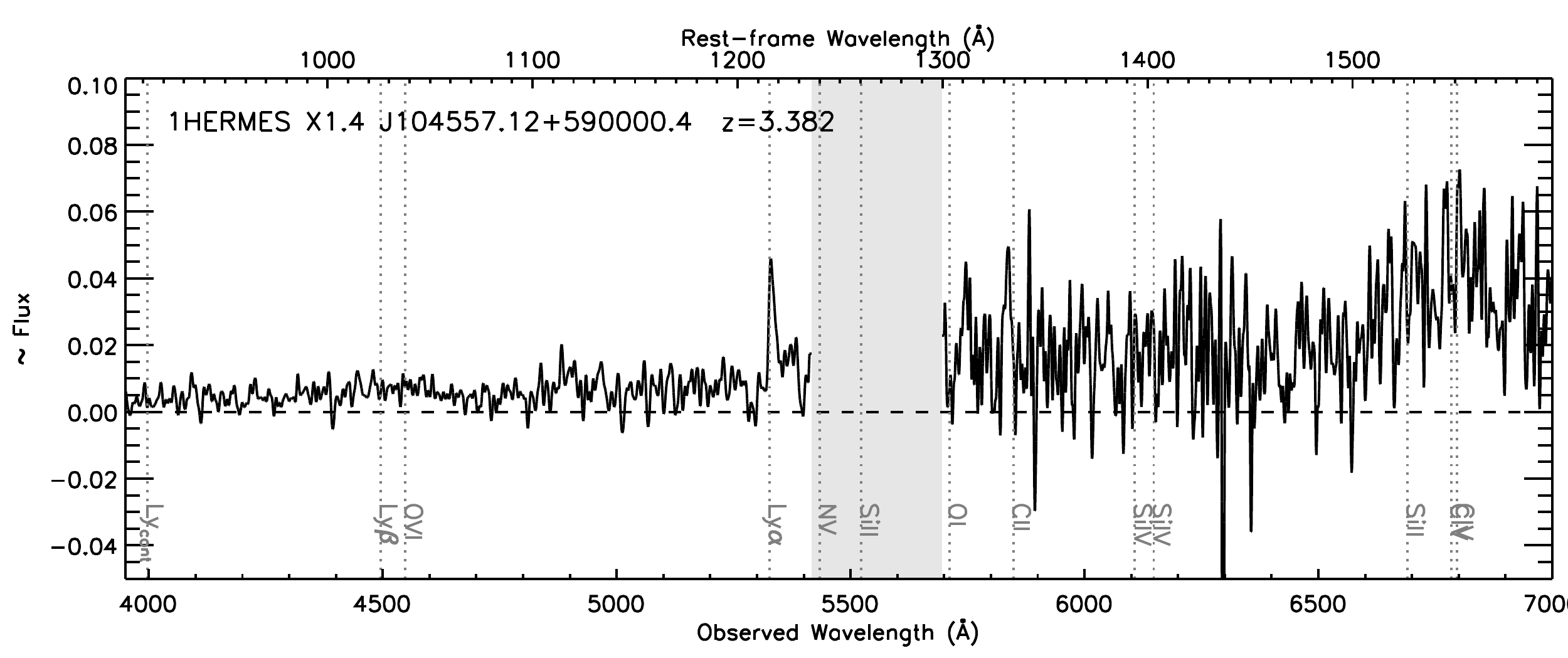}\includegraphics[width=0.58\columnwidth]{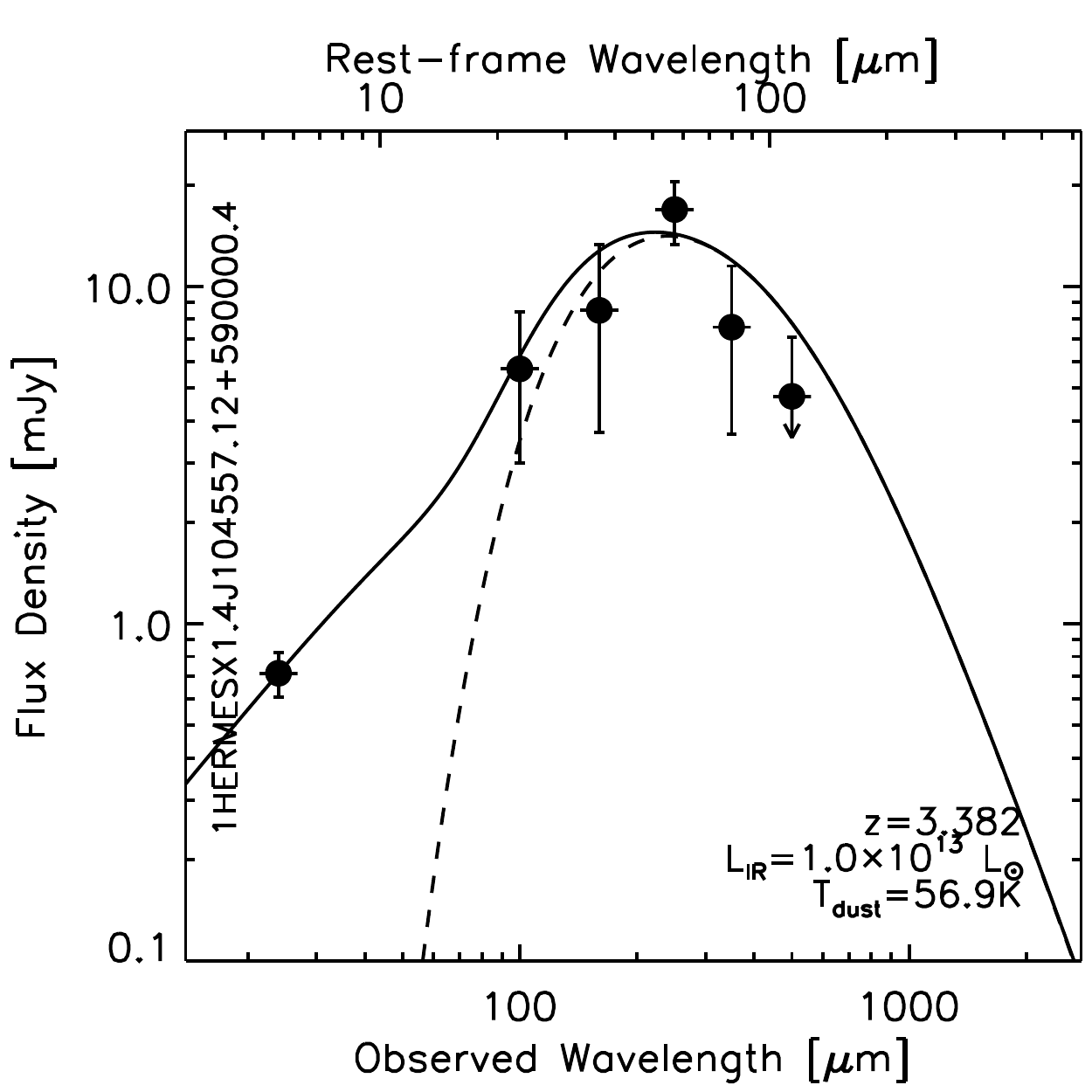}\\
\includegraphics[width=1.38\columnwidth]{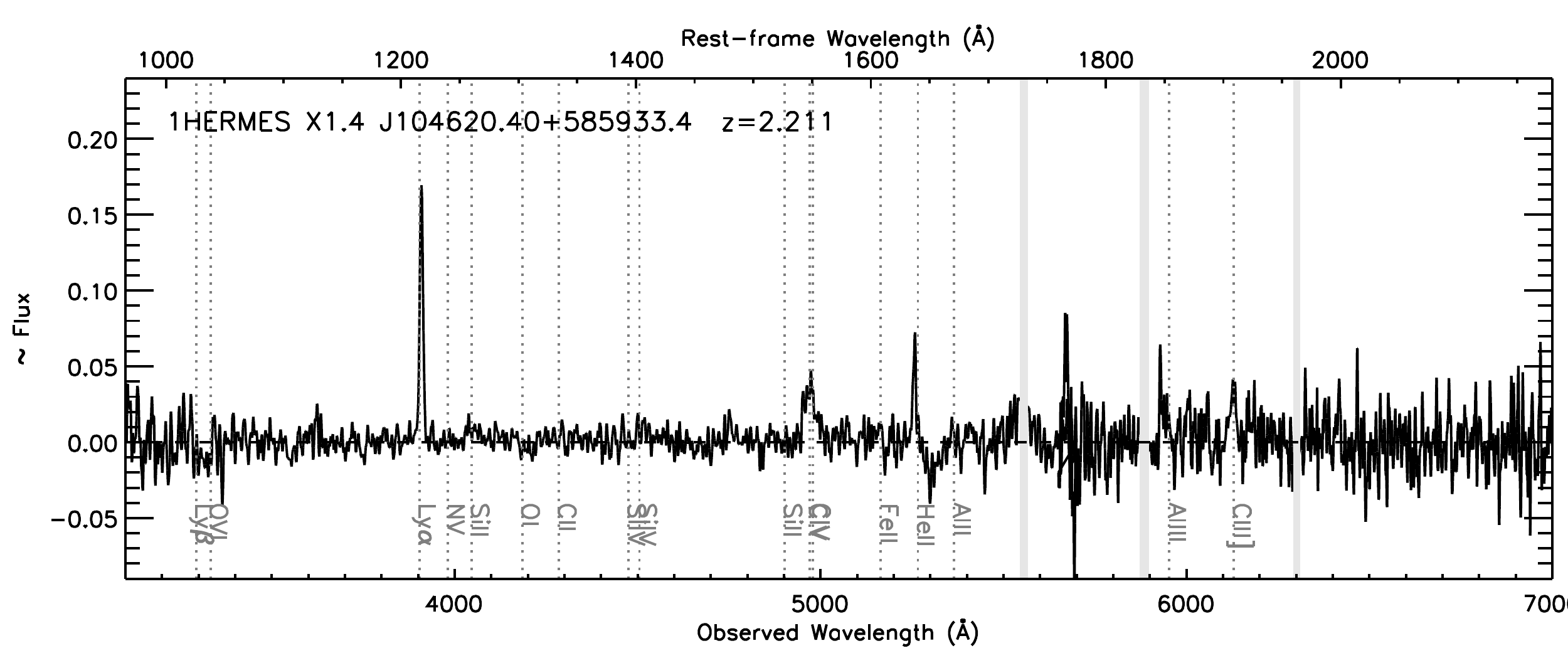}\includegraphics[width=0.58\columnwidth]{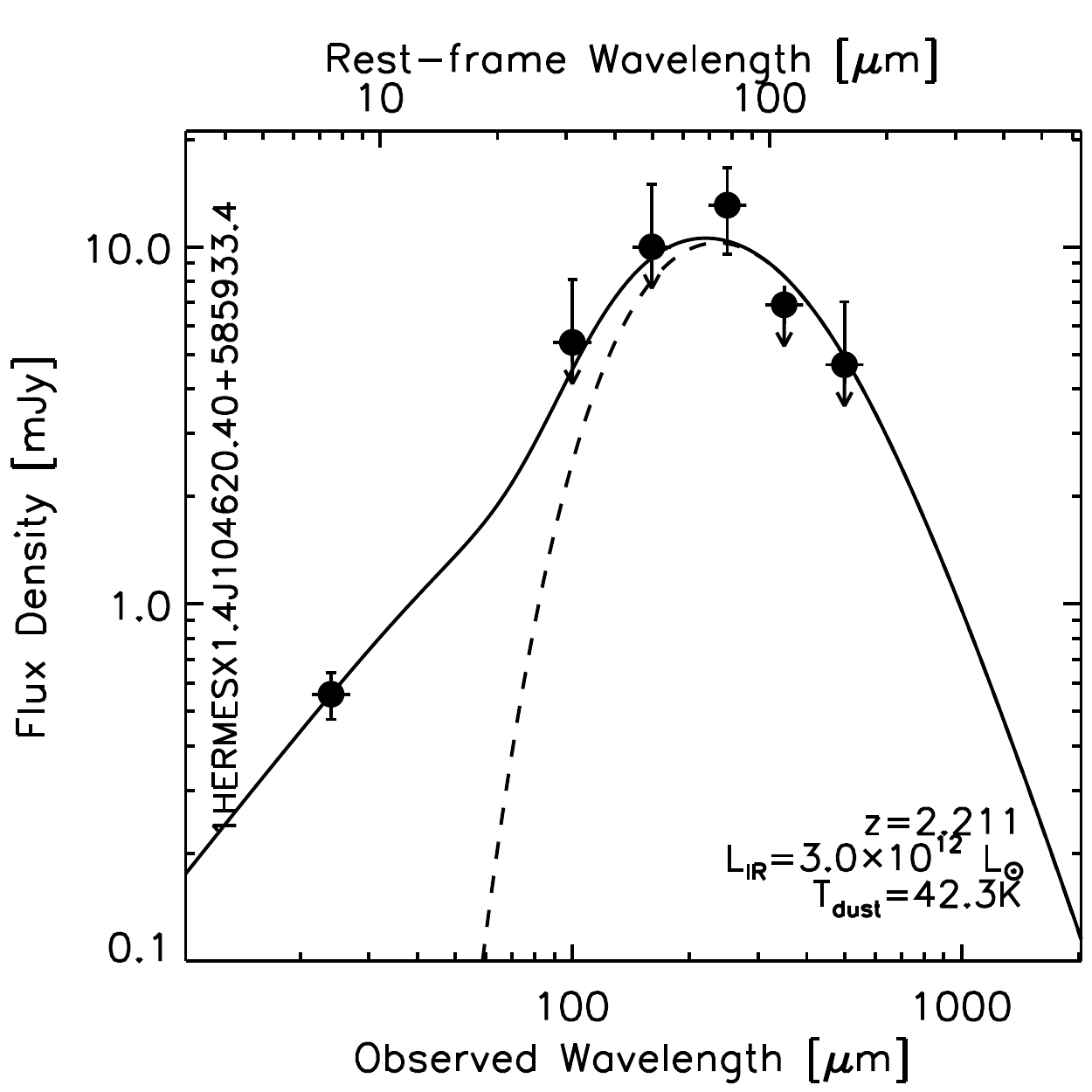}\\
\includegraphics[width=1.38\columnwidth]{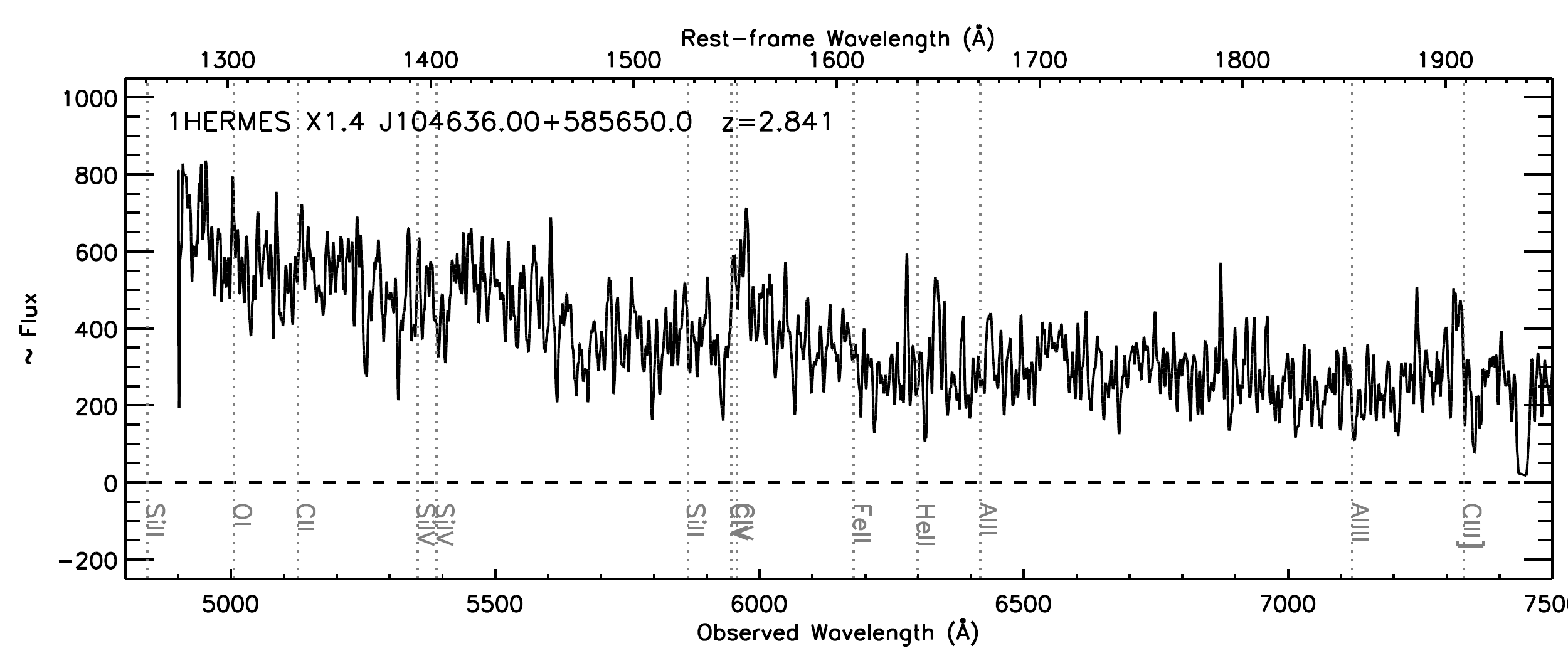}\includegraphics[width=0.58\columnwidth]{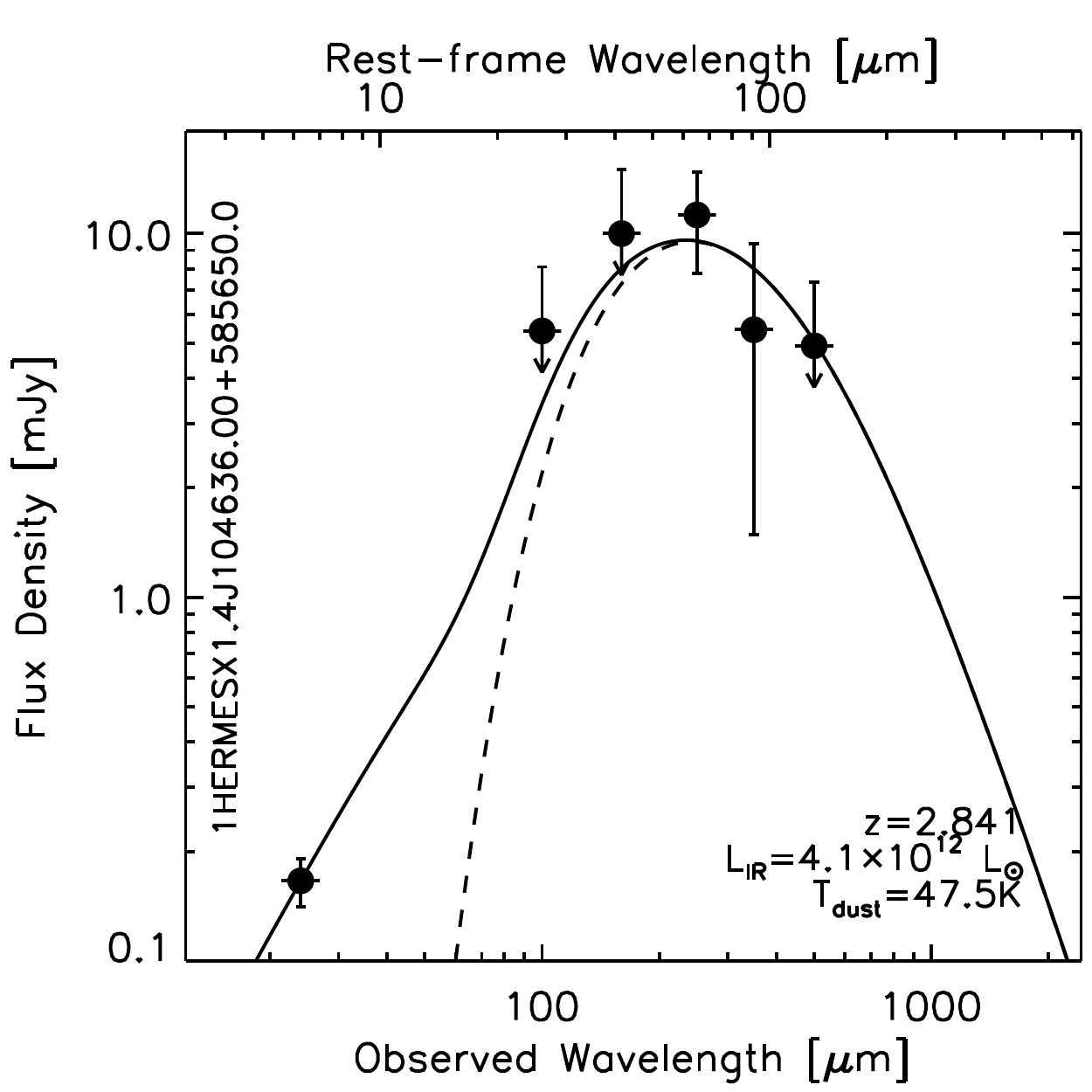}\\
\centerline{{\small Figure~\ref{fig:spectra} --- continued.}}
\end{center}
\clearpage
\begin{center}
\includegraphics[width=1.38\columnwidth]{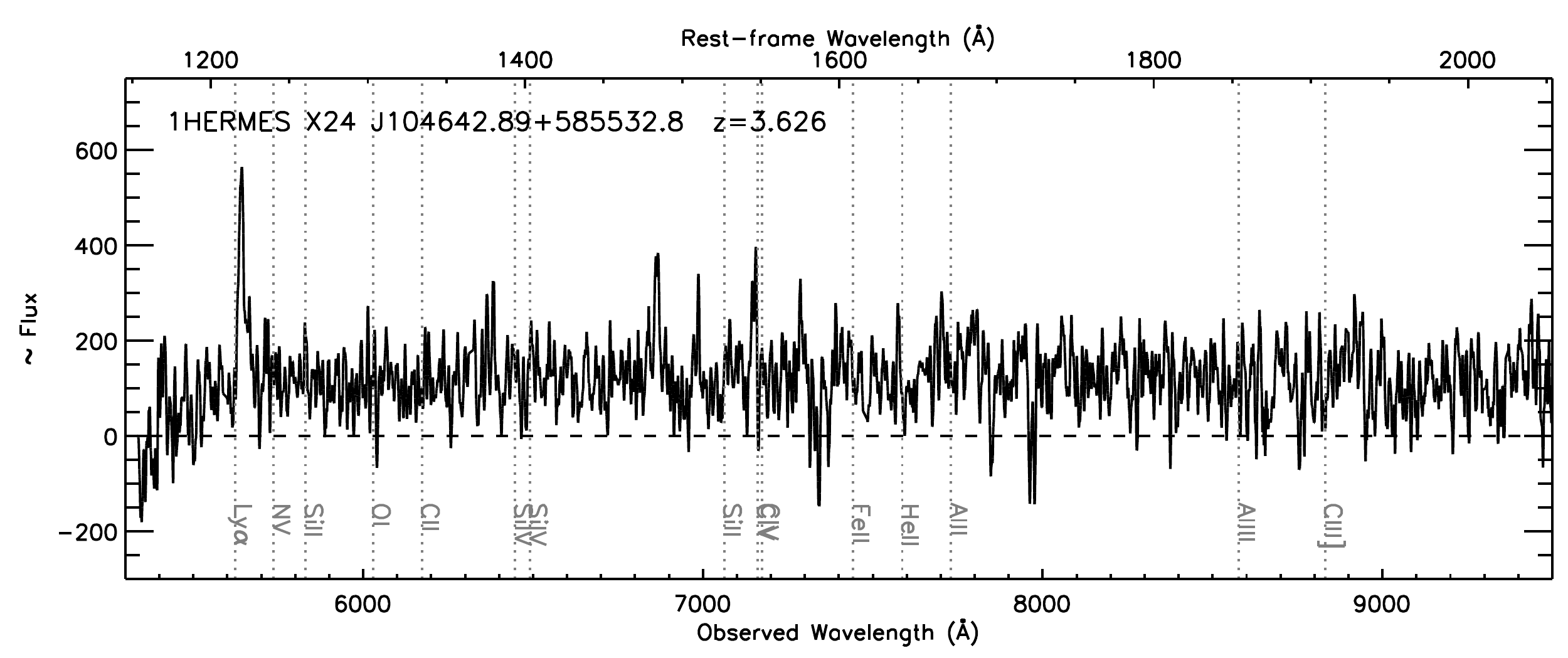}\includegraphics[width=0.58\columnwidth]{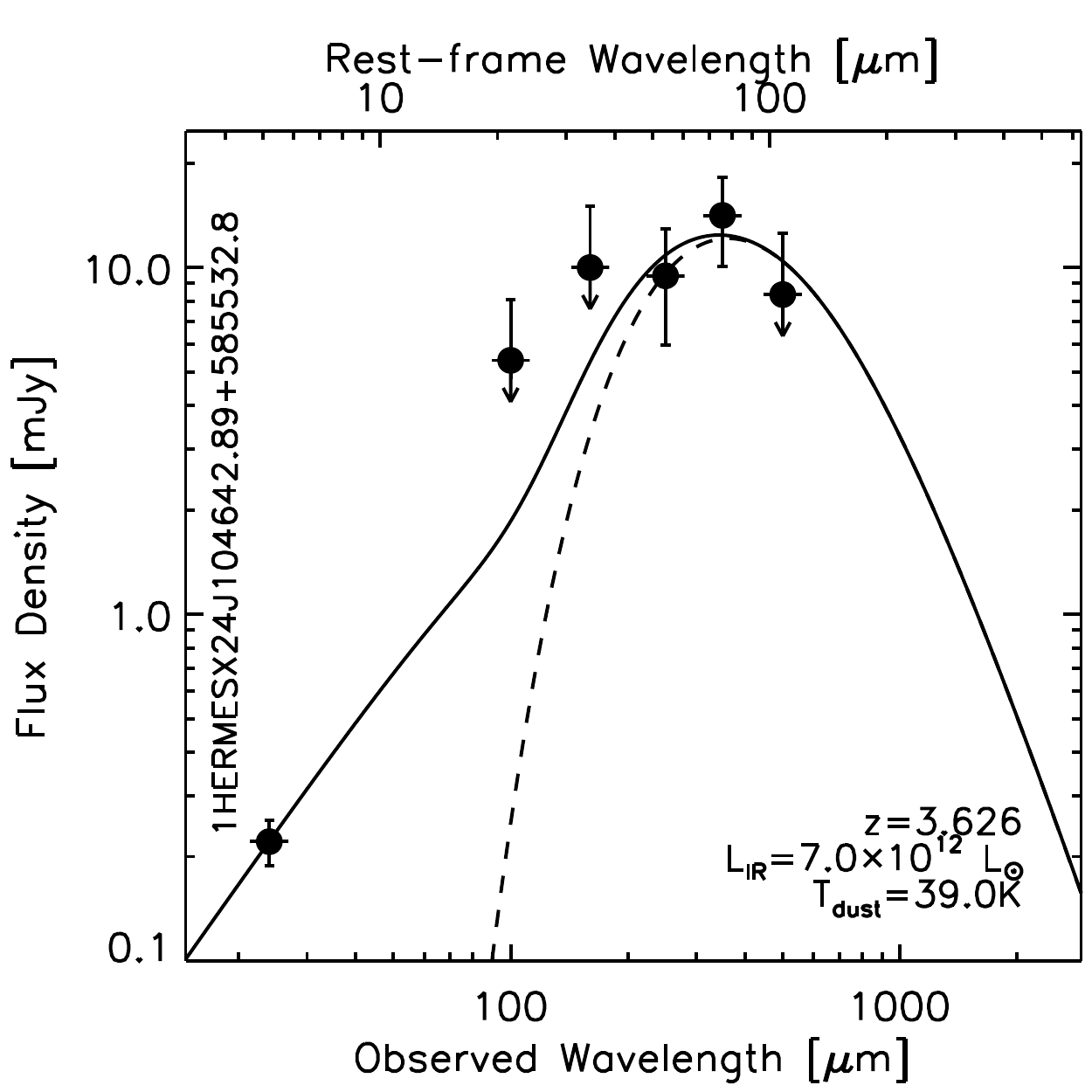}\\
\includegraphics[width=1.38\columnwidth]{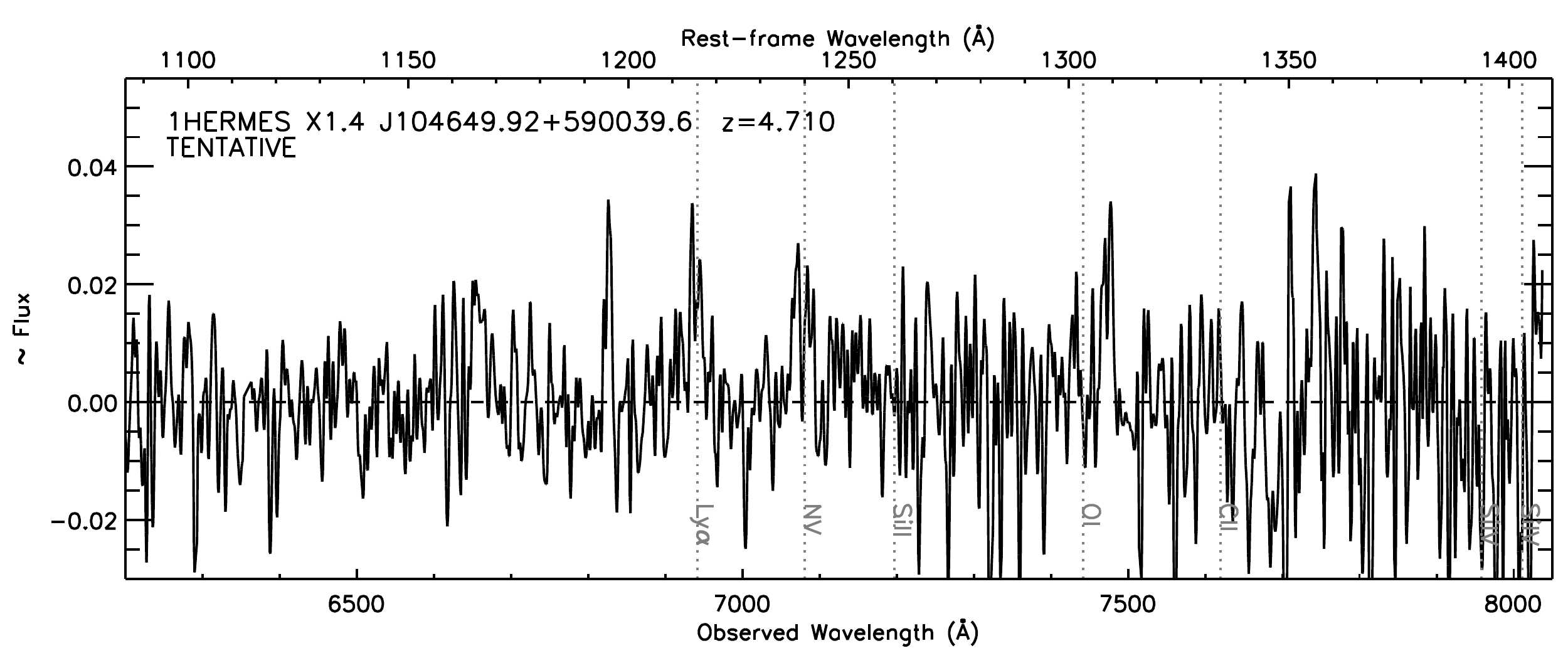}\includegraphics[width=0.58\columnwidth]{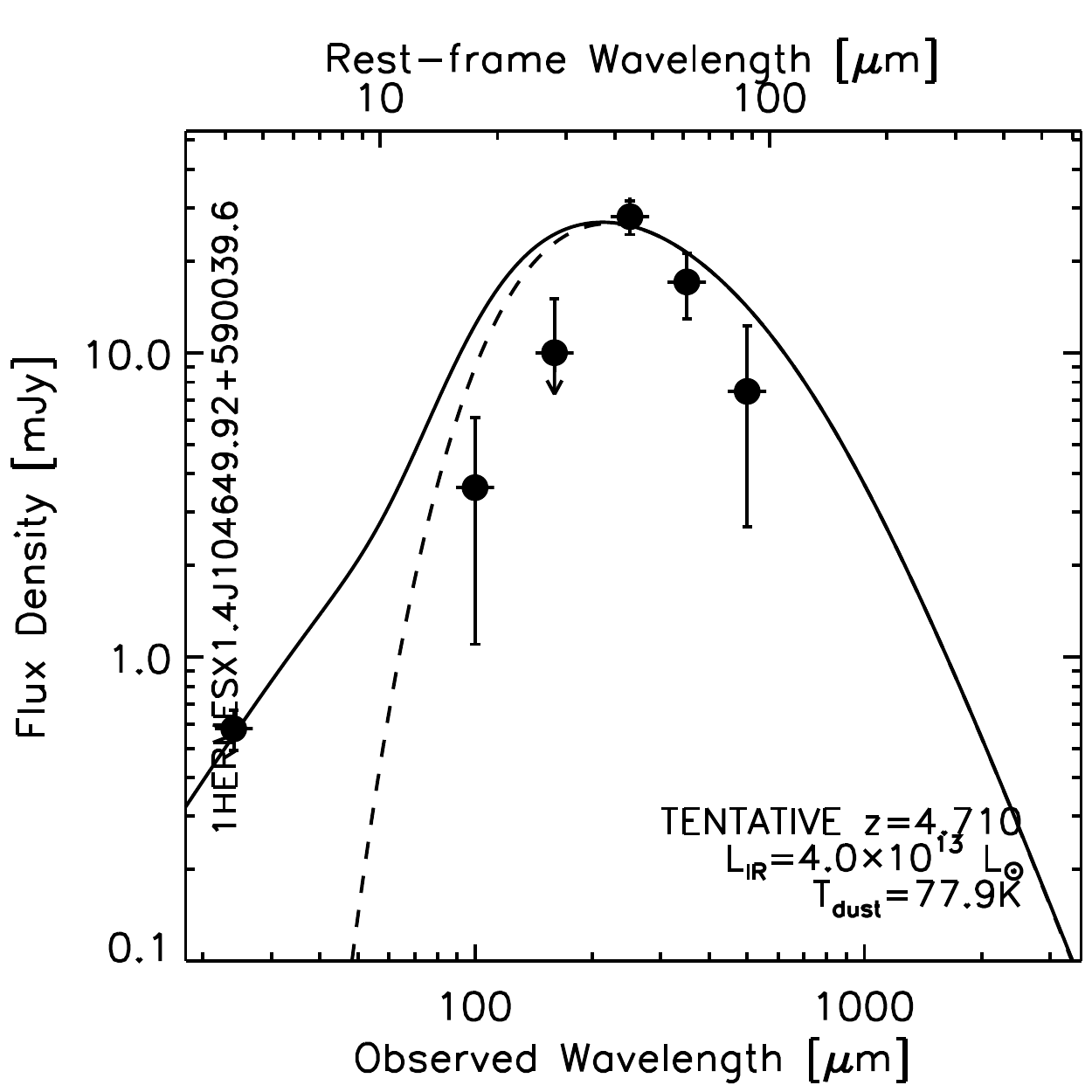}\\
\includegraphics[width=1.38\columnwidth]{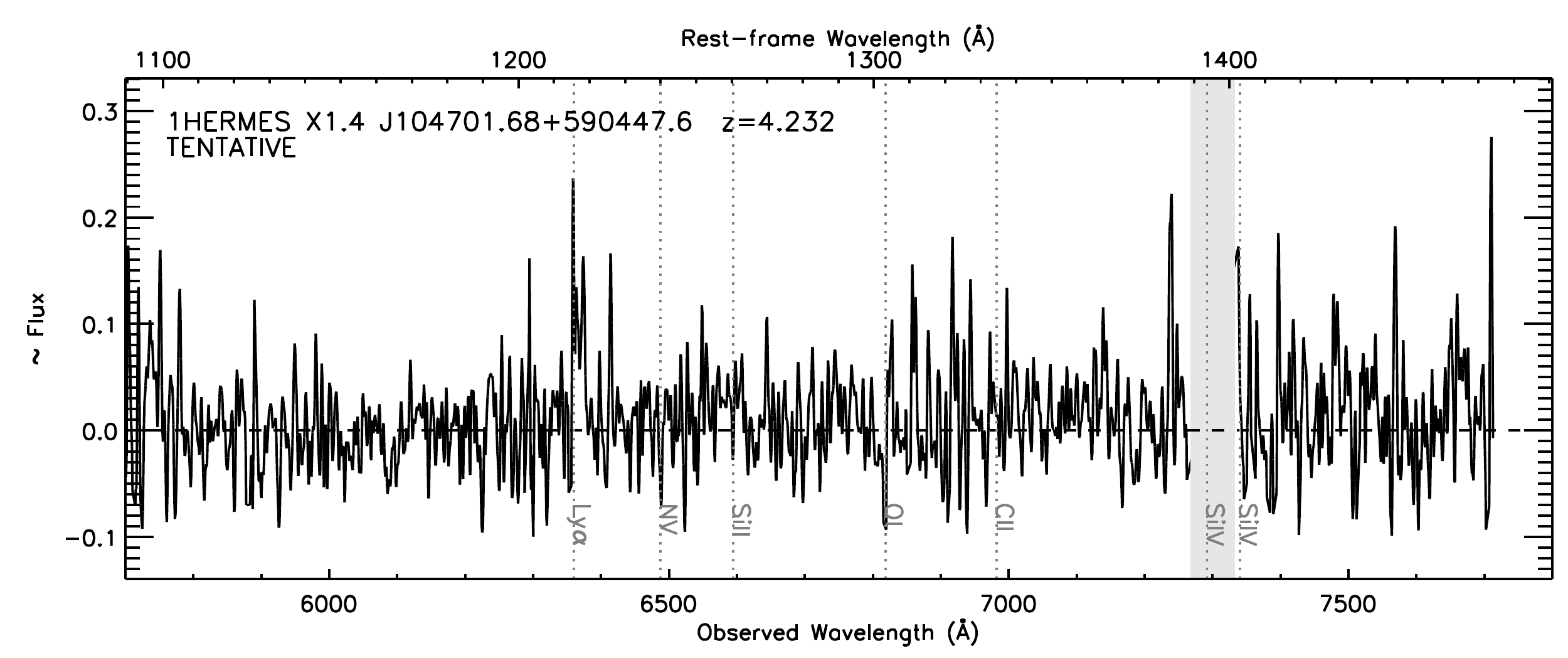}\includegraphics[width=0.58\columnwidth]{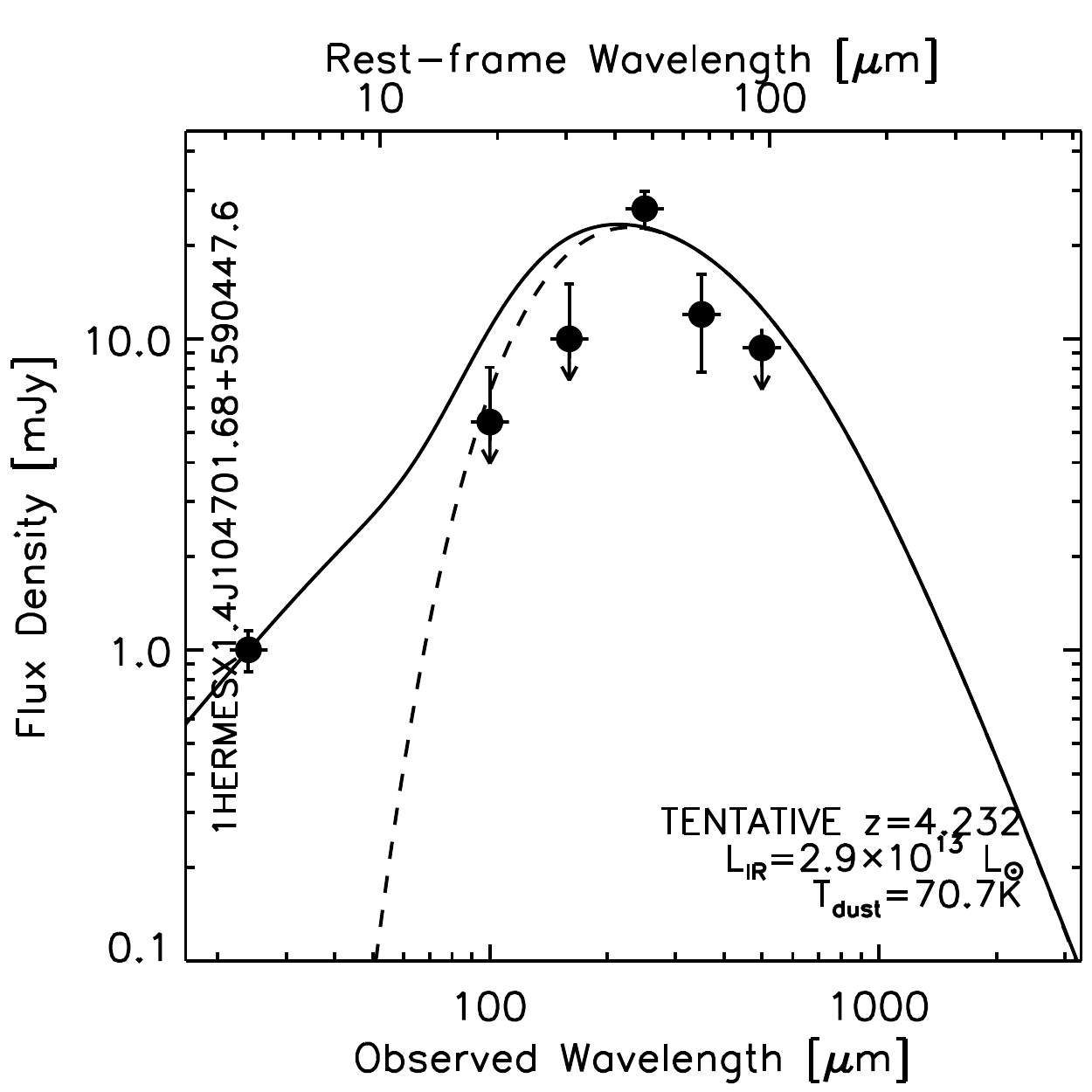}\\
\includegraphics[width=1.38\columnwidth]{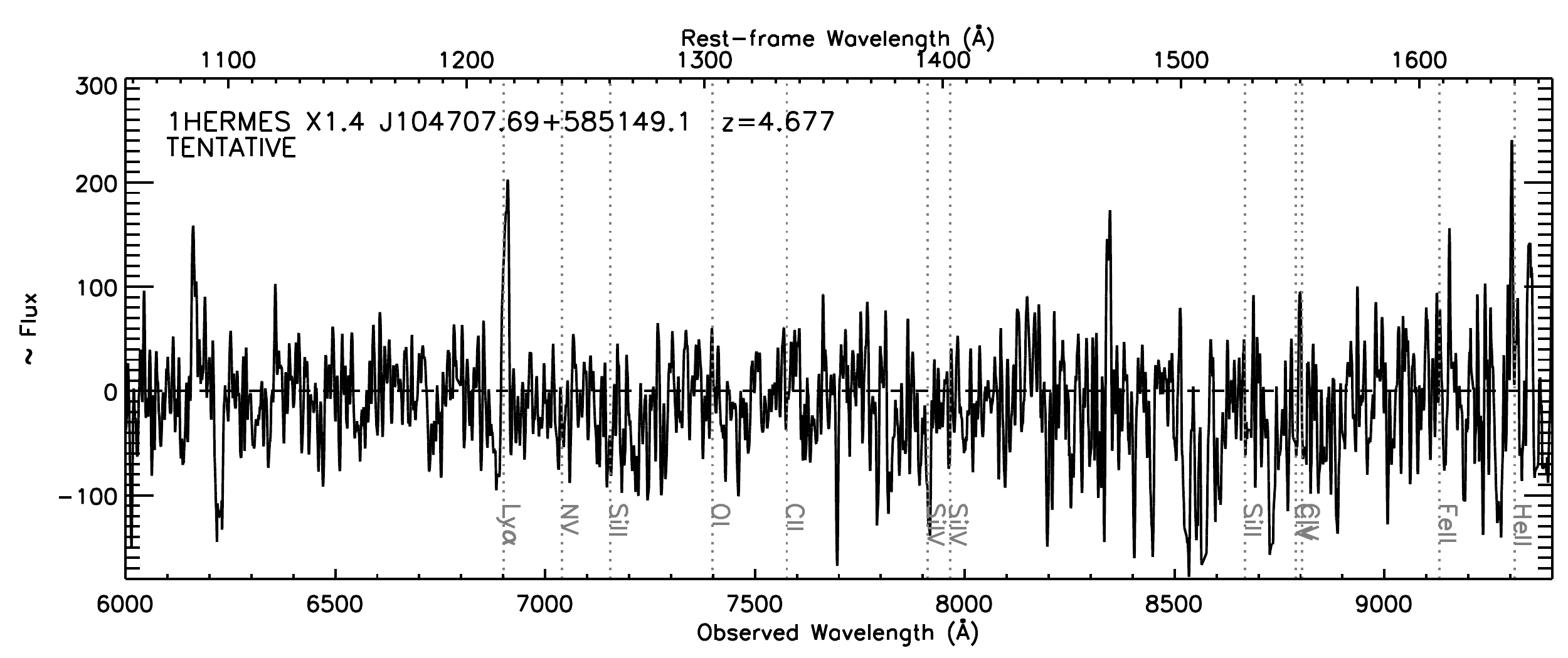}\includegraphics[width=0.58\columnwidth]{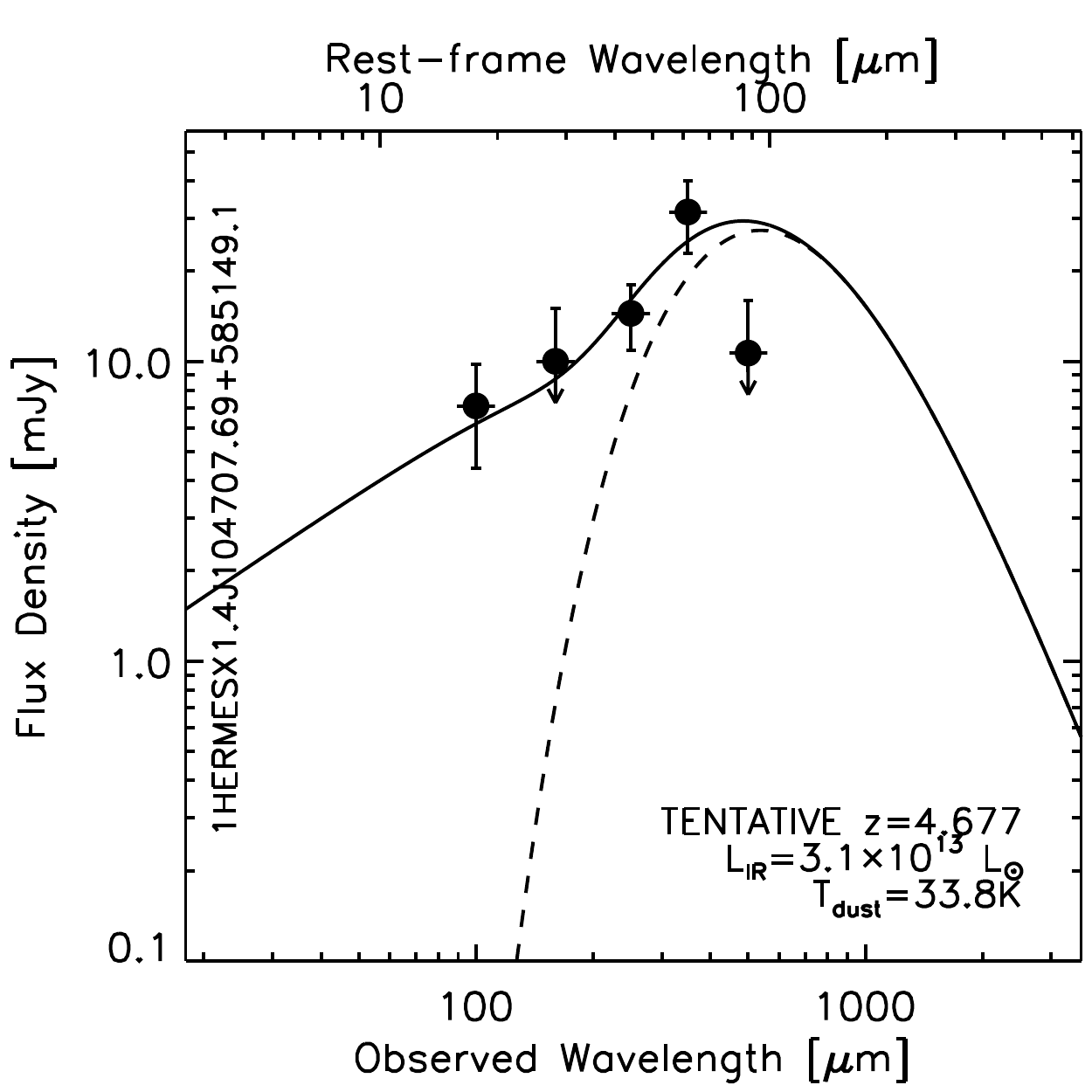}\\
\centerline{{\small Figure~\ref{fig:spectra} --- continued.}}
\end{center}
\clearpage
\begin{center}
\includegraphics[width=1.38\columnwidth]{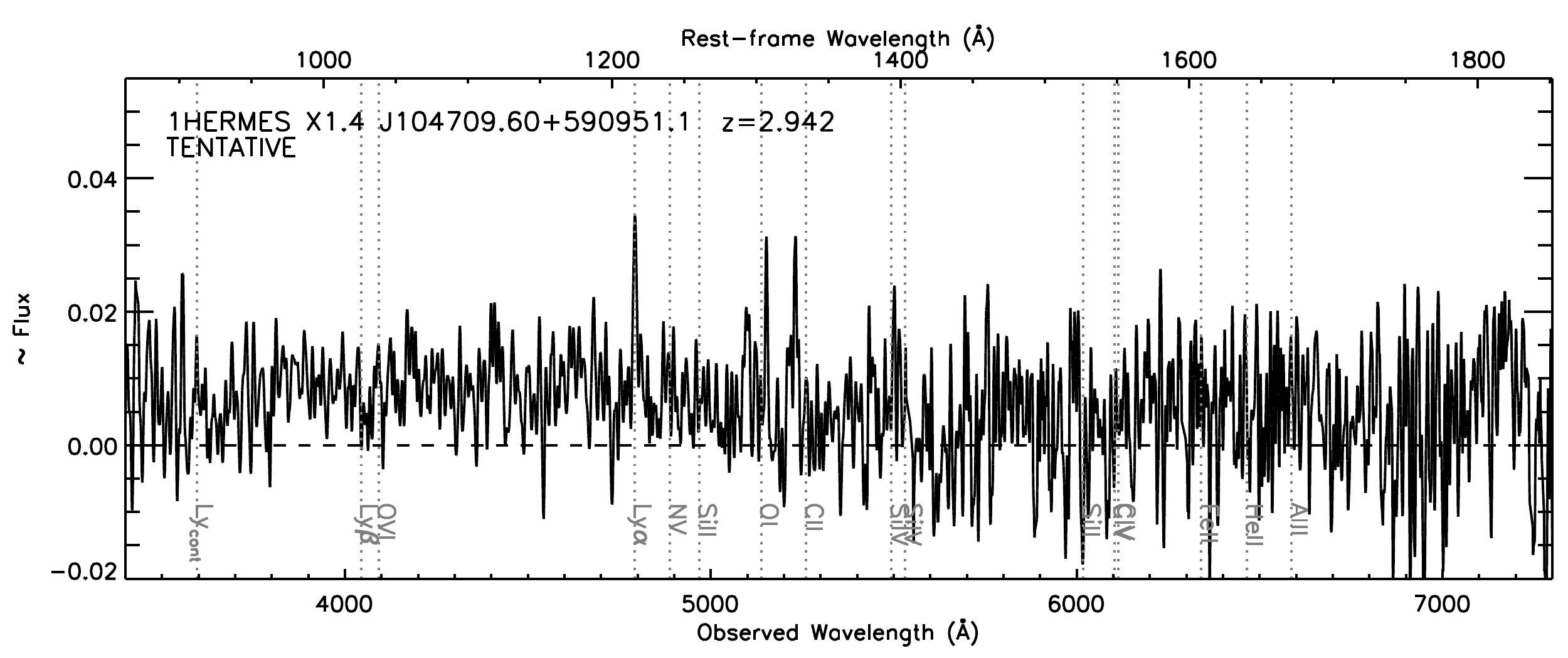}\includegraphics[width=0.58\columnwidth]{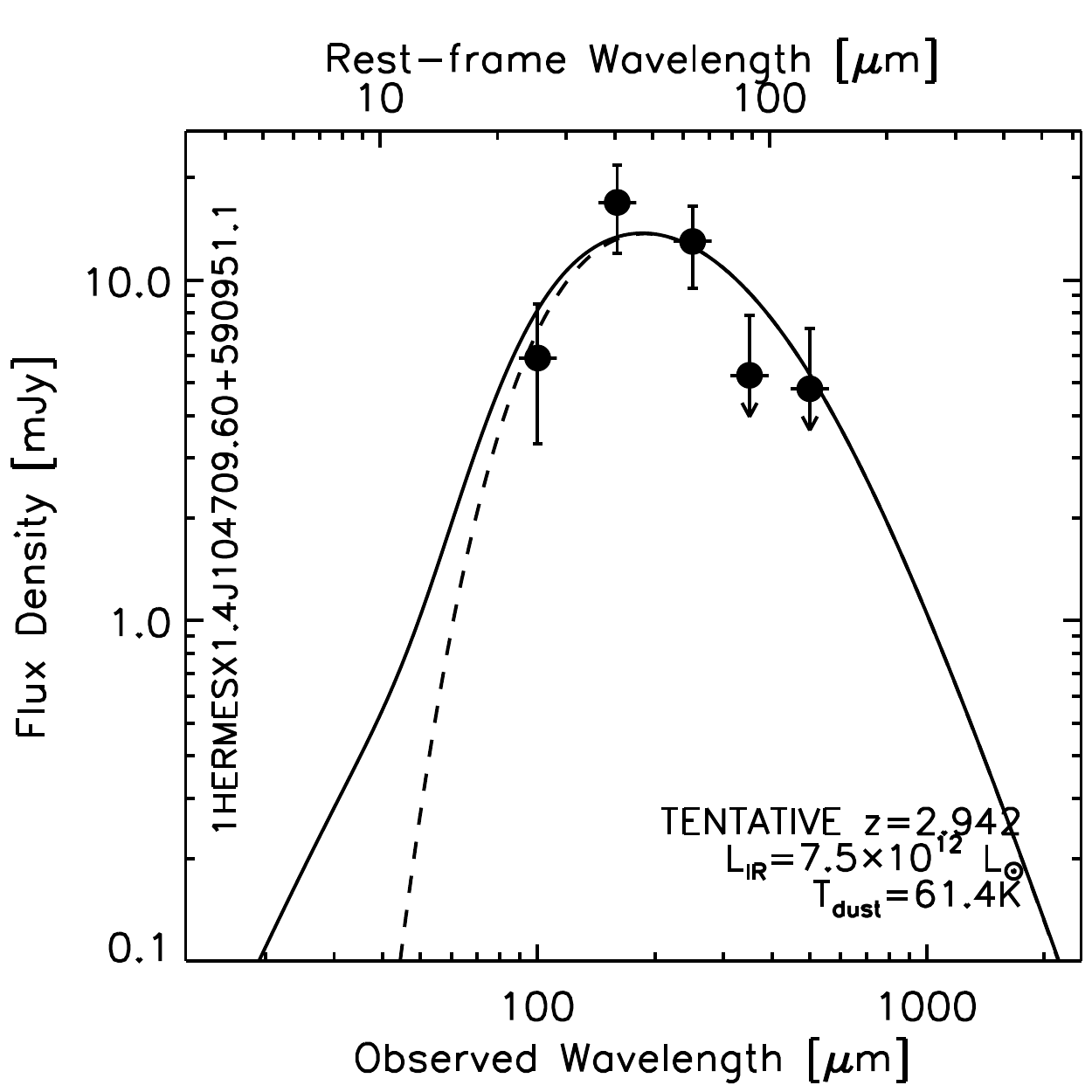}\\
\includegraphics[width=1.38\columnwidth]{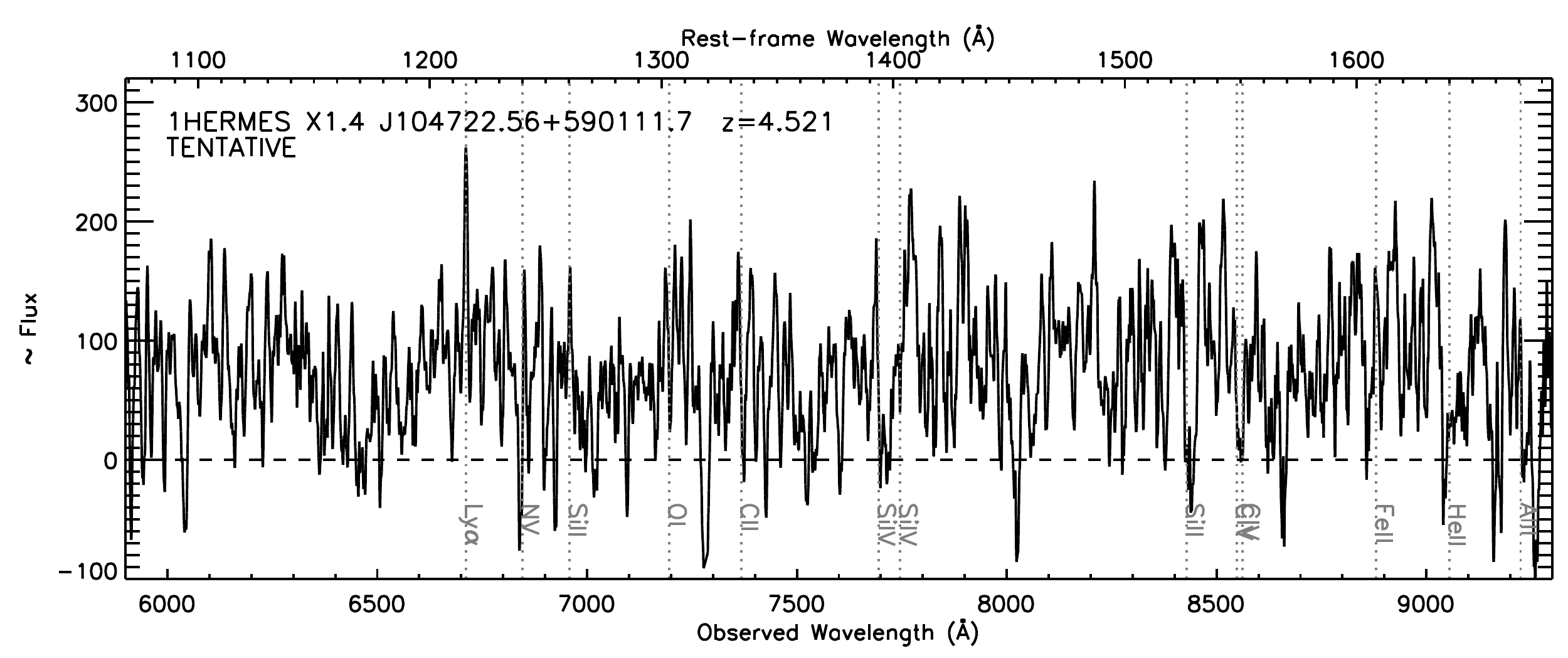}\includegraphics[width=0.58\columnwidth]{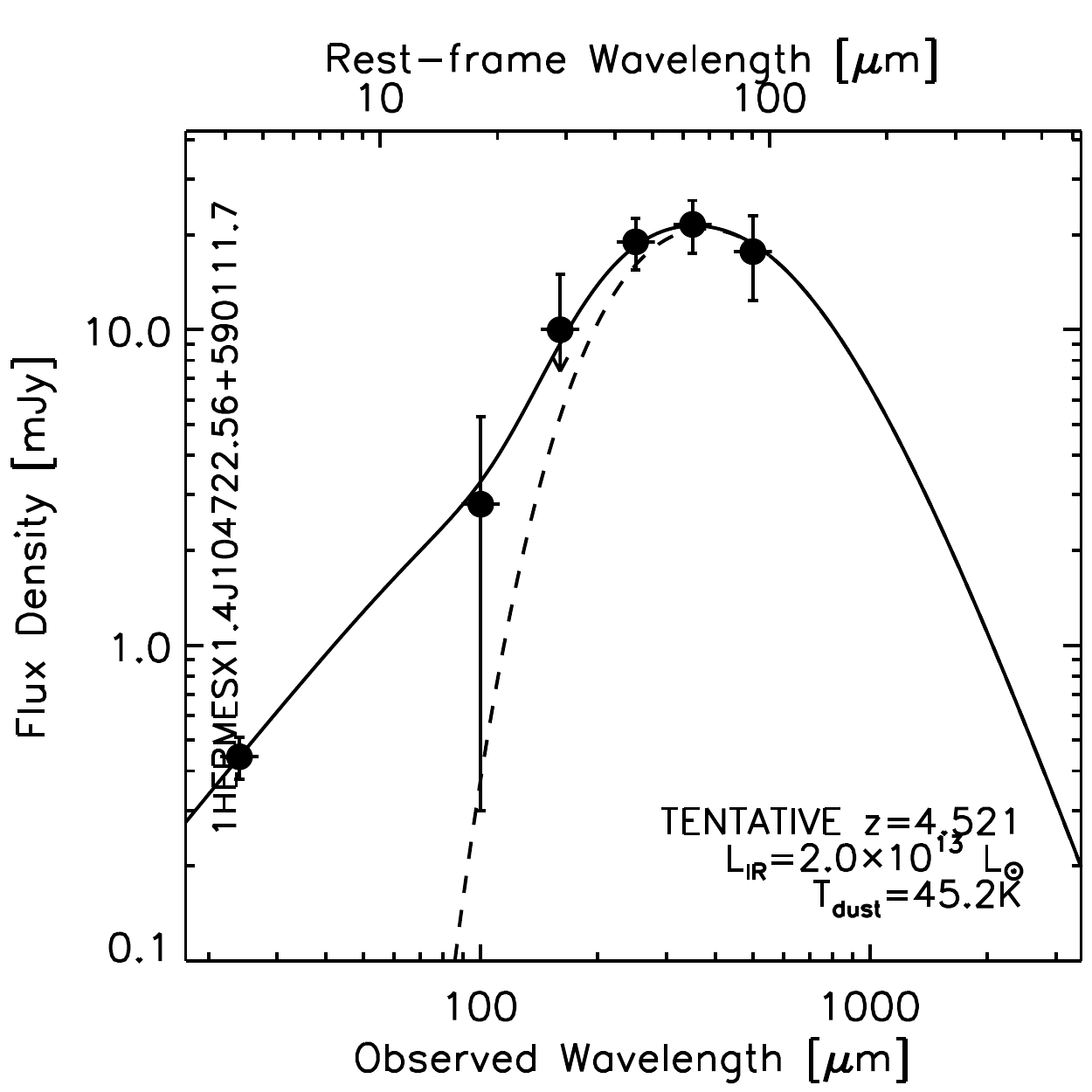}\\
\includegraphics[width=1.38\columnwidth]{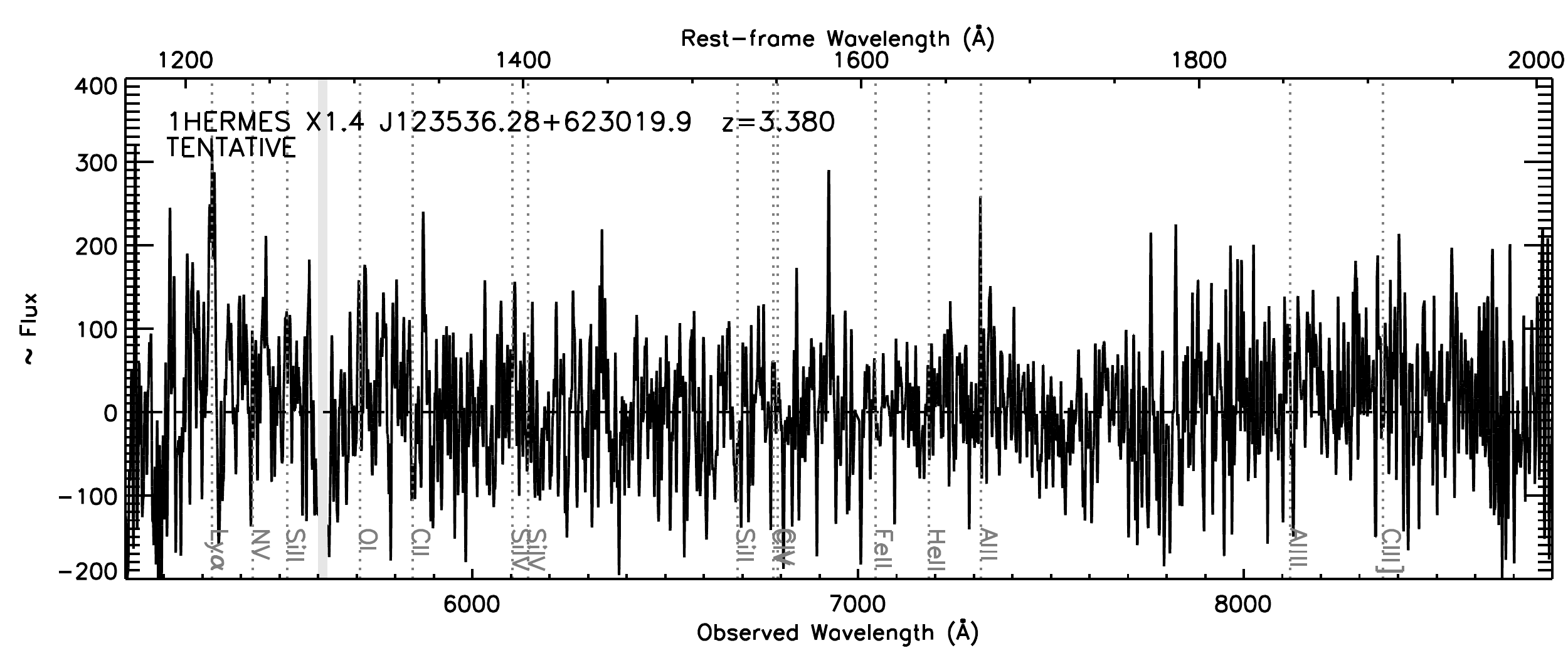}\includegraphics[width=0.58\columnwidth]{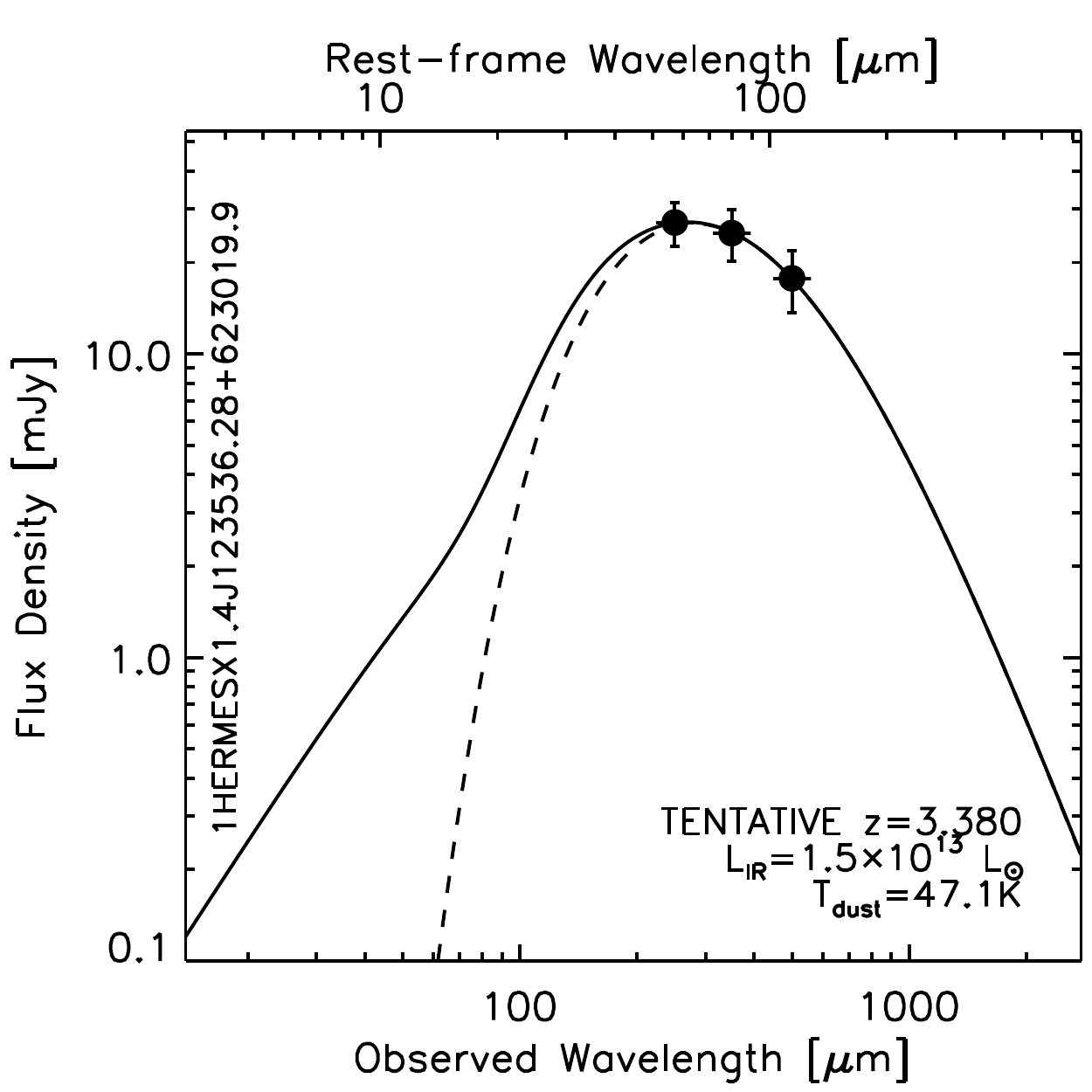}\\
\includegraphics[width=1.38\columnwidth]{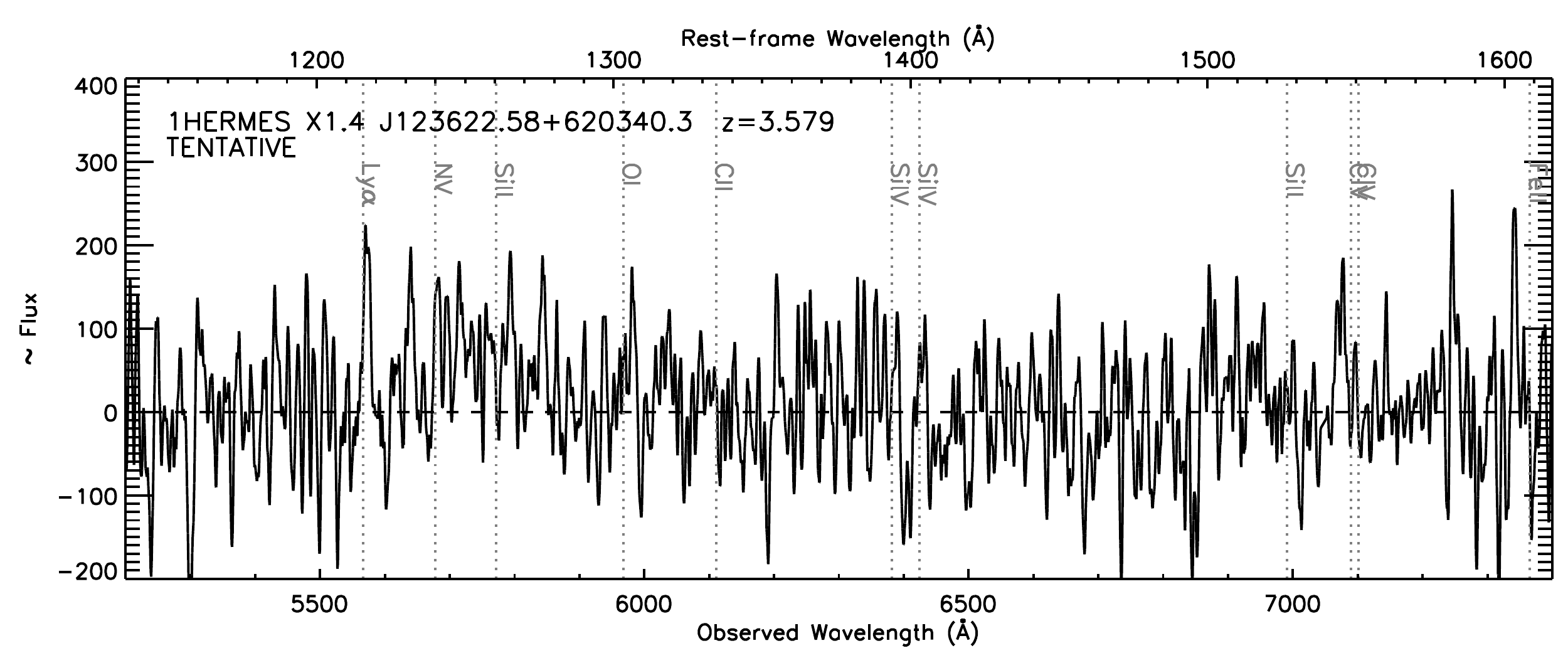}\includegraphics[width=0.58\columnwidth]{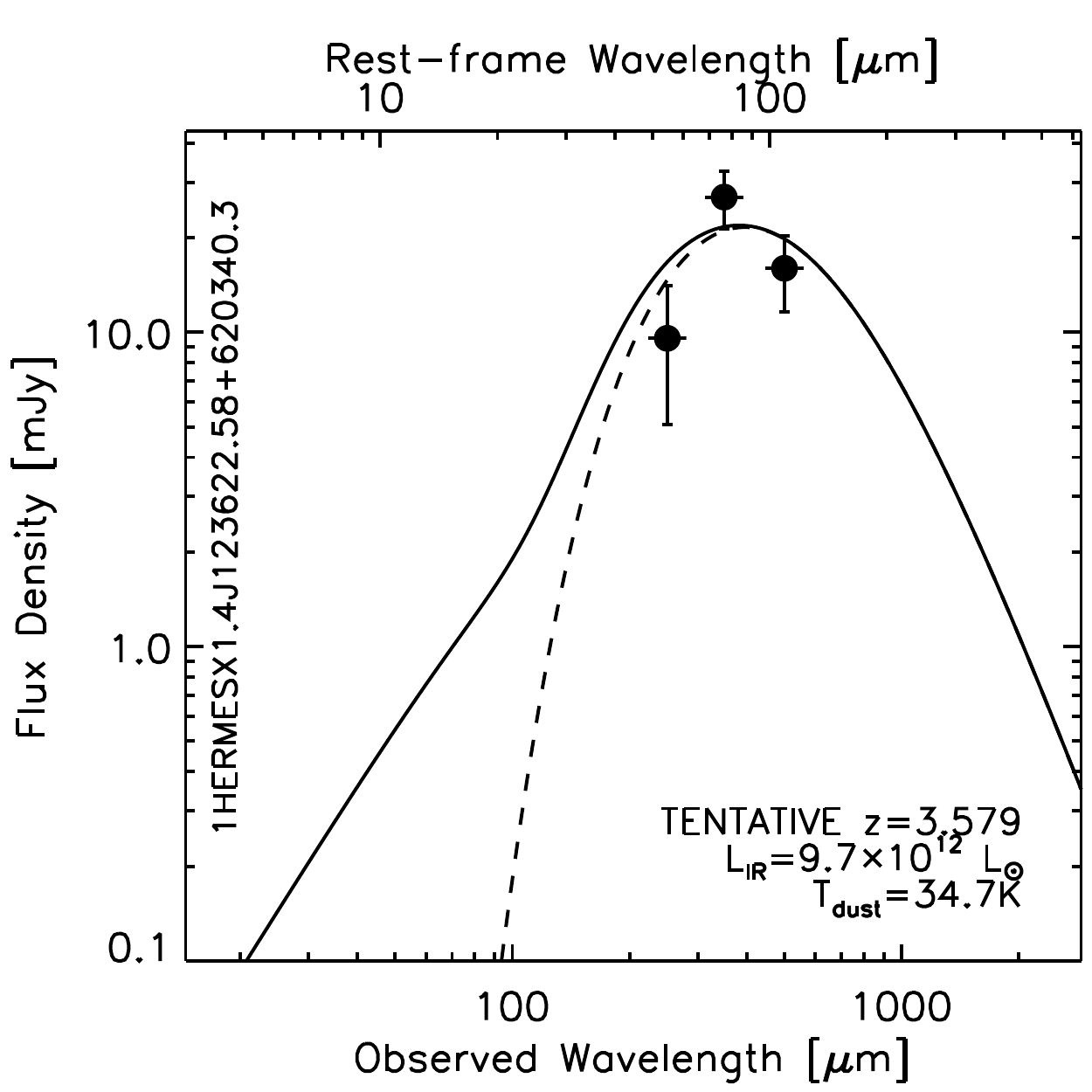}\\
\centerline{{\small Figure~\ref{fig:spectra} --- continued.}}
\end{center}
\clearpage
\begin{center}
\includegraphics[width=1.38\columnwidth]{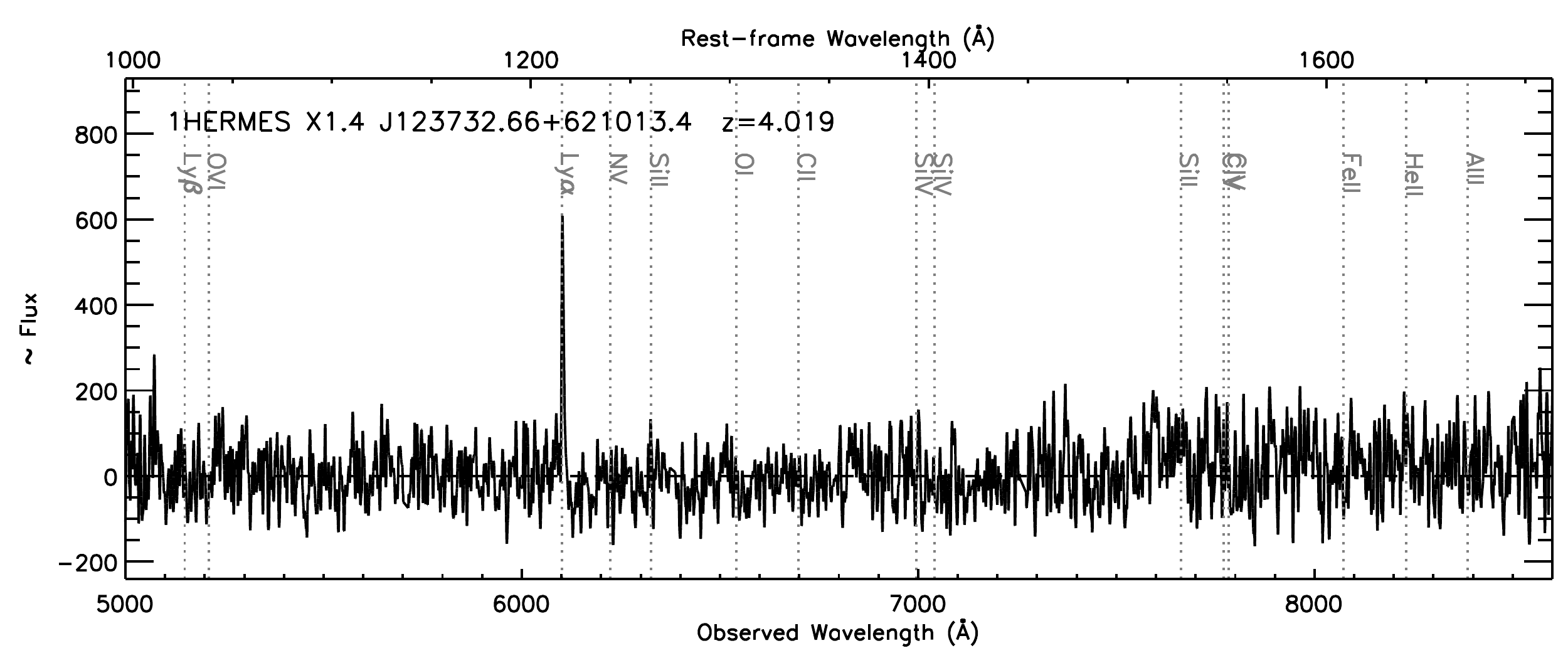}\includegraphics[width=0.58\columnwidth]{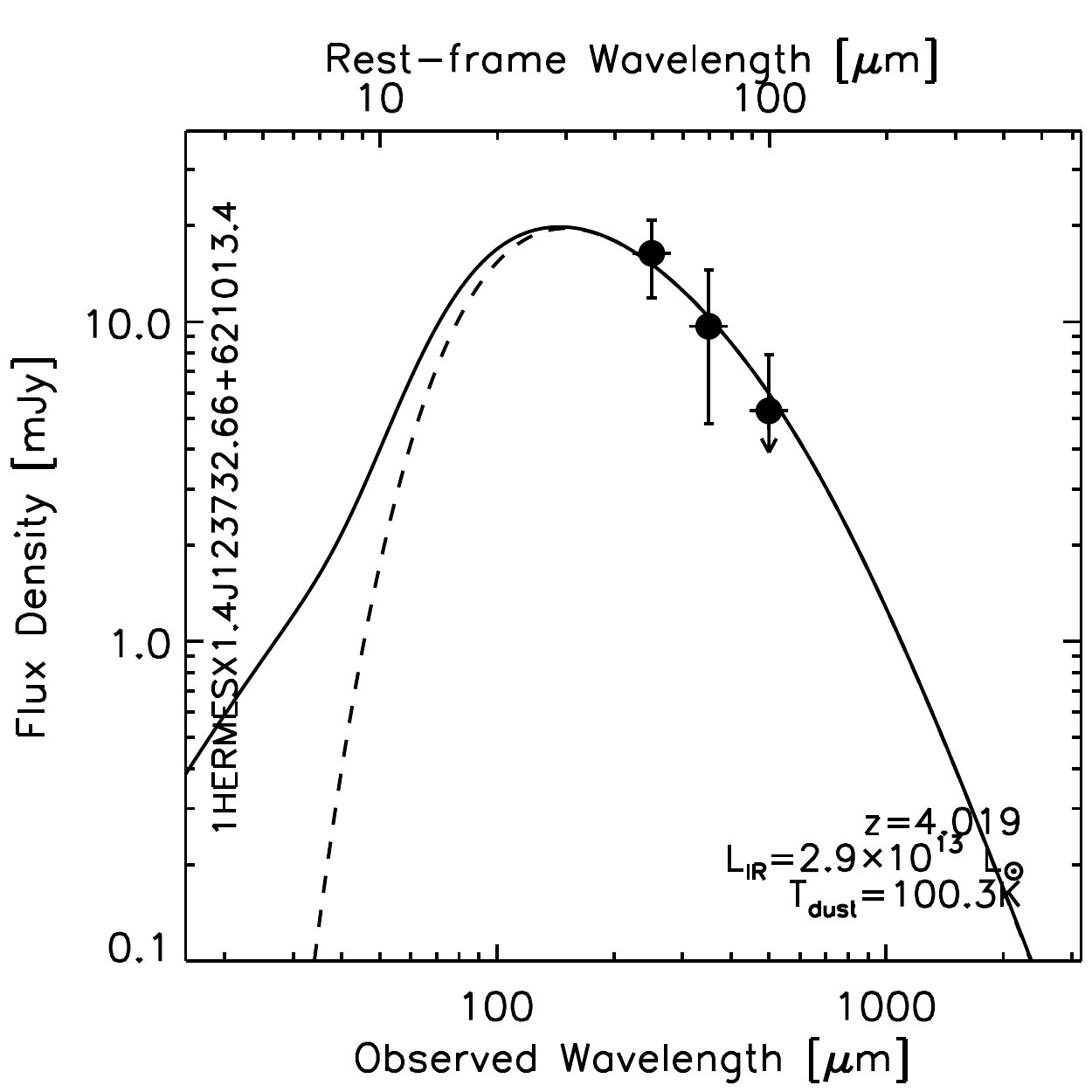}\\
\includegraphics[width=1.38\columnwidth]{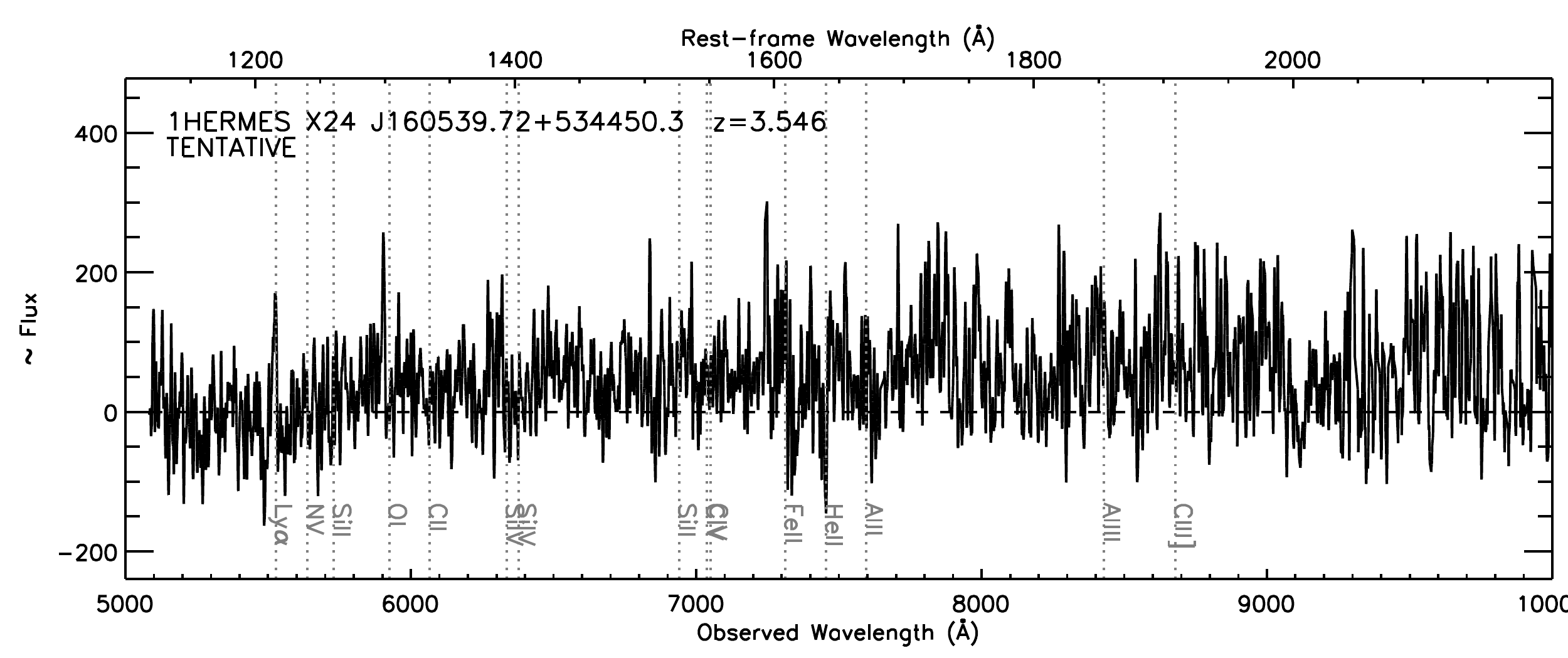}\includegraphics[width=0.58\columnwidth]{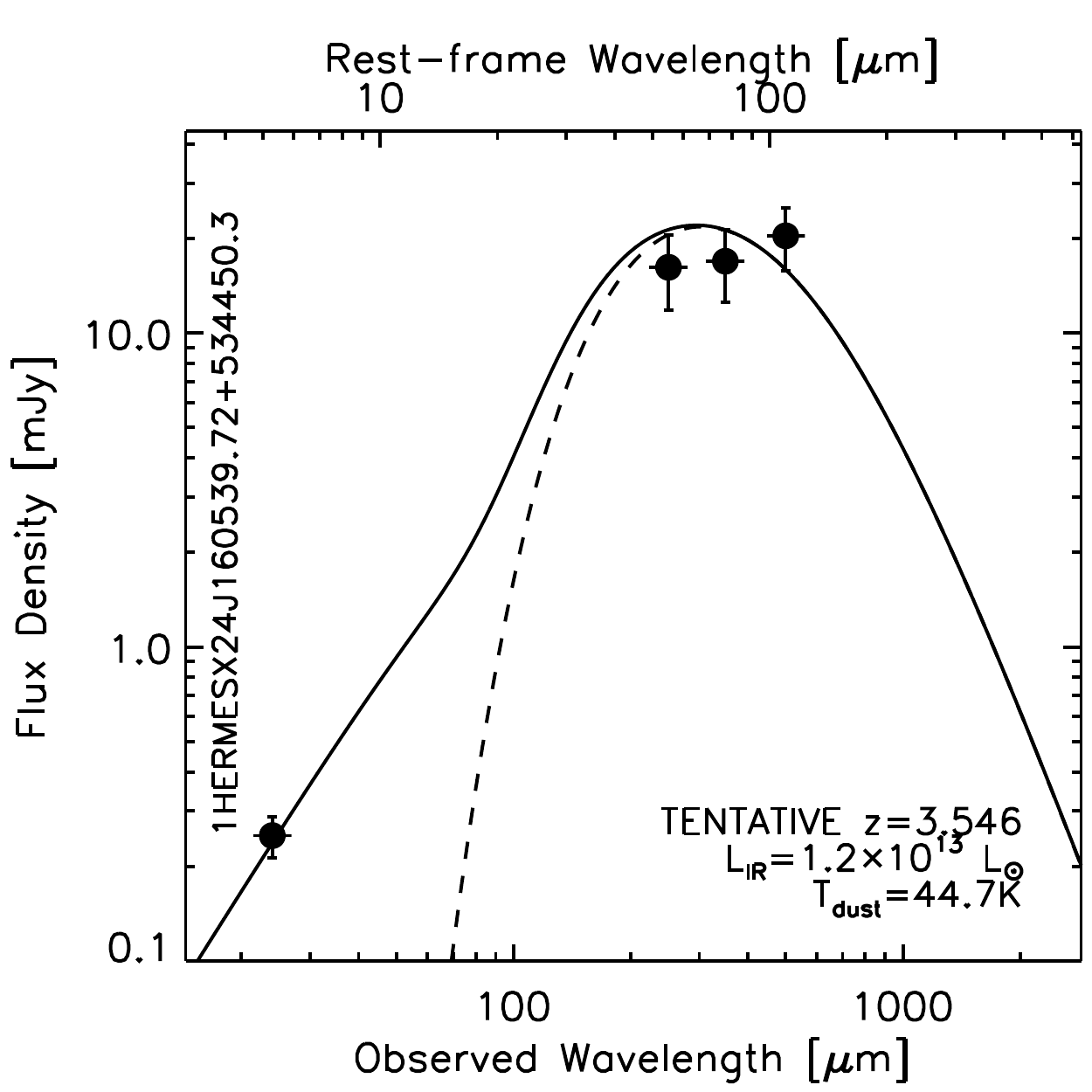}\\
\includegraphics[width=1.38\columnwidth]{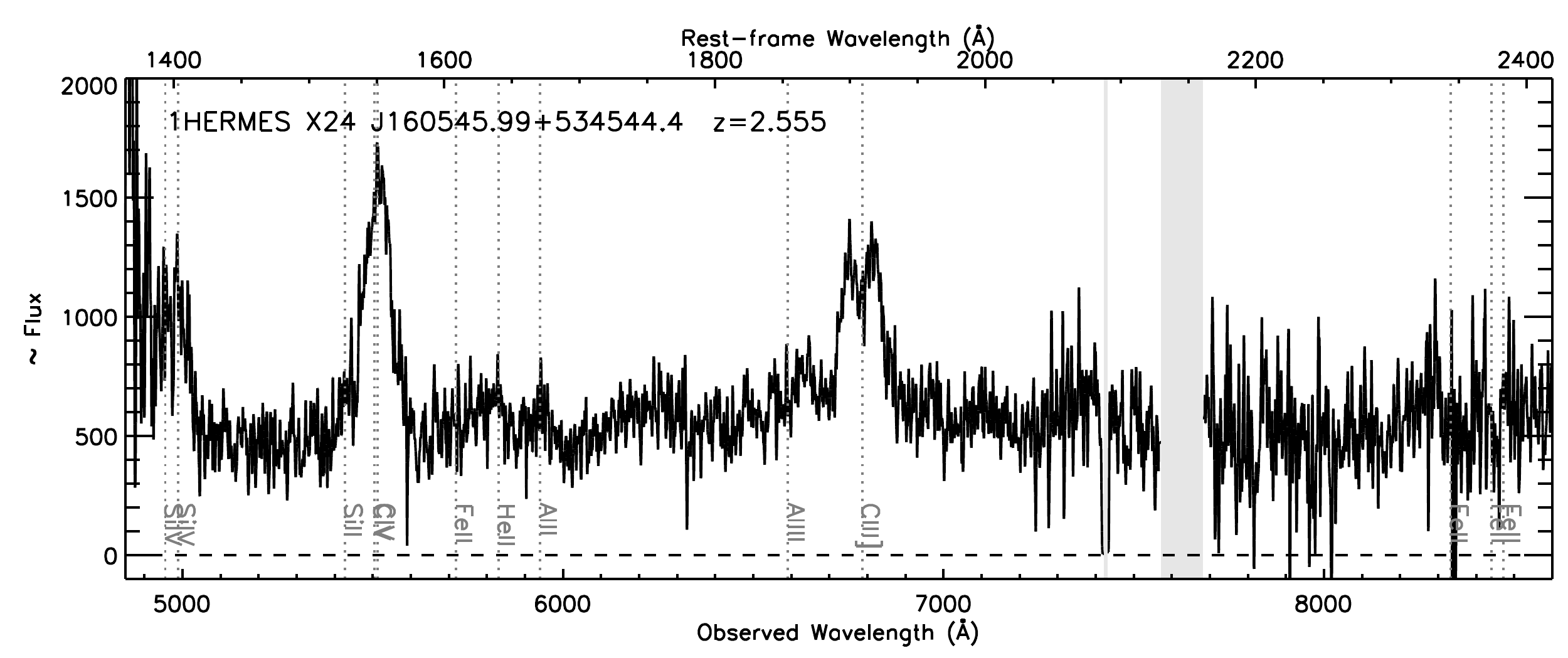}\includegraphics[width=0.58\columnwidth]{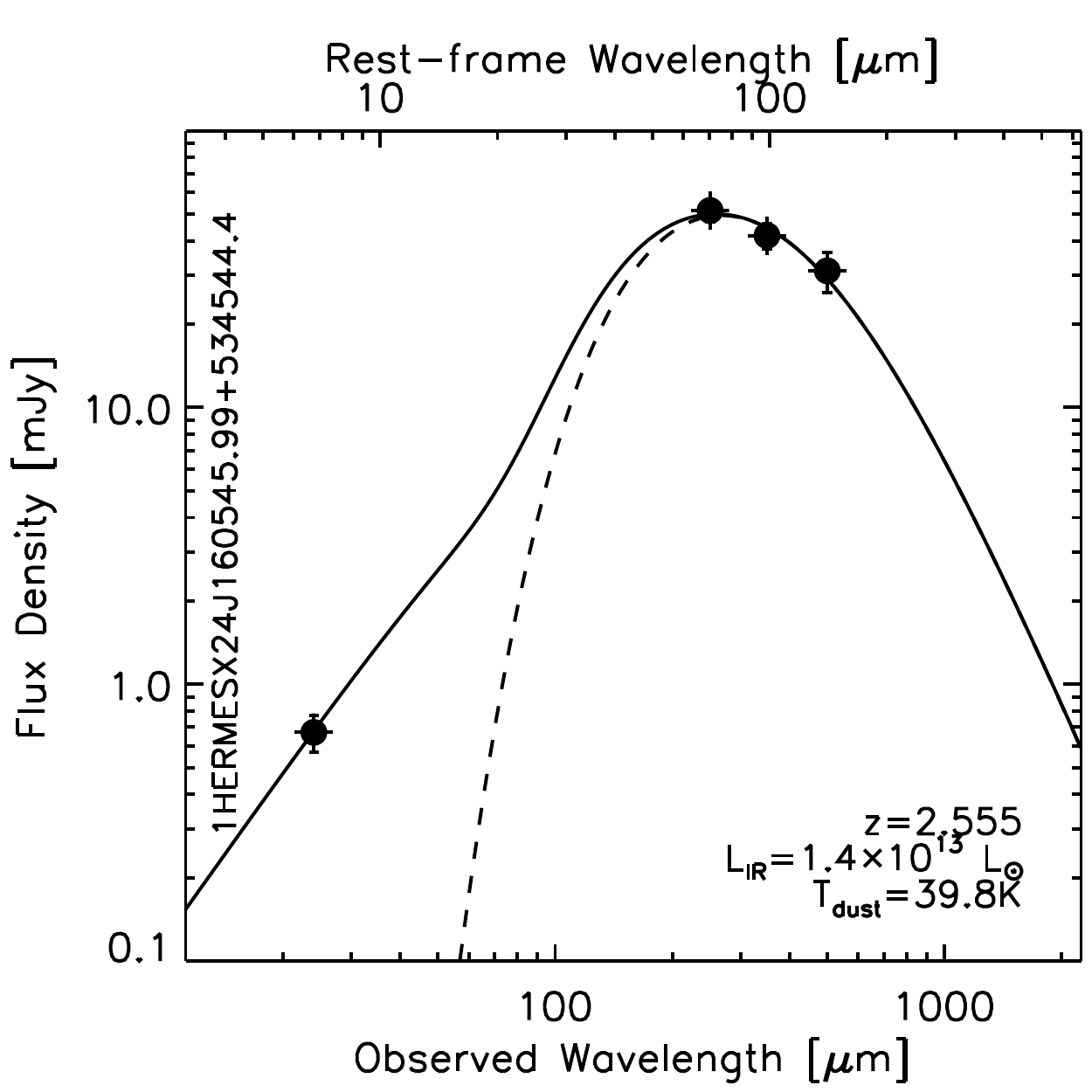}\\
\includegraphics[width=1.38\columnwidth]{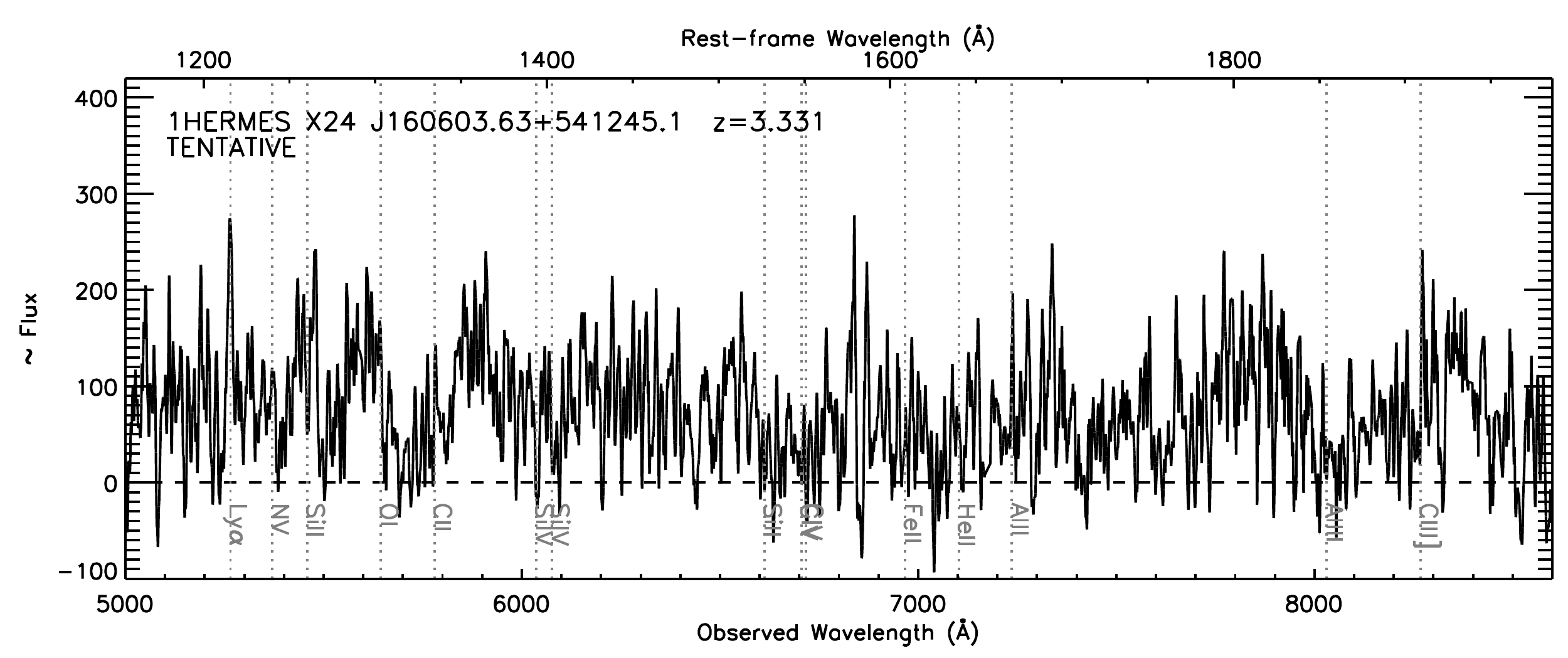}\includegraphics[width=0.58\columnwidth]{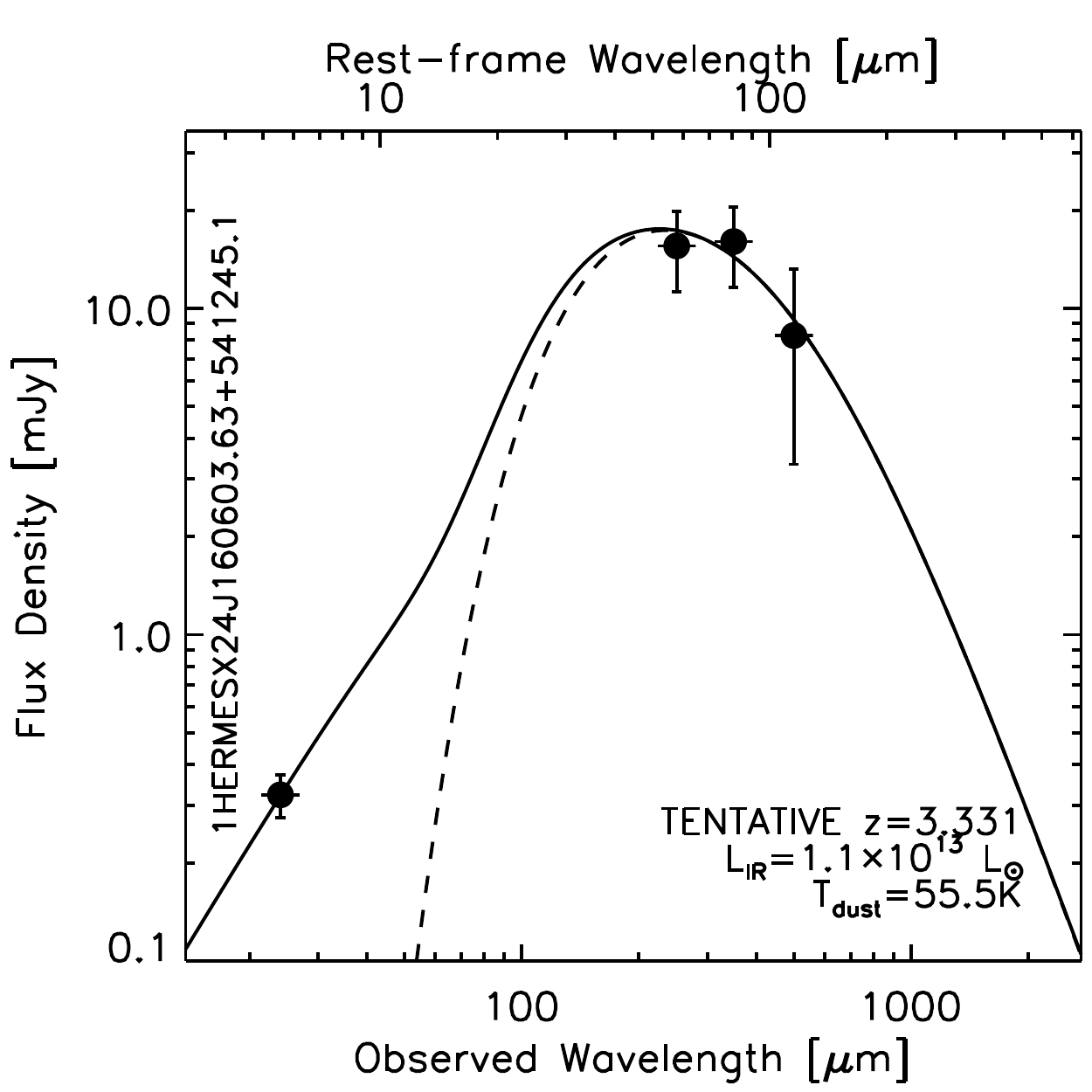}\\
\centerline{{\small Figure~\ref{fig:spectra} --- continued.}}
\end{center}
\clearpage
\begin{center}
\includegraphics[width=1.38\columnwidth]{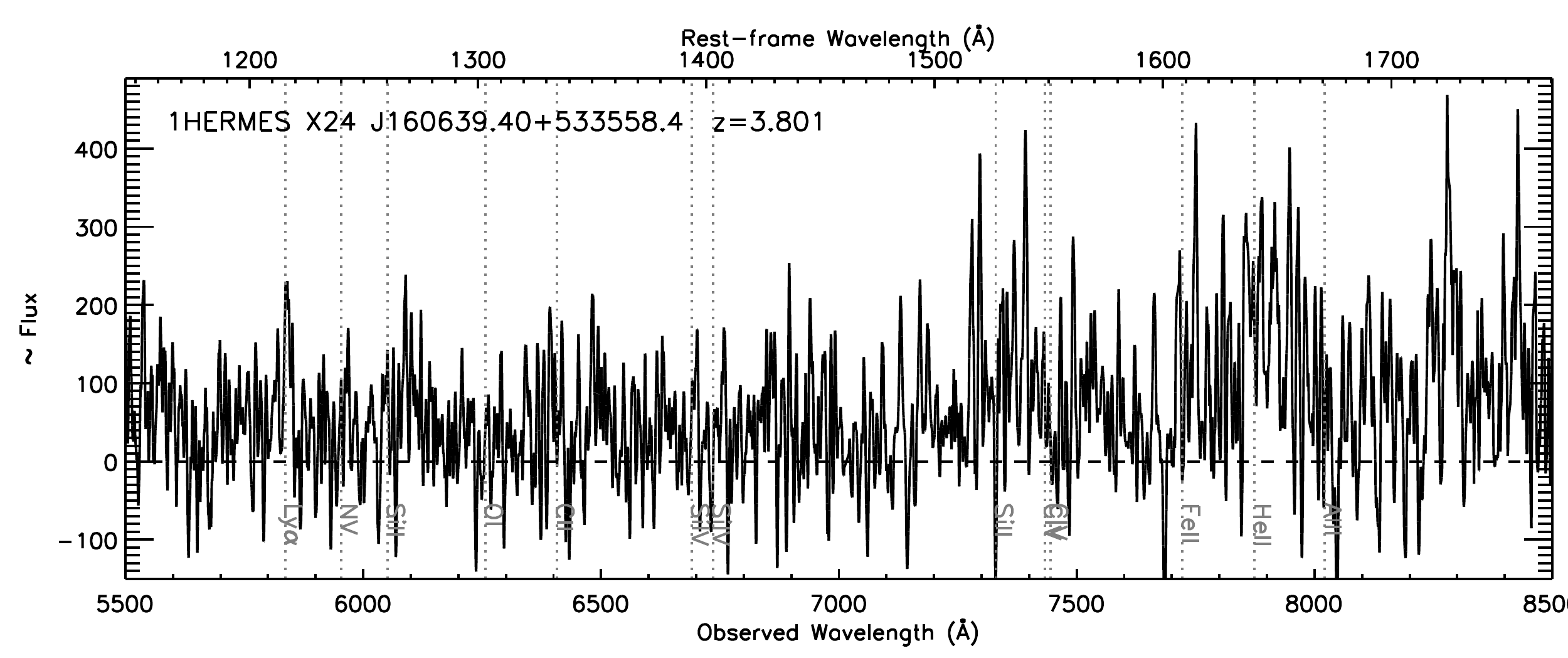}\includegraphics[width=0.58\columnwidth]{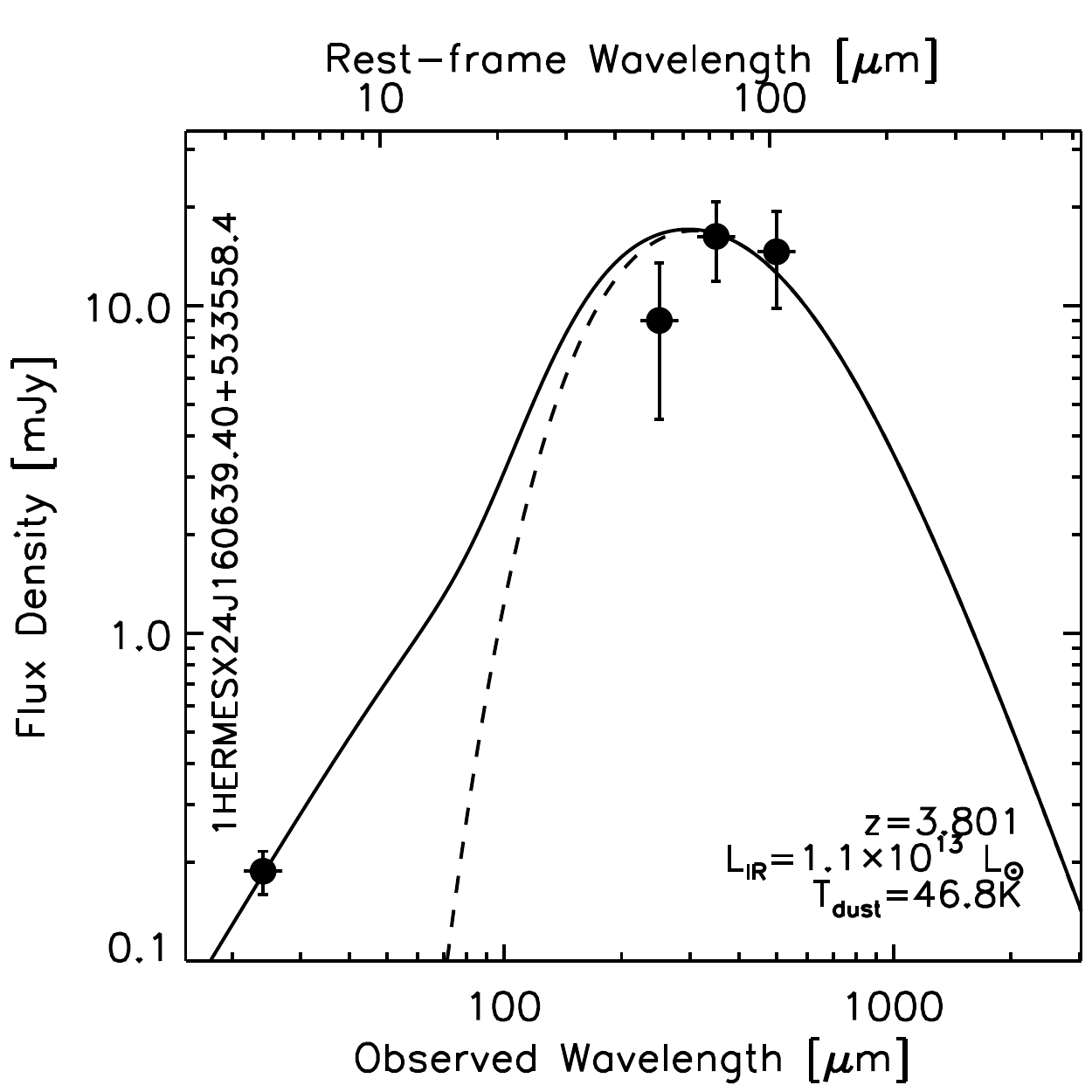}\\
\includegraphics[width=1.38\columnwidth]{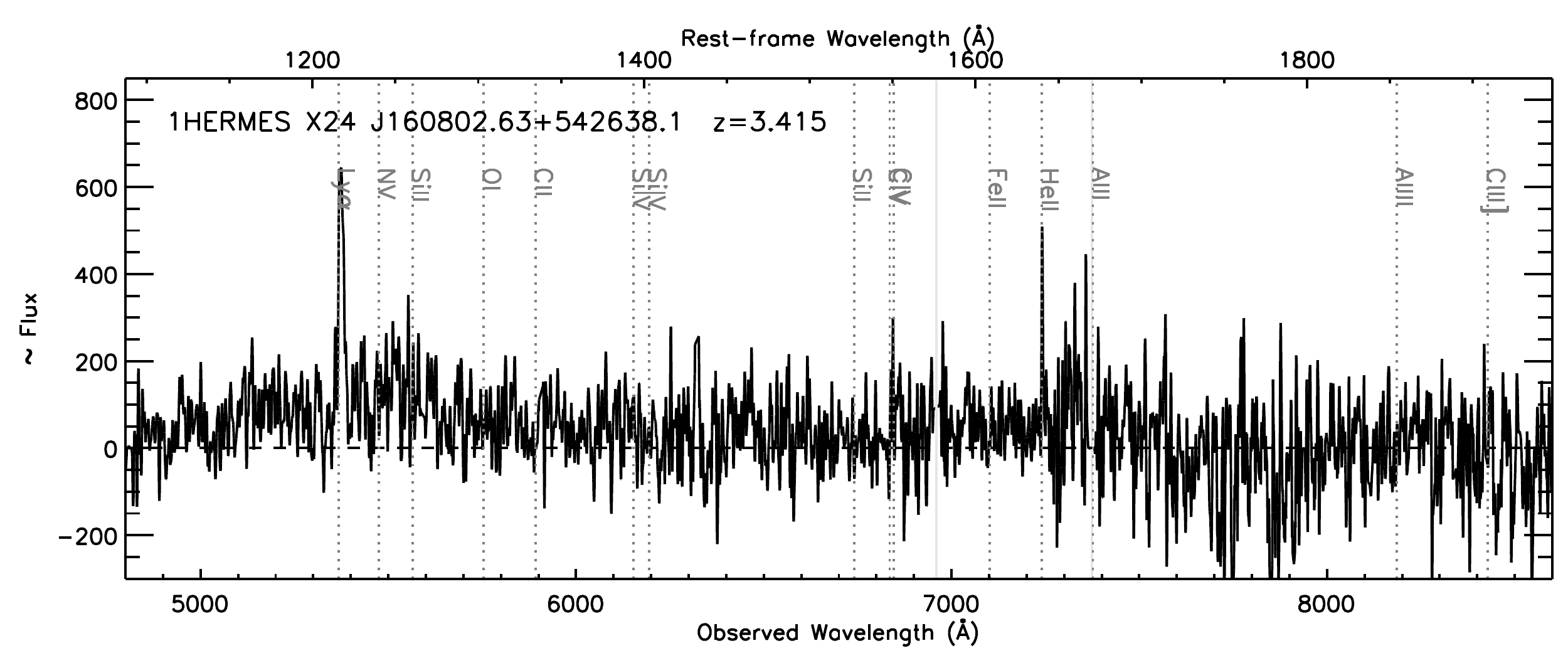}\includegraphics[width=0.58\columnwidth]{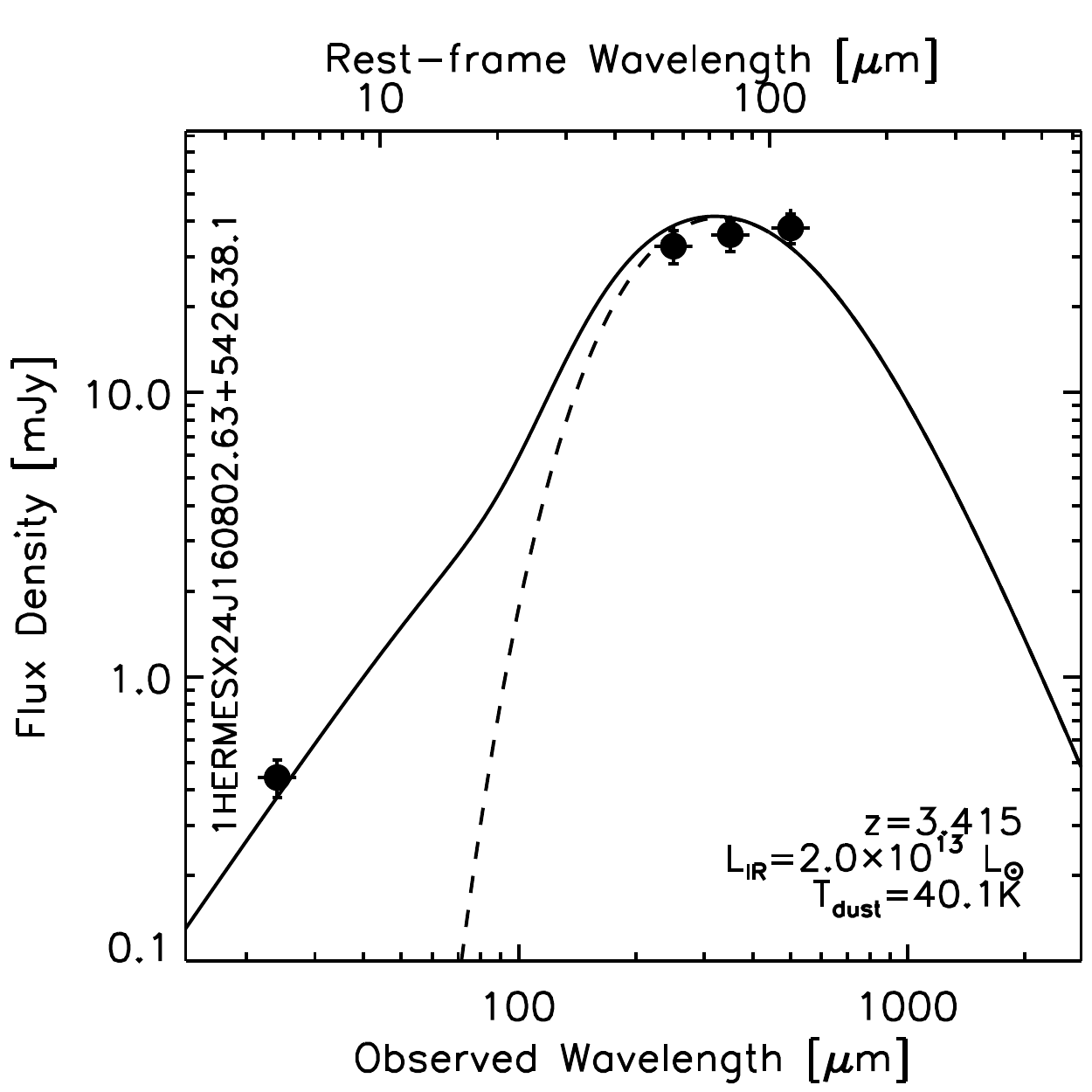}\\
\includegraphics[width=1.38\columnwidth]{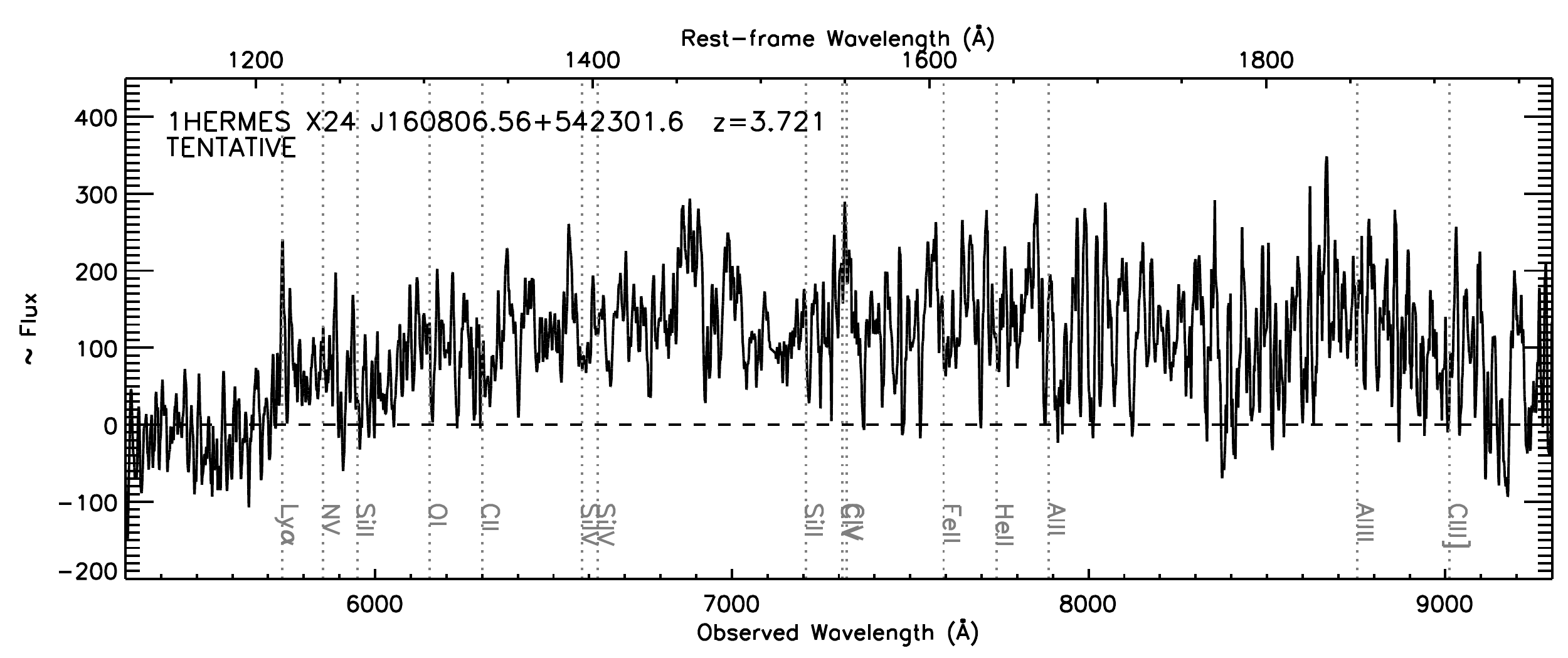}\includegraphics[width=0.58\columnwidth]{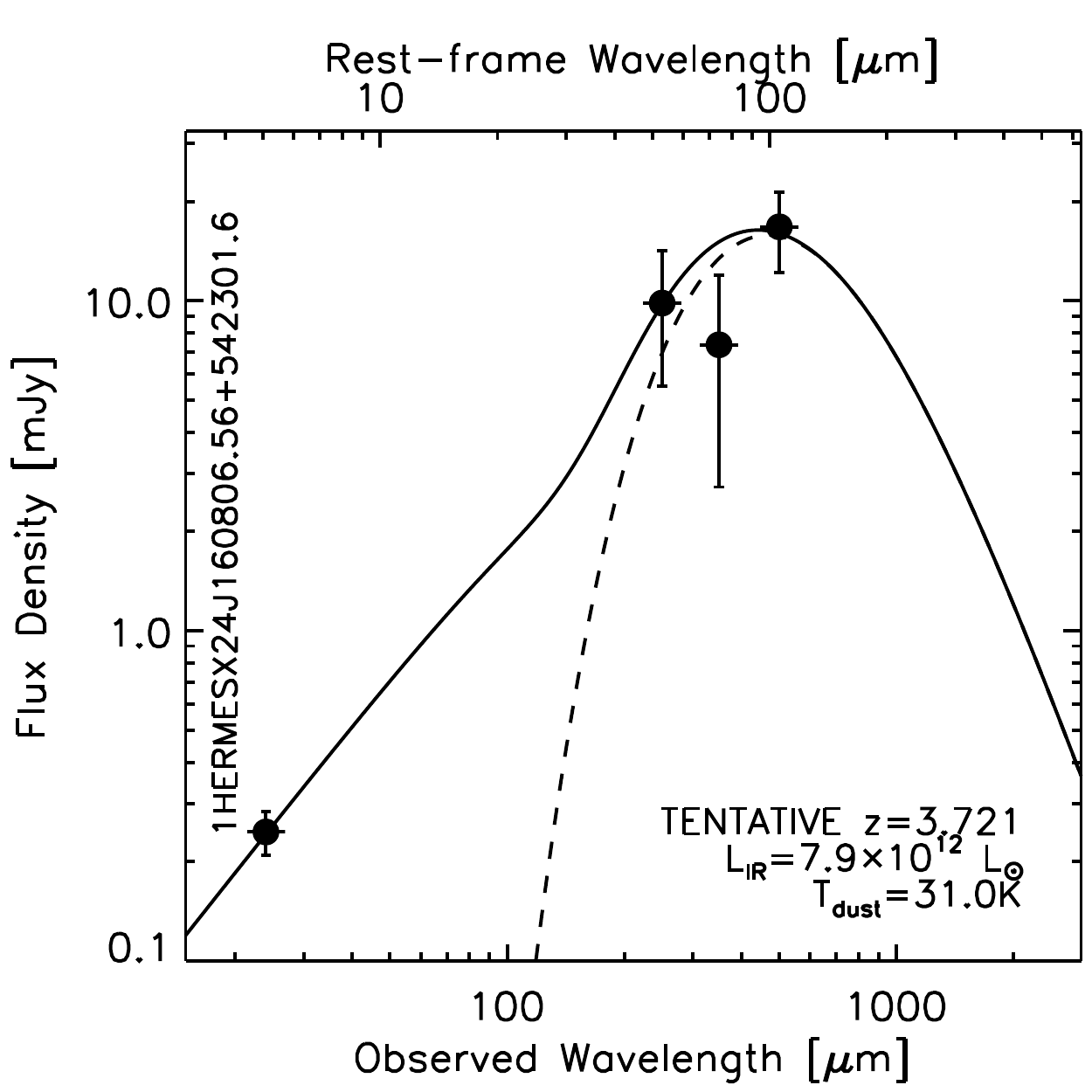}\\
\includegraphics[width=1.38\columnwidth]{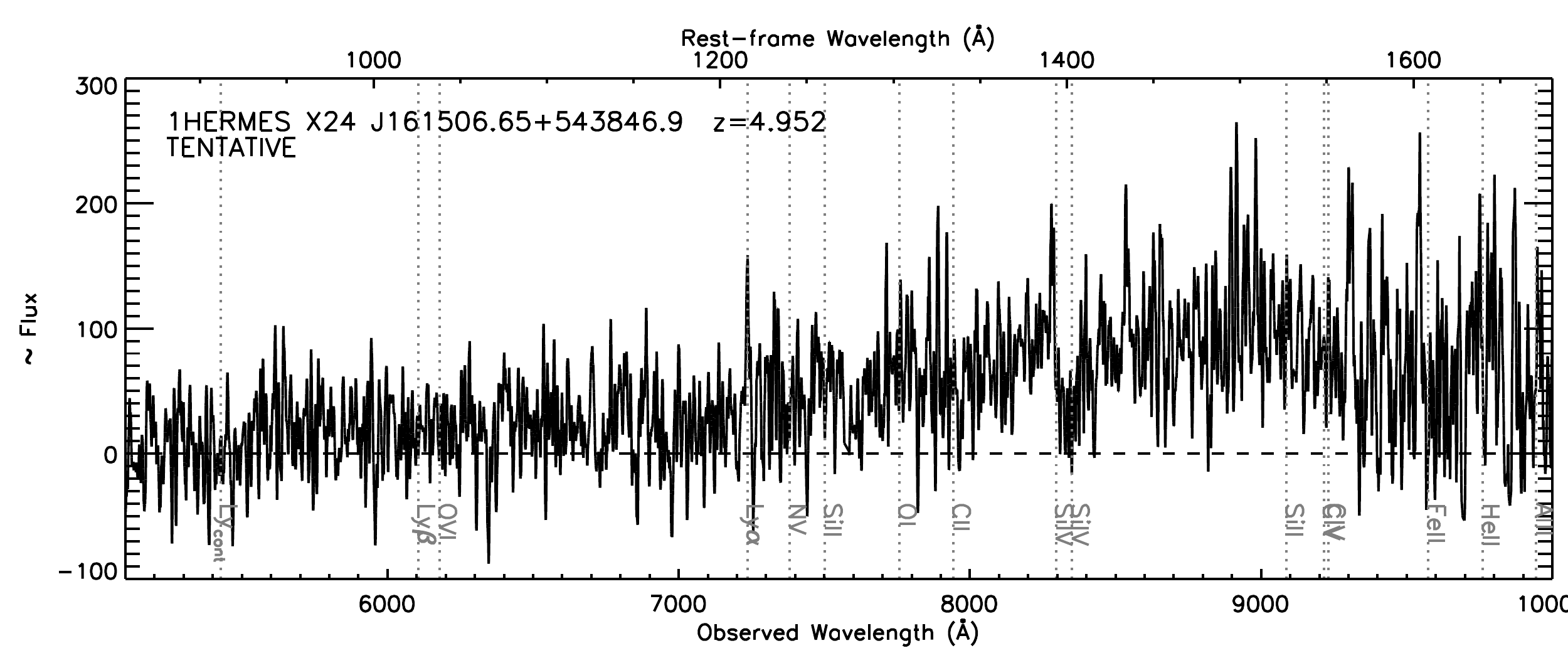}\includegraphics[width=0.58\columnwidth]{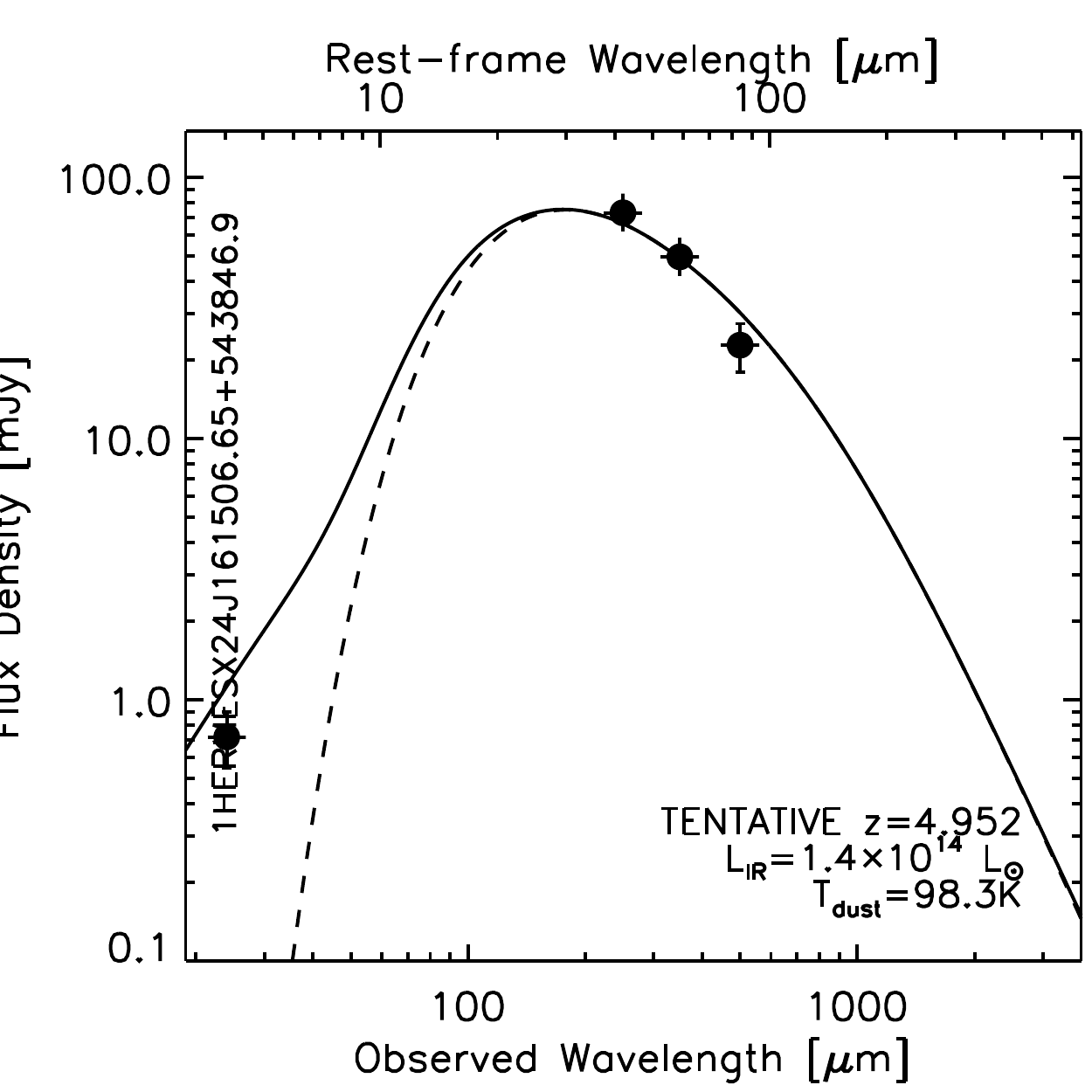}\\
\centerline{{\small Figure~\ref{fig:spectra} --- continued.}}
\end{center}
\clearpage
\noindent Fifteen of the 36 sources do not have photometric
redshifts$-$even though some of them lie in areas of very deep
ancillary data$-$a testament to their dusty and optically obscured
nature.  The sources with fewer identifiable spectral signatures in
our spectra are less optically luminous than those with multiple
features.  Those which are optically fainter will have less reliable
photometric redshifts and are more likely to be categorized here as
tentative.

\subsection{Alternate Redshift Identifications}

\begin{figure}
\centering
\includegraphics[width=0.99\columnwidth]{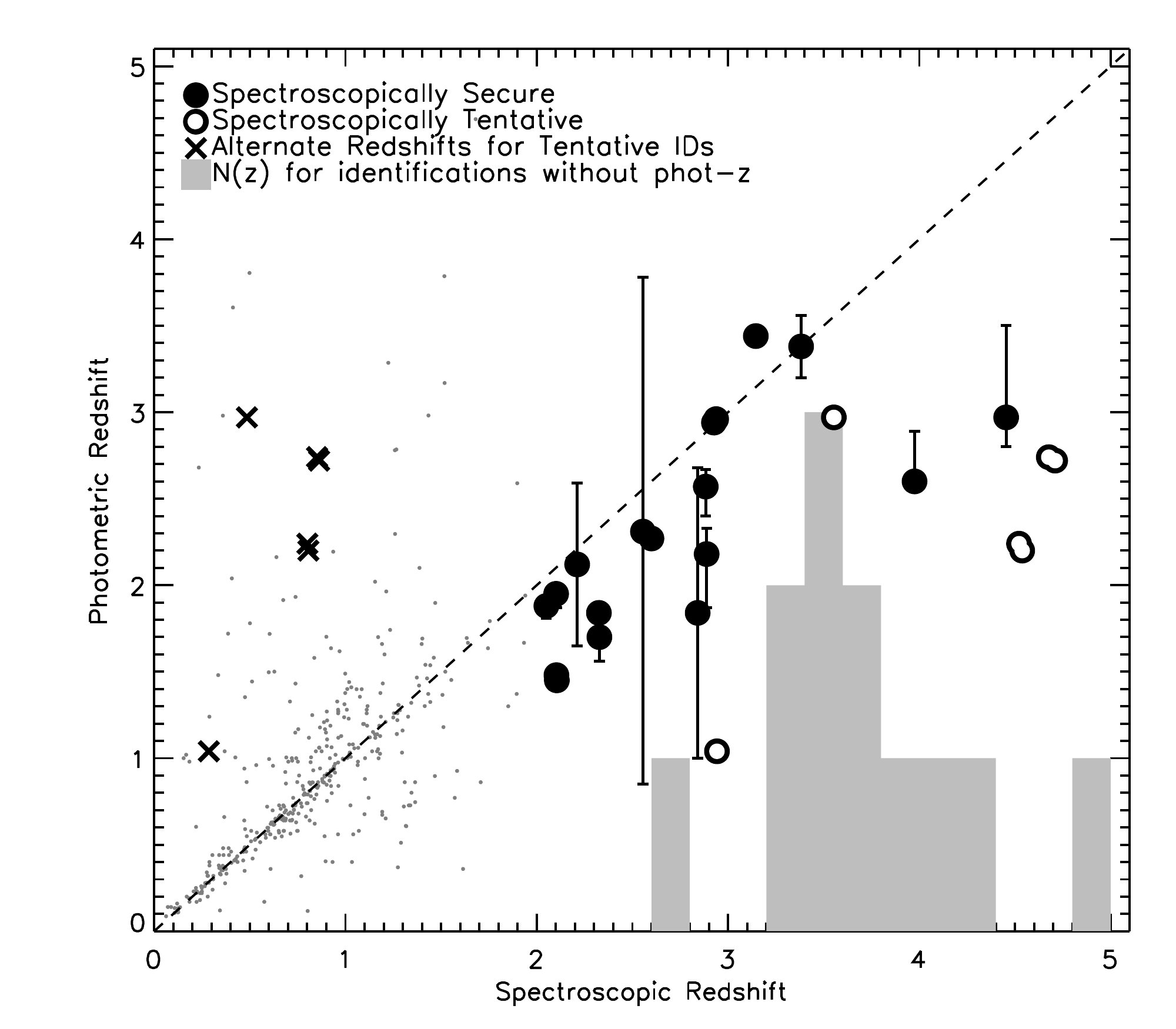}
\caption{Spectroscopic redshift against photometric redshift for our
  {\sc Spire}-selected galaxy sample.  Sources with identifications
  based on multiple spectroscopic features (emission lines, absorption
  lines, breaks) are shown as black circles, while identifications
  made from a single-feature (i.e. \lya\ emission) are open circles.
  The photometric and spectroscopic redshifts for the full sample of
  767 HSGs (C12) is shown as small gray points.  Since the single-line
  source emission lines could have been mis-identified, we also mark
  the corresponding spectroscopic redshifts if \oii\ is assumed and
  find generally poor agreement with the photometric redshift (black
  crosses).
  The distribution in spectroscopic redshifts for sources with $no$
  photometric redshifts is shown in gray; there is no obvious trend
  with redshift.
}
\label{fig:photzspecz}
\end{figure}

Figure~\ref{fig:photzspecz} plots photometric redshift against
spectroscopic redshift where photometric redshifts are available.  The
distribution in spectroscopic redshift of sources without photometric
redshift is the gray histogram.  Alternate redshifts, assuming the
emission lines are actually \oii\ instead of \lya, are shown as
crosses on Figure~\ref{fig:photzspecz}.  The \oii-implied
spectroscopic redshifts are less consistent with photometric redshifts
than the \lya-implied spectroscopic redshifts.  This makes sense in
the context of the probable enhanced \lya-to-continuum ratios in these
galaxies.

While our identified \lya\ emission lines could also be
\mgii\ (2798\AA), \mgii\ would be suggestive of a strong AGN.  With
AGN, other features (e.g. \ciii\,1909\AA\ and \oii) would also be
detectable. Given the expected relative strengths of \mgii\ relative
to \oii, \ciii, \civ, and \lya, \mgii\ is unlikely to be identified as
the only discernible feature.  In the case of \oii, both \oiii\ and
\hb\ emission lines and/or Ca\,H \&\ K absorption, or the
4000\AA\ break features are expected.  Also at wavelengths $>$6000\AA,
the \oii\ doublet is resolvable with LRIS and DEIMOS.  In comparison
to our lower redshift sample (C12), for which there are $\sim$500
\oii\ identified lines at the same wavelengths, the sources in this
paper have none of the above signatures which would point to a
mis-identification as \oii.

Also note that if the \lya\ identified lines were actually \oiii\ or
\hb\ they would sit at even lower redshifts than if \oii.  Because
there are no sufficiently bright emission lines between
\lya\ (1216\AA) and \oii\ (3727\AA) for starbursts, it is
straightforward to segregate between $z>2$ and $0<z<1$ sources using
photometric redshift.

We inspect the optical images of each source which does not have a
photometric redshift to judge the plausibility of our
identifications. All three GOODS-N sources are $u$-band drop-outs.
Both 1HERMES\,X1.4\,J123622.58 and 1HERMES\,X1.4\,J123732.66 are also
$b$-band drop-outs (the two higher redshift sources), and all three
have optical photometry which is consistent with their spectroscopic
identifications.  Of the eight LHN sources without photometric
redshifts, 1HERMES\,X1.4\,J104557.12 is compact and has $i\approx24.5$,
1HERMES\,X1.4\,J104701.68 and 1HERMES\,X1.4\,J104709.6 have
$i\approx24.3$, 1HERMES\,X1.4\,J104722.6 and 1HERMES\,X1.4\,J104707.7
have $i\approx25$, and 1HERMES\,X1.4\,J104620.4,
1HERMES\,X24\,J104642.9, and 1HERMES\,X1.4\,J104649.9 are $i$-band
drop outs.  These magnitudes do not rule out the possibility that these are
low-redshift sources, but indicates consistency between our high-$z$
spectroscopic identifications and photometry.  In COSMOS, only one
source, 1HERMES\,X24\,J100133.36+023726.9, has no photometric redshift
since it drops out in all images.
  Elais-N1 sources have much shallower photometric coverage than the
  other fields, hence more sources without photometric redshifts.
  While all sources are detectable in wide $i$-band imaging,
  multi-band imaging is not available across the whole field.  None of
  the sources are sufficiently bright or extended at $i$-band to be a
  convincing $z<1$ identification.

Note that 1HERMES\,X24\,J161506.65+543846.9, the highest redshift
source in our sample at $z=4.95$ has an odd assortment of photometric
measurements, dropping out in all wavebands (including $z$-band)
except $i$-band, where it has a magnitude of 22.8 (AB).  The low
photometric redshift ($z_{\rm p}=1.94$ is likely caused by this
peculiar optical SED, but is also perfectly consistent with \lya\ in
$i$-band at $z\sim5$, and enhanced \lya-to-continuum ratios
\citep{neufeld91a}.  We also note that this source is classified as
tentative.


\section{Sample Characteristics}\label{sec:results}

The importance and context of the $2<z<5$ HSG population can only be
judged with a basic understanding on the physical characteristics of
the sample.  Here we measure those physical characteristics, compare
them to the properties of other galaxy populations, and assess the
impact of infrared selection biases on our interpretation.

\subsection{SED fits, luminosities and dust temperatures}

The infrared photometry summarized in Table~\ref{tab_fullsample} is
fit with a FIR spectral energy distribution (SED) consisting of a
coupled single dust temperature modified blackbody and mid-infrared
power law such that
\begin{equation}
\label{eq1}
S(\lambda) =
N_{\rm bb}\,\frac{(1-e^{-(\frac{\lambda_{0}}{\lambda})^{\beta}})(\frac{c}{\lambda})^{3}}{e^{hc/\lambda\!kT}-1}
+
N_{\rm pl}\,\lambda^{\alpha}\,e^{-(\frac{\lambda}{\lambda_{\rm c}})^2}
\end{equation}
where S$(\lambda)$ is in units of Jy and $T$ is the galaxy's
characteristic ``cold'' dust temperature (the dust temperature
dominating most of the infrared luminosity and dust mass).  The
emissivity index is represented by $\beta$, and $\lambda_{0}$ is the
wavelength at which optical depth is unity \citep[here fixed at
  $\lambda_{0}=200$\,\um, as described in ][]{conley11a}.  The slope
of the mid-infrared powerlaw component is given by $\alpha$, and
$\lambda_{\rm c}$ is the wavelength where the gradient of the modified
blackbody is equal to $\alpha$.  $N_{\rm bb}$ and $N_{\rm pl}$ are the
coefficients of the modified blackbody and power laws respectively.
$N_{\rm pl}$ is a fixed function of $N_{\rm bb}$, $\alpha$, and $T$
such that the powerlaw and modified blackbody are continuous at
$\lambda_{\rm c}$.  This SED fitting method is described fully in
\citet{casey12a}, and is also discussed in C12 as applied to the
low-redshift population. It is given in related forms in
\citet{blain02a}, \citet{blain03a}, \citet{younger09a}, and
\citet{conley11a}.

To reduce the number of free parameters, we fix $\alpha\,=\,2.0$ for
sources without {\sc Pacs} photometry and $\beta\,=\,1.5$ for all
\citep[the measured values found for local IRAS and some distant
  ULIRGs][]{casey12a}.  These leaves two free parameters: $N_{\rm
  bb}$, which effectively scales with $L_{\rm IR}$, and $T$, the
temperature of the modified blackbody.  We remeasure dust temperatures
for each galaxy by determining the wavelength where the flux density
peaks and convert that to a dust temperature via Wien's Law.  This
provides a more consistent measure of dust temperature which can be
used in comparisons between SEDs fit with alternate techniques, using
model templates or direct fits
\citep{chary01a,dale02a,blain03a,siebenmorgen07a,draine07a,kovacs10a}.
We compute infrared luminosities by integrating the above best-fit
spectral energy distribution between 8 and 1000\um.  The SED fits are
shown alongside the sources' optical spectra in
Figure~\ref{fig:spectra}, and their infrared luminosities and dust
temperatures are given in Table~\ref{tab_fullsample}.

The luminosities of this sample range from
3.2$\times$10$^{12}$--6.3$\times$10$^{13}$\lsun, implying infrared
star formation rates of 500--9000\sfr (with one outlier at
1.6$\times$10$^{14}$\lsun, 26000\sfr, whose redshift is tentative).
Most galaxies in the sample have SFRs an order of magnitude beyond the
extreme activity seen in ULIRGs \citep[which have
  $SFR\approx$\,200--1000\sfr\ by the scaling given
  in][]{kennicutt98a}.  These starbursts are amongst the most extreme
star forming galaxies seen in the Universe \citep[amongst other HyLIRG
  populations,][]{rowan-robinson00a,bridge12a}.  

The conspicuously high star formation rates (e.g. above
$\sim$1000\sfr) might lead us to believe that AGN contaminate the FIR
luminosity or rather, that there is potential variation in star
formation laws at high redshift.  There has been some recent
discussion of whether or not the \citet{kennicutt98a} scaling between
IR luminosity and SFR holds under `extreme' conditions or at
high redshifts \citep{swinbank08a}.  Assuming a modified IMF
would produce more modest SFRs than the default Salpeter IMF.  While
this might change our interpretation and change the star formation
rates we measure here, we use the Kennicutt scaling for SFRs in this
paper to be consistent with literature work.

\begin{figure}
\centering
\includegraphics[width=0.99\columnwidth]{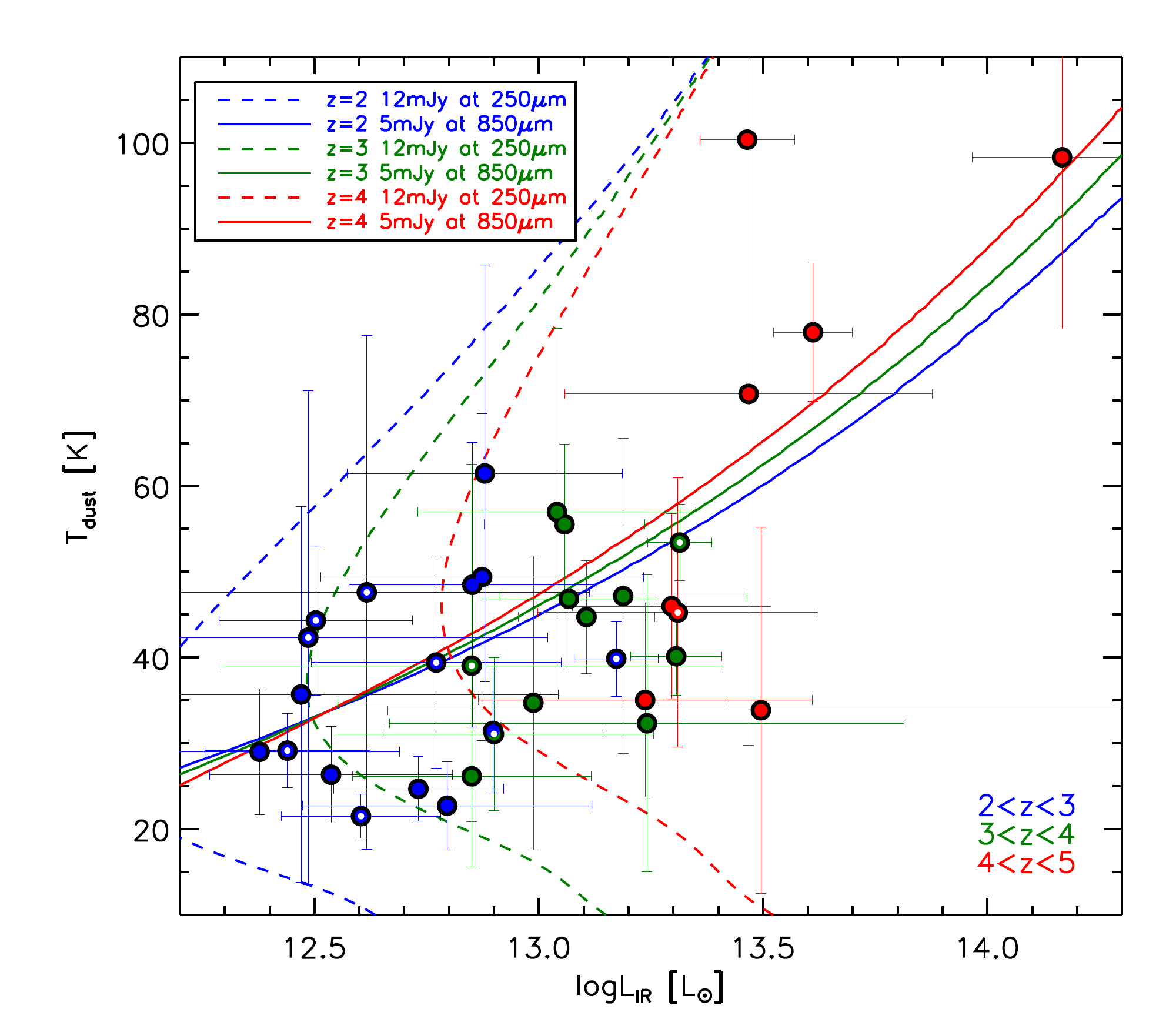}
\caption{Infrared luminosity against dust temperature for the sample,
  color coded by redshift interval: $2<z<3$ (blue), $3<z<4$ (green),
  and $4<z<5$ (red). Overplotted are typical lower luminosity limit
  boundaries$-$a function of dust temperature$-$at a given wavelength,
  redshift, and flux density limit.  A 12\,mJy flux density limit is
  assumed for the {\sc Spire} bands (dashed lines) and a 5\,mJy limit
  is assumed at 850\,\um\ (solid lines). This illustrates how half of
  the {\sc Spire}-selected sample have dust temperatures too warm to
  be 850\,\um-detectable.  Sources with AGN optical signatures are
  marked with small white dots at their centers.}
\label{fig:lirtd}
\end{figure}
Figure~\ref{fig:lirtd} shows the infrared luminosity against dust
temperature for the HSG sample.  There is a noticable absence of very
warm sources at lower luminosities.  Similarly, there are very few
cold sources at high luminosities.  This is primarily a consequence of
selection effects in the {\sc spire} bands.  Warm-dust galaxies are
selected against in the {\sc Spire} bands, even at these
high-redshifts, due to the sensitive variation of infrared flux
density measurements with dust temperature; the dashed lines
illustrate the lower luminosity detection limits as a function of dust
temperature, for a galaxy at $z=2$, $z=3$, or $z=4$.  This
dust-temperature selection bias is even more exaggerated at 850\,\um\.
The luminosity detection limits for 850\um\ selection are shown as
solid lines on Figure~\ref{fig:lirtd}, nearly bisecting the {\sc
  Spire} population so that about half would be 850\,\um-undetected).
Before {\it Herschel}, the cold-dust temperature bias of submillimeter
observations was the focus of many studies looking for the elusive
``warm-dust'' SMGs
\citep{blain04a,chapman04a,casey09b,casey11a,chapman10a,magdis10a,chapin11a}.

\begin{figure}
\centering
\includegraphics[width=0.99\columnwidth]{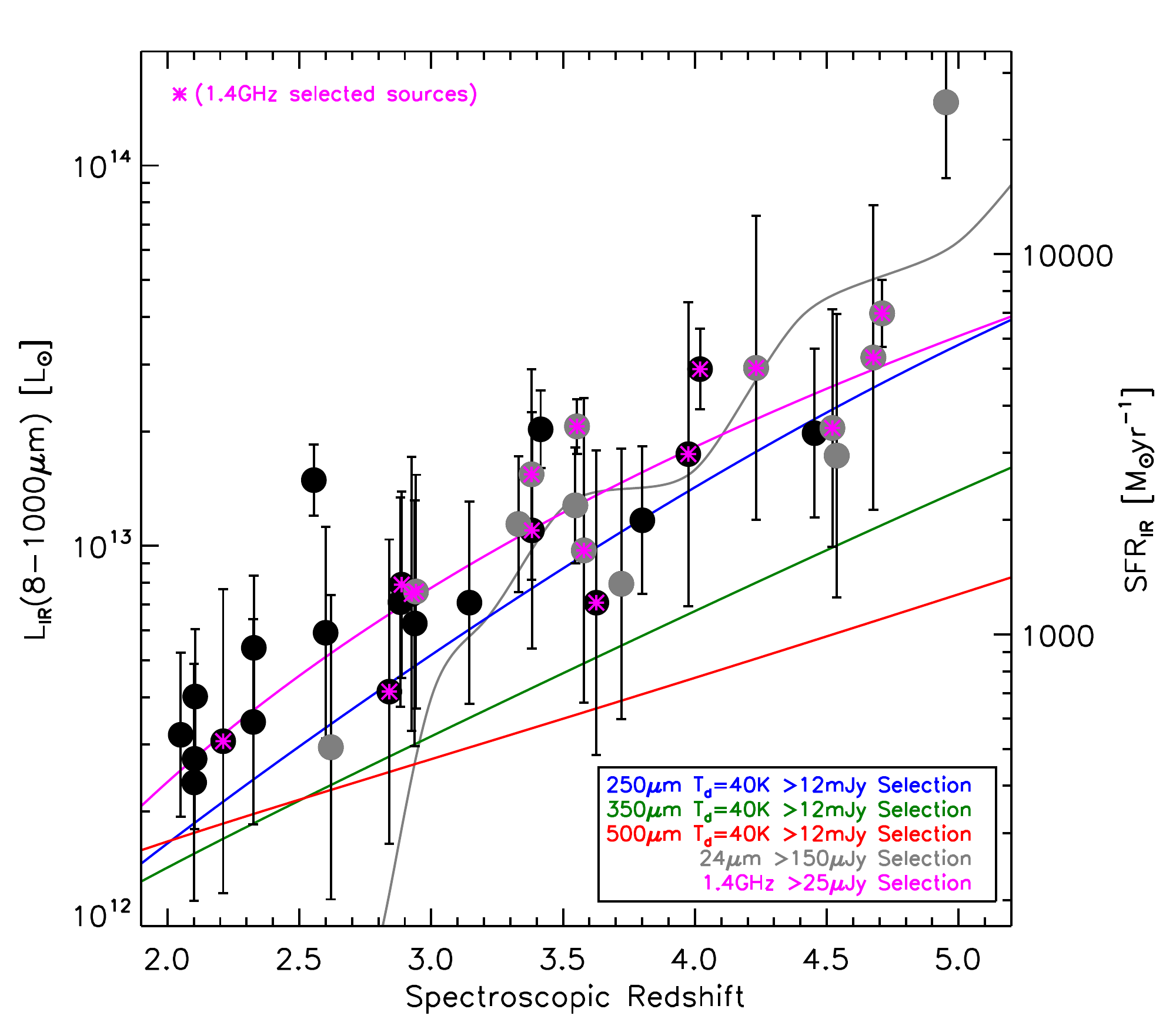}
\caption{Integrated infrared luminosity (8--1000\um) against
  spectroscopic redshift for our sample (black sources are secure and
  gray sources are tentative, as in Table~\ref{tab_fullsample}).
  Luminosities span 2.4$\times$10$^{12}$--4.0$\times$10$^{13}$\lsun,
  with one outlier at 1.4$\times$10$^{14}$\lsun.  The detection
  boundaries of each selection wavelength are also illustrated as a
  function of redshift.  The {\sc Spire} detection boundaries assume
  flux densities $>$12\,mJy and an SED dust temperature of 40\,K.  The
  radio boundary assumes that the FIR/radio correlation
  \citep{helou85a} holds and a $>$40\,\uJy\ radio detection limit.
  The 24\,\um\ limit assumes a 150\,\uJy\ detection limit along with a
  typical 24\,\um--$L_{\rm IR}$ scaling \citep[e.g.][]{le-floch05a}.
  Radio sources are marked by a magenta asterisk, while
  24\,\um\ sources are black. }
\label{fig:lirz}
\end{figure}

Figure~\ref{fig:lirz} plots HSGs' infrared luminosities against
redshift relative to the detection limits of the selection
wavelengths: the three 250, 350, and 500\,\um\ {\sc Spire} bands,
24\,\um\ and 1.4\,GHz.  The detection limits in the {\sc Spire} bands
depend on the assumed SED shape and dust temperature which is not
uniform for all {\sc Spire}-selected galaxies.  We use the
distribution of measured dust temperatures in Figure~\ref{fig:lirtd}
to set limits on the detection boundaries in $L_{\rm IR}$--$z$ on
Figure~\ref{fig:lirz}.  In other words, we measure that the mean dust
temperature for 250\um-selected galaxies to be 40\,K, for
350\um-selected galaxies as 39\,K, and for 500\um-selected galaxies as
37\,K.  Differences on the scale of a few degrees are negligible for
these illustrative boundary lines in Figure~\ref{fig:lirz}, so we
adopt a 40\,K SED for all three selection boundaries.  Most of the
fields which have radio data have a detection limit \simgt40\,\uJy, so
we construct the radio detection boundary based on the FIR/radio
correlation for starbursting galaxies \citep{helou85a,condon92a} with
$q_{\rm IR}$ evolving as in \citet{ivison10b}.  The 24\,\um\ detection
boundary is the least certain as it scales with $L_{\rm IR}$; many
recent works note up to $\sim$1\,dex disagreement between extrapolated
24\,\um\ infrared luminosities and direct measurements
\citep[e.g.][]{le-floch05a,le-floch09a,elbaz11a}.  This tells us that,
although we can use the 24\,\um\ detection boundary shown in
Figure~\ref{fig:lirz} as a rough guide, it should not be concerning
that 24\,\um-selected sources fall below the line by \simlt0.3\,dex.

Although very few of these sources have existing 850\,\um\ data, this
work (especially Figure~\ref{fig:lirtd}) suggests that half of the
{\sc Spire}-selected population would be undetectable in the original
{\sc Scuba} 850\,\um\ surveys $even$ at these redshifts.
Specifically, we estimate that only 31/36 sources would have $S_{\rm
  850}>2$\,mJy (86\%), while only 21/36 sources have $S_{\rm
  850}>5$\,mJy (58\%).  The statistics of the temperature-bias
selection effect are discussed in more detail in C12, as they relate
to the lower redshift population where the statistics are more robust.

\subsection{FIR/radio correlation}

For the 23 adio-detected sources, we investigate the FIR/radio
correlation for starburst galaxies \citep{helou85a,condon92a}.
Measuring the FIR/radio correlation in this sample is useful for
checking that our sample is roughly consistent in FIR/radio given
expectation from our measured redshifts (i.e. it is another
reassurance on tentative identifications in particular).  This
correlation is measured via the ratio of FIR luminosity to radio
luminosity such that
\begin{eqnarray}
q_{\rm IR}&=&\log\left ( \frac{1.02\times10^{18}\,L_{\rm FIR}}{4\pi\,D_{\rm L}^{2}}\,\left [\frac{{\rm cm}^{2}}{L_{\odot}} \right] \right) \\
 & & - \log\left (1\times10^{-32}\,S_{\, 1.4}\,(1+z)^{\alpha-1}\right [\mu\!{\rm Jy}^{-1}] ) \nonumber
\end{eqnarray}
where $L_{\rm FIR}$ is the far-infrared luminosity measured in the
range 40--120\,\um\ given in \lsun, $D_{\rm L}$ is the luminosity
distance in cm, S$_{\rm 1.4}$ is the 1.4\,GHz flux density in \uJy,
and $\alpha$ is the synchrotron slope, here set to 0.75
\citep{ibar10a,ivison10a} and defined such that
$S_{\nu}\,\propto\,\nu^{-\alpha}$.

Twenty-three of the 36 galaxies in our sample are radio detected
(64\%), and their measured $q_{\rm IR}$ ranges from 0.7 to 2.1 with
mean value $\langle\!q_{\rm IR}\rangle\!=1.58\pm0.35$.  Note that the
infrared luminosity component in $q_{\rm IR}$ is $L_{\rm
  FIR}(40-120)$, not $L_{\rm IR}(8-1000)$.  If $L_{\rm IR}(8-1000)$ is
used instead, luminosities and $q$ increase by 0.40$\pm$0.15\,dex and
the scatter in $q$ grows; the increased scatter is caused by the
contribution from the mid-infrared flux to $L_{\rm IR}$.  Only one
galaxy in our sample is 'radio-loud' and indicative of AGN (the source
at $z=3.579$); this is consistent with the observed \civ\ emission in
its rest-UV spectrum.  Since we observe the rest of the radio-detected
sample to agree within uncertainties with previous measures of $q_{\rm
  IR}(z)$ in previous samples \citep{ivison10a,kovacs10a,magnelli12a}
lends additional credence to our redshift identifications.  While each
of the literature samples have different measures for $q$ (ranging
1.3--2.2), all with uncertainties on the order of $\sim$0.15--0.25,
the overall trend of an evolving $q_{\rm IR}$ is consistent between
samples.


\subsection{Composite Ultraviolet Spectra}

Since the signal-to-noise on individual galaxy spectra shown in
Figure~\ref{fig:spectra} is quite low for most sources (except in the
detection of \lya), we construct a composite rest-frame ultraviolet
spectrum which can serve two purposes: it validates bulk redshift
identification by way of detecting lower S/N spectral features around
\lya, and it begins to shed light on the intrinsic rest-frame
ultraviolet emission properties of extremely infrared luminous
starbursts.  Unfortunately, larger samples are necessary to perform
the latter analysis; in this work, our primary goal is to help
validate our redshift identifications through cross-correlation to the
composite spectrum for sources with only single-line identifications
(e.g. \lya).

It is clear from Figure~\ref{fig:spectra} that these
infrared-starbursts exhibit a wide range in spectral properties, from
Lyman Break Galaxy (LBG) spectra, quasar spectra, starburst spectra,
to those with very steep to very shallow UV slopes.  We classify
sources as LBGs if they exhibit a steep cutoff in continuum flux at
rest-frame 1216\AA, as quasars if they exhibit broad, high-ionization
emission lines, and starbursts as narrow-emission line galaxies whose
emission line luminosity is more significant than continuum
luminosity.  Many sources are dust obscured and thus are noisy except
for the detection of \lya\ emission.  The construction of a composite
spectrum serves as a sanity check on the redshifts.  The detection of
lower S/N features in a composite spectrum does not {\it directly}
confirm that every galaxy added in is correctly identified, but it
does indicate that most of them are.  Since co-adding spectra from
higher S/N sources would wash out the low S/N features of the other
galaxies (in a sample of 36), we exclude the following sources from
the composites (excluded on the basis of detection of
non-\lya\ spectral features at $>$3$\sigma$ significance, including
continuum and \civ):
1HERMES X24 J033136.96 $-$275510.9,
1HERMES X24 J095917.28 +021300.4,
1HERMES X24 J095948.00 +024140.7,
1HERMES X1.4 J100008.64 +022043.1,
1HERMES X24 J100020.16 +021725.2,
1HERMES X24 J100036.00 +021127.6,
1HERMES X1.4 J100024.00 +021210.9,
1HERMES X1.4 J100111.52 +022841.3,
1HERMES X24 J100146.56 +024035.6,
1HERMES X1.4 J104620.40 +585933.4,
1HERMES X1.4 J104636.00 +585650.0,
1HERMES X24 J160545.99 +534544.4, and
1HERMES X1.4 J033151.94 $-$275326.9.
In other words, the composite spectra are only made up of
\lya\ single-line detections, those with low $\sim$2-3$\sigma$
\civ\ detections, and those without.

\begin{figure*}
\centering
\includegraphics[width=1.75\columnwidth]{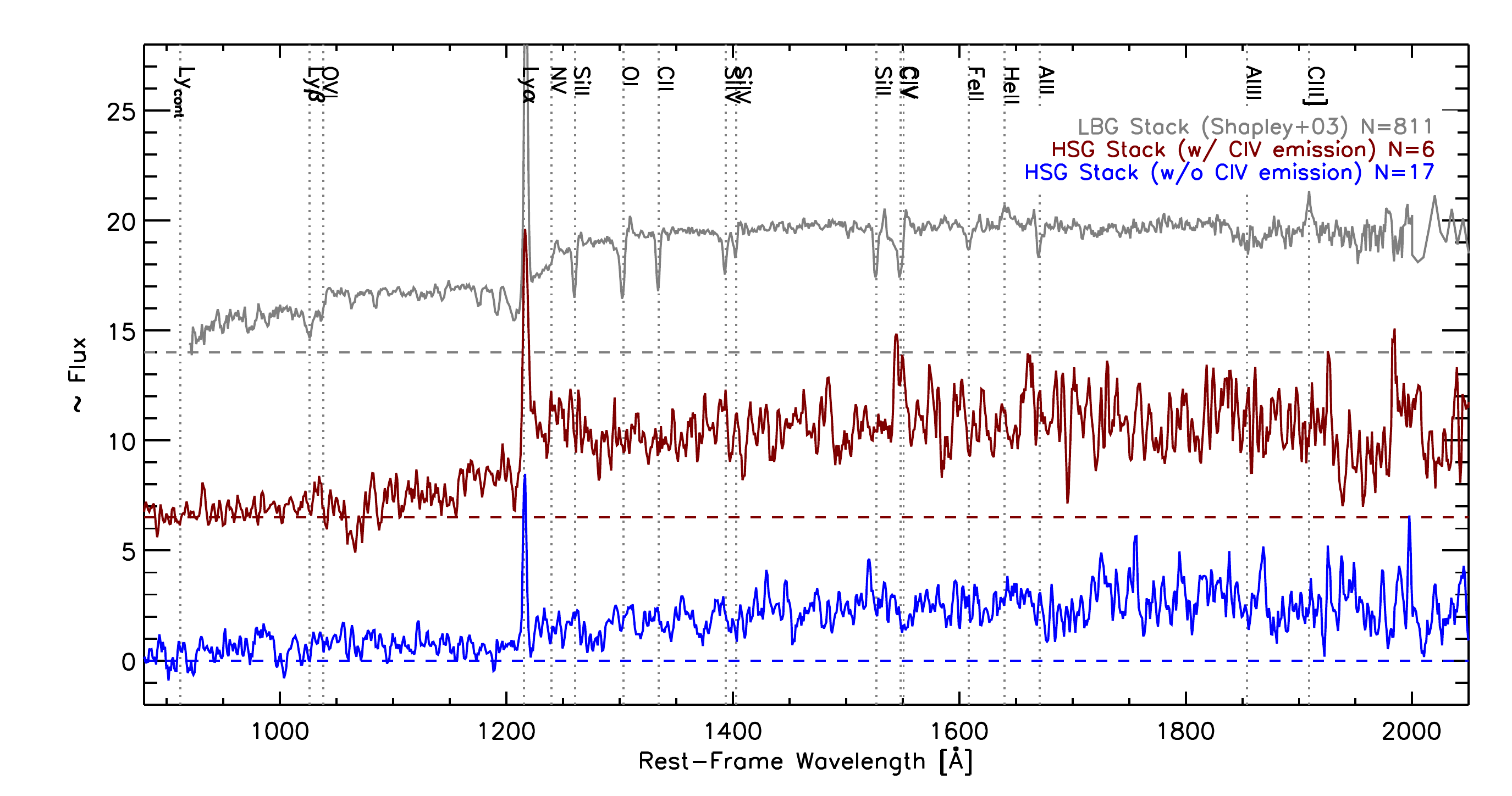}
\caption{A comparison of the rest-frame ultraviolet composite spectra
  for {\it Herschel}-selected galaxies (HSGs) with the composite of
  Lyman Break Galaxies (LBGs; top spectrum) compiled in
  \citet{shapley03a} from 811 individual galaxies.  The composite
  spectra of HSGs are split into those with \civ\ emission (middle
  spectrum) and \civ\ absorption (bottom spectrum) and {\it exclude}
  all sources with high-S/N ($>$3$\sigma$) spectral features which are
  not \lya\ emission.  There are six galaxies in the \civ\ emission
  stack and 17 galaxies in the \civ\ absorption stack, both having a
  median redshift of $\langle z\rangle=$3.6.  \lya\ and the Lyman
  break, \siiv\ absorption and \civ\ emission/absorption are detected
  in both composites.  More observations of similar sources is needed
  to enhance the S/N of the composites, which exhibit a wide range of
  spectral properties (as seen here and in Figure~\ref{fig:spectra}).
  The spectra are arbitrarily offset in flux with marked zero-points.}
\label{fig:restuvstack}
\end{figure*}

Two different composites are constructed based on the detection or
non-detection of \civ\ emission at this low S/N level.  Since most
starbursts are expected to show absorption in \civ, the co-addition of
sources with and without \civ\ might easily yield a null result and no
absorption or emission.  Six sources are co-added in the \civ\ emission
composite (these are the remaining sources for which \civ\ emission is
detected, as indicated in Table~\ref{tab_fullsample}, attributed to an
AGN).  The remaining 17 sources are co-added to form the composite
without \civ\ emission.  Each composite is constructed by scaling the
flux of each galaxy to an arbitrary fixed mean value in the wavelength
range 1330--1400\AA.  This wavelength range is chosen for its
proximity to \lya\ and absence of spectral signatures.  The two
composite rest-frame ultraviolet spectra are shown in
Figure~\ref{fig:restuvstack}.

Both composites show a very high-S/N \lya\ line, \lya\ break,
\siiv\ absorption, and either \civ\ emission or absorption.  The
composite LBG spectrum from \citet{shapley03a} is shown for comparison,
although the relative sample sizes should be contrasted.  The
\civ\ emission composite, by design, consists of galaxies with
non-negligible AGN emission; as a result, the width of the \lya\ line
is broader in the \civ\ emission composite than in the
\civ\ absorption composite.

We test for consistency between the composite spectra and individual
source spectra through cross-correlation in the off-\lya\ wavelength
regions (note that for each galaxy's spectrum, we measure the
cross-correlation with a composite {\it excluding} that source).  This
provides an indication of sources which might be contaminating the
composite rather than boosting its signal-to-noise.  All individual
sources in the \civ\ composite have correlations $>0.4$ at
$\Delta\lambda=0$ offsets. The individual source spectra making up the
other composite are of lower signal-to-noise, and have
cross-correlations ranging 0.2--0.7.  However, four sources
(1\,HERMES\,X24\,J095830.24+015633.2,
1\,HERMES\,X1.4\,J104707.69+585149.1,
1\,HERMES\,X1.4\,J104649.92+590039.6 and
1\,HERMES\,X1.4\,J123536.28+623019.9) have almost no correlation with
the composite (0.05--0.15) which is caused by no continuum detection
(since only the off-\lya\ spectra are considered in the
cross-correlation test).  Three of these sources are considered
`tentative' in their spectroscopic identifications in
Table~\ref{tab_fullsample}. The fourth source,
1\,HERMES\,X24\,J095830.24+015633.2, is secure as judged by the
quality of the Ly$\alpha$ detection and the inconsistency of this line
being incorrectly identified as \oii.

\subsection{Spectral Signatures of AGN}

Since many of the sources in our sample have clear AGN features in
their optical spectra, one might think that the infrared luminosities
are contaminated by significant AGN heating rather than starburst
heating.  Typically, the presence of an AGN warms dust to temperatures
\simgt\,100--200\,K.  In this sample the majority of galaxies have
dust temperatures \simlt\,70\,K.  Furthermore, two QSOs and the 13
sources with \civ\ detections have dust temperatures in the 30--50\,K
range, perfectly consistent with star formation dominated infrared
emission.  While there is still potential for AGN contribution to
$L_{\rm IR}$, the lack of correlation with dust temperature indicates
that the effect is small \citep[\simlt25\%, the nominal contribution
  of mid-infrared powerlaw emission to $L_{\rm IR}$,][]{casey12a}.
This is consistent with prior measures of SMGs with AGN ranging
$\sim$15--25\%\ \citep{swinbank04a,pope08b,menendez-delmestre09a,laird10a,coppin10a}.
Note that one study, \citet{alexander05a}, could be interpreted to
disagree with this work (finding $\sim$75\%\ of SMGs have AGN),
however a minority of the sources in that data have AGN which dominate
the sources' bolometric luminosity.  

We can draw some basic conclusions from the \civ\ emission and
spectral types in our sample to infer the overall AGN content of $z>2$
HSGs.  For sources of sufficient signal-to-noise ($>$5$\sigma$ in
continuum), we can assess AGN spectral signatures source by source.
Of the 20 galaxies which meet this S/N cut, there are three Lyman
Break Galaxies (LBGs), three quasars, seven starbursts with AGN
(e.g. \civ\ emission), and seven ``pure'' starbursts (see
Table~\ref{tab_fullsample} for details).  Of the remaining single-line
identifications, none show AGN signatures.  Of the 36 sources, 10 have
AGN signatures, three of which are obvious quasars.  Although very
qualitative, this analysis implies an AGN fraction of
$\sim$\,25\%\ for the $z>2$ HSG sample.  From the composite spectra
(which was constructed from only the lower luminosity sources), our
statistics agree by construction; in other words, 6 out of 23 sources
had \civ\ emission, or $\sim$\,26\%.  In a series of detailed studies
on the multiwavelength properties of 850\,\um-selected SMGs,
\citet{alexander05a}, \citet{pope08b} and
\citet{menendez-delmestre09a} also measure AGN fractions
$\approx25$\%\ for similarly luminous $z\sim2$ starbursts.

Interestingly, the sources exhibiting AGN signatures in the optical do
not show hotter dust temperature SEDs in the infrared.  One might
expect higher dust temperatures in the infrared with the presence of
an AGN heating the surrounding material to temperatures
$\sim$100--200\,K, exceeding normal heating from star formation,
$\sim$30--50\,K.  The sources with AGN signatures are marked with
small white dots in Figure~\ref{fig:lirtd}.  The observation that the
AGN does not seem to have significant impact on infrared luminosity
or dust temperature is not surprising if you consider that the star
formation activity is at least an order of magnitude more luminous.

Although the selections of the SMG population and the HSG population
differ, finding 1/4 with AGN might suggest that HSGs are similar in
most ways to SMGs without any enhanced AGN activity, despite slightly
warmer overall dust temperatures and brighter mid-infrared fluxes in
comparison (described in the next section).  However, further detailed
work on these samples is needed before any conclusion is drawn as to
the evolutionary nature of these {\it Herschel}-selected galaxies
relative to classic 850\,\um-selected SMGs.

\subsection{Composite Infrared Spectra}

\begin{figure}
\centering
\includegraphics[width=0.99\columnwidth]{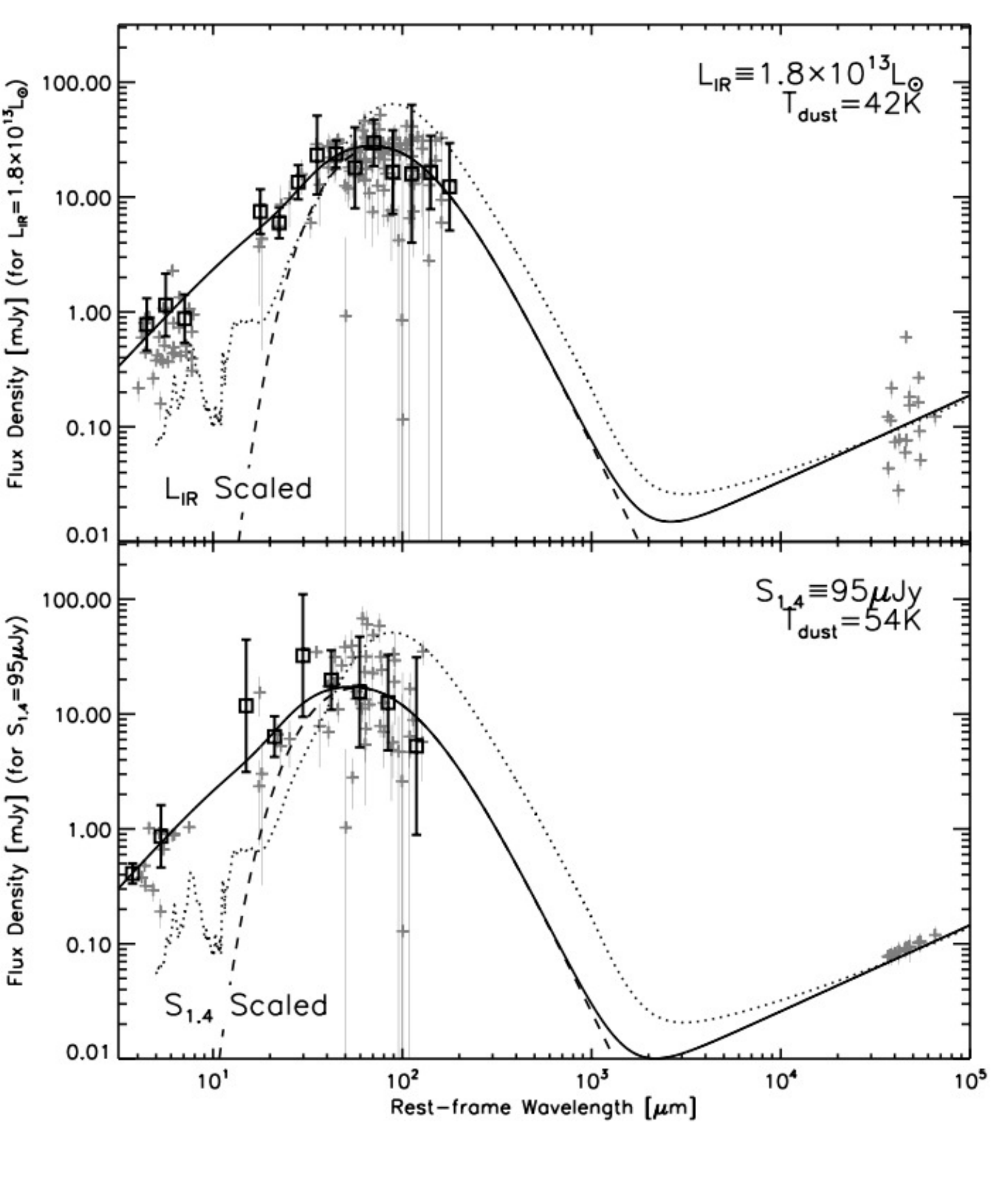}
\caption{The mean infrared and radio SEDs for the sample re-normalized
  to the mean IR luminosity of the sample,
  1.8$\times$10$^{13}$\lsun\ ($top$) and to mean radio flux density,
  95\,\uJy\ ($bottom$). 
  The mean normalized flux density in $\log(\lambda)=$0.1 bins is
  shown as black squares, from rest-frame $\approx$4--150\um.
  Best-fit SEDs (solid black lines) are generated as described in the
  text according to Equation~\ref{eq1} with fixed $\beta=$1.5.  They
  comprise a cold-dust modified blackbody fit (dashed line) and a
  mid-infrared power law representative of warm dust emission.  Radio
  synchrotron emission is added onto this best-fit infrared SED by
  assuming the FIR/radio correlation holds with a synchrotron slope of
  $\alpha=$0.75.  The composite SMG SED described in \citet{pope08b}
  is shown as a dashed line.  Both the luminosity-scaled SED and
  radio-scaled SED appear to have a 24\,\um\ excess relative to the
  SMG expectation.  The radio-scaled SED has a hotter characteristic
  dust temperature than the luminosity-scaled SED (both of which are
  uncertain by $\sim$3\,K), likely driven by the bias against
  colder-dust galaxies of similar flux densities (less likely to be
  radio detected).
}
\label{fig:meansed}
\end{figure}

Figure~\ref{fig:meansed} combines all {\sc Spire}, {\it Spitzer} and
{\sc Pacs} (where available) infrared photometry for all galaxies in
our sample from rest-frame $\approx$\,40 to 150\um\ and radio data.
This includes observations at 24\um, 70\um, 100\um, 160\um, 250\um,
350\um, and 500\um.  At these redshifts, the {\sc Spire} bands probe
the Wien-side of the thermal dust emission peak.  Mean SEDs are fit
using the modified blackbody plus power law method described by
Equation~\ref{eq1} for photometric data which are scaled to the mean
infrared luminosity of the sample, $\langle L_{\rm
  IR}\rangle=1.8\times10^{13}$\lsun, and then separately, scaled to
the mean radio flux density of the sample, $\langle
S_{1.4}\rangle=95$\,\uJy\ (or rest-frame S$_{\rm
  1.4}=283$\,\uJy\ assuming $\alpha=0.75$).  While all 36 galaxies are
used in the former SED fit (top panel of Figure~\ref{fig:meansed}),
the radio-scaled SED fit only has contributions from radio-detected
galaxies.

There are two notable aspects of these mean SED fits seen in
Figure~\ref{fig:meansed}; the first is the difference between the
observed 24\,\um\ flux densities relative to predictions from a
850\,\um-selected SMG template spectrum \citep{pope08b}, and second is
the difference in dust temperatures between luminosity- and
radio-scaled SEDs.  The issue of the discrepancy of mid-infrared
emission relates to the ongoing discussion of suppression of PAH
emission in infrared starburst galaxies \citep[e.g.][]{elbaz11a},
where it is suggested that most normal galaxies have a fixed ratio
($\equiv {\rm IR}8$) between 7.7\um\ emission (or
$\approx$8\um\ emission) and total integrated infrared luminosity,
$L_{\rm IR}$ and that sources with enhanced infrared emission are
called infrared starbursts.  However, as \citet{hainline09a} point
out, the mid-infrared portion of the spectrum does not lend itself to
simple interpretation in terms of what is or is not AGN dominated or
starburst dominated.  SMGs, known to be extreme starbursts, do in fact
have enhanced infrared emission relative to PAH strength
\citep[e.g. see][]{pope08b,menendez-delmestre09a}, however in this
sample$-$which is on average more distant than most 850\,\um-selected
SMGs$-$we see mid-infrared flux densities $\sim$2--5 times the SMG
expectation, more consistent with the measured IR{\it 8} value for
most ``main sequence'' galaxies.  What does this suggest about the
{\sc Spire} sample's evolutionary histories?  Is PAH emission simply
not suppressed in these distant starbursts, or could the
24\,\um\ ``excess'' be due to AGN heating? Or could these high
redshift infrared luminous galaxies be ``main sequence'' secularly
evolving galaxies?  While this might be a selection bias based on
24\um\ or 1.4\,GHz detectability, the radio-selected galaxies (bottom
panel) still show a mid-infrared excess above expectation from the SMG
template.

The differences in dust temperatures between the two mean SED fits
(42$\pm$3\,K vs. 54$\pm$4\,K for luminosity-scaled and radio-scaled,
respectively) is traceable to a radio selection bias.  For two
galaxies with similar {\sc Spire} flux densities, one with a warm
temperature ($\sim$60\,K) and one with a cold temperature
($\sim$30\,K), the galaxy with the warm temperature is going to have a
much higher integrated infrared luminosity and therefore much brighter
1.4\,GHz detection at these redshifts.  Therefore, when we consider
just radio-detected galaxies, the average dust temperature increases
due to the exclusion of cold, non-detectable galaxies.

As discussed earlier, we estimate that 16--43\%\ of these galaxies
would be formally undetected at 850\,\um\ at $<$\,2--5\,mJy.  In other
words, 16--43\%\ of HSGs at $z>2$ are consistent with the
Submillimeter Faint Radio Galaxy (SFRG, formerly optically faint,
``OFRG'') selection and not SMG selection
\citep{chapman04a,casey09b,chapman10a,magdis10a,casey11a,casey11b}.
The composite infrared SEDs from Figure~\ref{fig:meansed} support this
conclusion, since the range of observed 850\,\um\ flux densities from
the best-fit SEDs is in the $1<S_{\rm 850}<10$\,mJy range, not as
luminous at long wavelengths as the 850\,\um-selected composite
\citep{pope08b}.

\section{SFRD Implications and Discussion}\label{sec:discussion}

To place these $z>2$ {\sc Spire}-selected galaxies in context with
other high$-z$ infrared galaxies and lower luminosity galaxies, we
estimate their contribution to the cosmic star formation rate density
\citep[SFRD;][]{madau96a,madau98a,hopkins06a}.  The SFRD contribution
allows a direct comparison of the importance of infrared-luminous
galaxies to the build-up of stellar mass in the Universe over a range
of epochs.  At lower redshifts $z$\simlt1, ultraluminous infrared
galaxies are very rare and contribute little to the SFRD
\citep{sanders03a} but towards $z\approx1$, the importance of ULIRGs
grows, and it is estimated that LIRGs and ULIRGs ($L_{\rm
  IR}>$10$^{11}$\lsun) could contribute as much as $\sim$1/2 of the
total SFRD \citep[see results from {\it Spitzer};][ and work in
  C12]{le-floch05a}.  At $z>1$, the contribution of infrared-luminous
sources is much more difficult to measure, limited by small numbers of
SMGs \citep{chapman05a,wardlow11a} or complex selection biases or
extrapolations from the mid-infrared
\citep{caputi07a,magnelli11a,capak11a}.  The $2<z<5$ galaxies in this
paper provide a unique sample to make this measurement, due to their
well characterized selection over a relatively large sky area,
$\sim$\,1\,deg$^2$.

To arrive at SFRD estimates, we first compute the infrared luminosity
function using a 1/$V_{\rm max}$ method, where each source is
associated with the maximum volume in which it could be detected at
its given luminosity, $L_{\rm IR}$ \citep{schmidt68a}.  The number
density of sources with luminosity between $L$ and $L+\Delta L$ is
given as $\Phi_{z}(L)\Delta L=\sum 1/(V_{i}(L)\times c_{i})$
in units of $h^{3}\, $Mpc$^{-3}\, $logL$^{-1}$.  Here $c_{i}$ is a
completeness estimator which corrects for sample incompleteness at the
selection wavelength, e.g. in this case, at 250--500\um.  C12 presents
a detailed discussion of this completeness factor as a function of
selection wavelength flux density which varies field to field (based
on the prior source catalog depths).  Sources with flux densities
$>$15\,mJy will be more than 90\%\ complete (e.g. $c_{i}>0.9$) in all
fields.

The completeness estimator's effect on the integrated SFRD is small
compared to the uncertainty in the luminosity function itself from
small number counts.  Note, however, that this correction only
pertains to one of the many sources of incompleteness of this sample;
it is far more difficult to quantify and correct for incompleteness
with respect to sources missing from the prior catalog list
(e.g. 1.4\,GHz or 24\,\um\ faint), which is particularly a problem at
$z>2$, as well as spectroscopic incompleteness, i.e. the galaxies
which have no detectable emission lines or are too optically obscured
to be identified.  

$V_{i}(L)$ is the maximum volume in which source $i$ could reside and
still be detectable by our survey.  Since the detection limits of {\sc
  Spire} alter between field catalogs, the detection limit determining
the maximum volume is calculated source by source.  This luminosity
detection limit is determined much like the curves in
Figure~\ref{fig:lirz}.  For example, a source might have its highest
$S/N$ at 350\um, then its luminosity detection limit, thus maximal
redshift limit $z_{\rm max}$, is determined by setting a 3$\sigma$
detection threshold at 350\um\ where sigma is the local confusion plus
instrumental noise in the 350\um\ map.  This $z_{\rm max}$ limit is
then found across the entire survey area probed to determine
accessible volume.  The assumed dust temperature is that measured for
the given source (since dust temperature does impact the steepness of
the luminosity limit with redshift).

\begin{figure}
\centering \includegraphics[width=0.99\columnwidth]{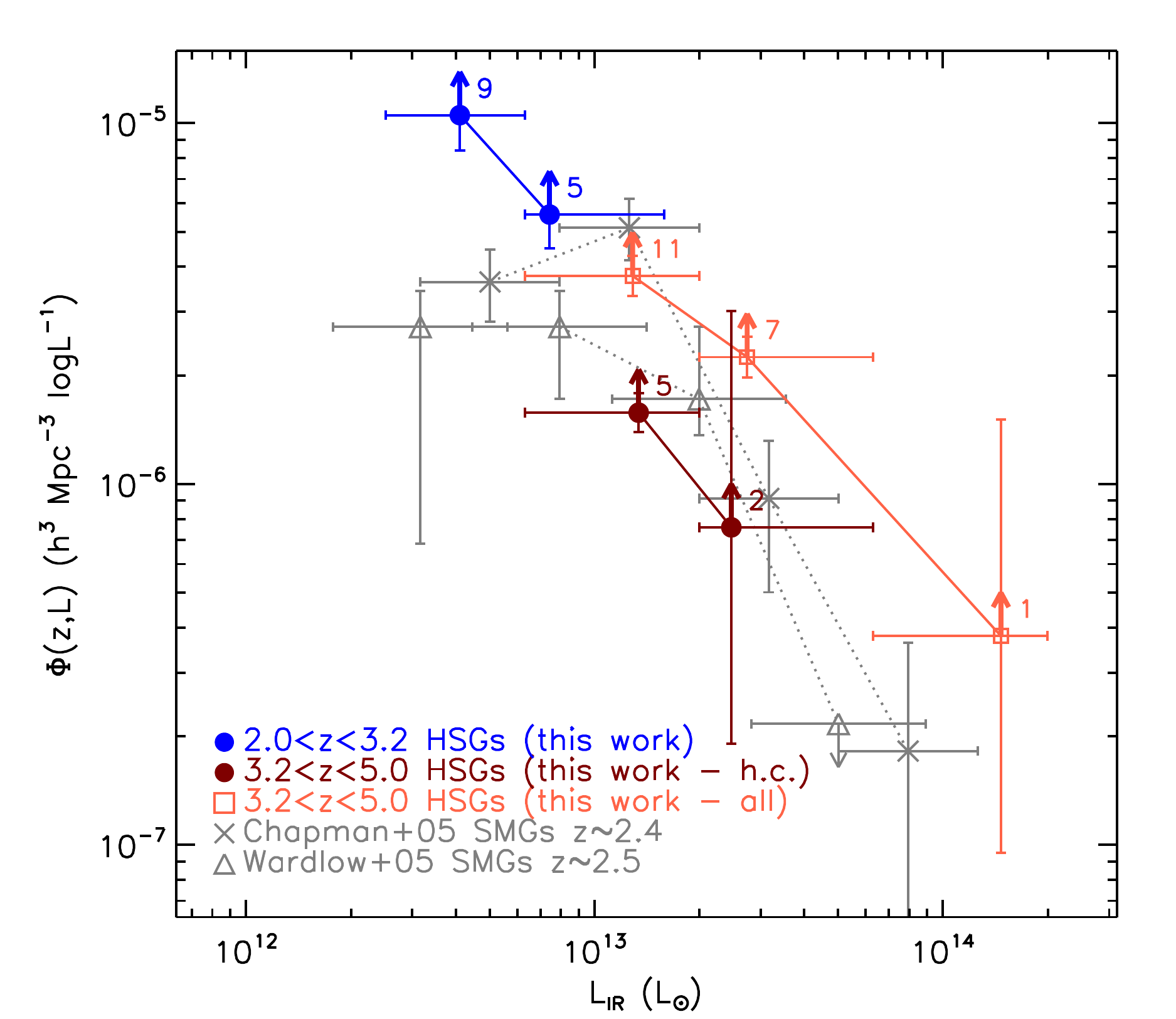}
\caption{The estimated luminosity function for 2$<z<$5 {\it
    Herschel}-{\sc Spire} selected galaxies compared to the luminosity
  function of 850\,\um-selected SMGs at $z\approx$2.5
  \citep{chapman05a,wardlow11a}.  Up arrows denote the fact that this
  survey is {\it spectroscopically incomplete} and that the
  incompleteness is not well quantified at $z>2$.  Numbers next to
  each point indicate how many galaxies from our sample contribute to
  that luminosity bin; the numbers are comparable to those in the
  $z\approx$2 SMG samples.  The luminosity function for the whole
  sample at 3.2$<z<$5.0 is shown in salmon, while the high confidence
  (h.c.) identifications' luminosity function is shown in dark red.}
\label{fig:lf}
\end{figure}

We split the luminosity function into two redshift bins: $2.0<z<3.2$
and $3.2<z<5.0$ with 15 sources in the former and 20 in the latter.
Since LRIS and DEIMOS have different wavelength coverage, DEIMOS
observations suffer from a redshift desert from $1.6<z<3.2$ that LRIS
observations do not, so we split the sample at $z=3.2$ and only
compute the density using LRIS observations between $z=2.0$ and $z=3.2$.
This excludes two sources from the calculation
(1HERMES\,X1.4\,J104636.00+585650.0 at $z=$2.841 and
1HERMES\,X24\,J160545.99+534544.4 at $z=$2.555) which were both
surveyed with DEIMOS and detected in the redshift desert due to strong
\civ\ emission caused by the presence of a quasar.  Since
high-redshift $z$\simgt2 sources are only detectable on the masks
observed in the best weather conditions, the effective area probed by
LRIS for this calculation is 0.13\,deg$^{2}$ over the range
$2.0<z<5.0$ and for DEIMOS 0.30\,deg$^{2}$ over the range
$3.2<z<5.0$.  The resulting {\it spectroscopically incomplete}
luminosity functions are shown in Figure~\ref{fig:lf}.

Since some of our identifications are less confident than others
(e.g. those marked with a $\dagger$ in Table~\ref{tab_fullsample}) we
also compute the luminosity function excluding tentative
identifications.  The result is seen in Figure~\ref{fig:lf}: while the
$2.0<z<3.2$ luminosity function remains the same (differing only by
two sources), the $3.2<z<5.0$ luminosity function drops by 11 sources
(salmon vs dark red LFs).

\begin{figure}
\centering
\includegraphics[width=0.99\columnwidth]{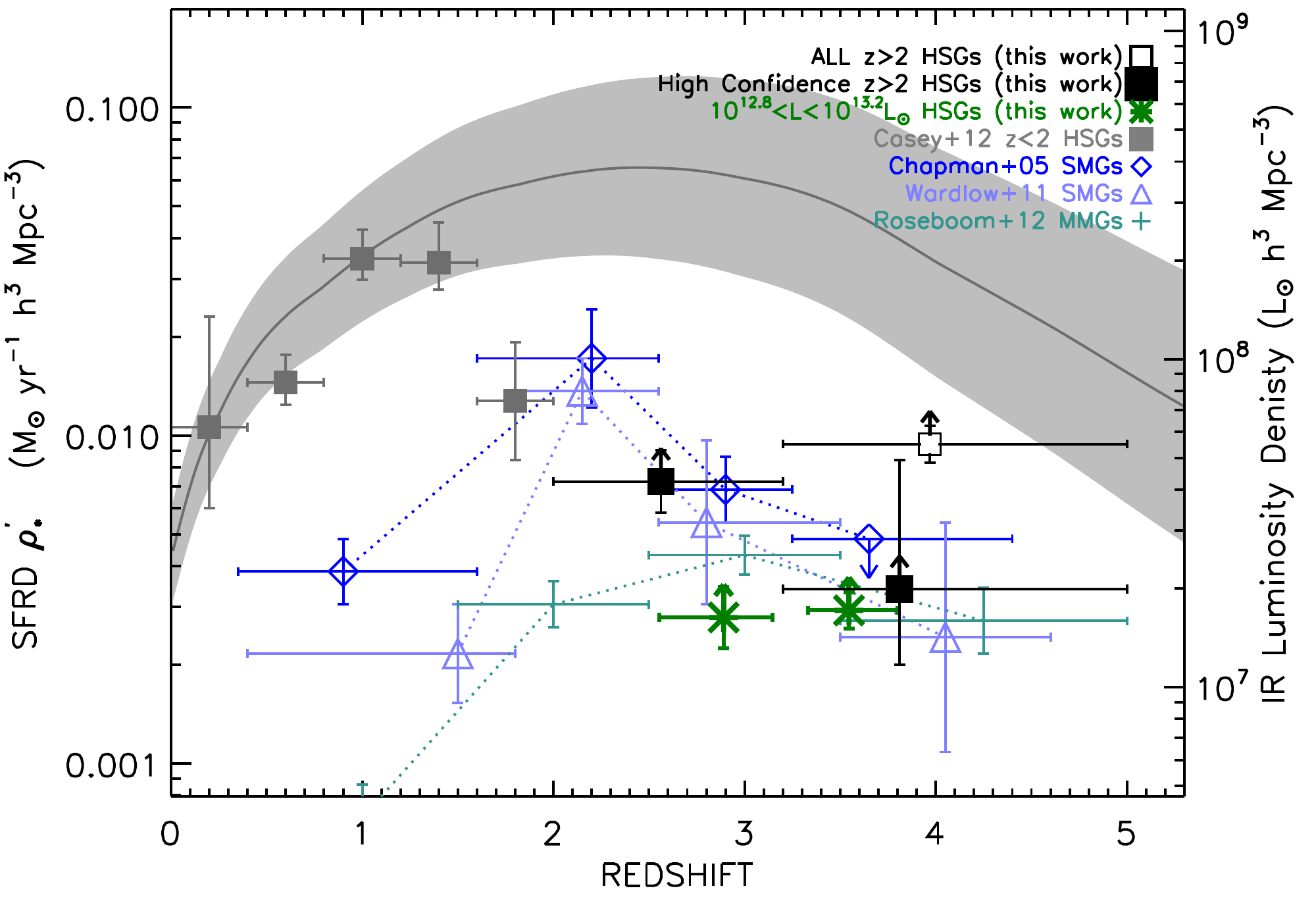}
\caption{Star formation rate density of {\it Herschel}-{\sc Spire}
  selected galaxies (black points) relative to 850\,\um-selected SMGs
  \citep{chapman05a,wardlow11a} and 1.2\,mm-selected MMGs
  \citep{roseboom12a}.  The compilation of SFRD measurements from
  \citep*{hopkins06a} is shown as a gray band, which is largely drawn
  from optical or rest-frame ultraviolet-selected galaxy populations
  corrected for dust extinction.  The luminosity limits of integration
  are 10$^{(12.4-13.4)}$\lsun\ at $z\sim2.5$ and
  10$^{(12.8-13.6)}$\lsun\ at $z\sim4$; the SFRD from sources sitting
  in the luminosity where the two redshift bins overlap
  (10$^{(12.8-13.2)}$\lsun) is shown in green.  The SFRD measurements
  for {\it Herschel}-selected galaxies at $z<$2 are shown as gray
  points.  The sharp drop in the SFRD of {\it Herschel}-selected
  samples at $z\sim$2 is caused by the redshifting of the SED peak
  such that more infrared-luminous galaxies are {\sc Spire}
  ``drop-outs'' and that only the warmest, most-luminous
  $>$10$^{13}$\lsun\ systems are detectable with {\sc Spire} at $z>2$.
  Note that we observe an increase in the infrared-luminous
  contribution to the SFRD from $z\sim2.6$ to $z\sim4$.  }
\label{fig:sfrd}
\end{figure}

The SFRD is then the luminosity weighted integral of the luminosity
function, or the raw summation of the luminosity (converted to SFR)
over accessible volume:
(7.0$\pm$2.0)$\times$10$^{-3}$\sfr\,h$^{3}$\,Mpc$^{-3}$ at
$2.0<z<3.2$ and
(5.5$\pm$0.6)$\times$10$^{-3}$\sfr\,h$^{3}$\,Mpc$^{-3}$ at $3.2<z<5.0$
(full sample) and
(2$^{+3}_{-1}$)$\times$10$^{-3}$\sfr\,h$^{3}$\,Mpc$^{-3}$ at $3.2<z<5.0$
(high confidence sample), shown in Figure~\ref{fig:sfrd} against other
comparison populations.  
These points are lower limits since they do not include any sources
which might be excluded from the prior catalogs at 24\,\um\ or
1.4\,GHz, which is speculated to be a non-negligible fraction
(\simgt\,20\%) at $z>2$ \citep[e.g.][, Smail \etal\ in
  prep]{magdis10a}.  Note also that the luminosity limits of the two
redshift bins differ: the $z\sim2.5$ bin covers
10$^{12.4-13.2}$\lsun\ while the $z\sim4$ bin covers
10$^{12.8-13.6}$\lsun. To assess luminosity evolution from $z\sim2-5$,
we compute the SFRD contributions in the overlapping luminosity range
of both redshift bins, 10$^{12.8-13.2}$\lsun, shown as green points in
Figure~\ref{fig:sfrd}.

Although these measurements of the {\it Herschel} contribution to the
SFRD at $2<z<5$ are lower limits due to our survey's incompleteness,
the effects of gravitational lensing and clustering could lead to an
overestimation.  Are these effects significant in this sample?  For
the former we use the conditional lensing probability as a function of
{\sc Spire} flux density \citep[Francesco De Bernardis, private
  communication,][]{wardlow12a}.  For {\sc Spire} flux densities
$S_{\rm 500}<80$\,mJy the distribution in number counts is dominated
by a Schecter function rather than the flat-sloped source counts at
$>80$\,mJy; galaxies with $S_{\rm 500}>80$\,mJy have a high
probability of being lensed by factors $>$\,2 while this model
predicts a mean lensing factor for this sample of $\langle\mu
\rangle<1.05$, which changes negligibly between $z=2$ and $z=5$.

The brightest source in our sample with $S_{\rm 250}=73.1$\,mJy,
1HERMES\,X24\,J161506.65+543846.9, has the highest probability of
being lensed (its expected lensing factor is $\langle\mu
\rangle=1.2$).  It is the highest redshift source in our sample at
$z=4.952$.  Due to its extreme luminosity compared to the rest of the
sample and its tentative spectroscopic identification, we exclude this
source from the star formation rate density (SFRD) measurement.

To assess the impact of clustering on the SFRD measurement, we need a
good grasp of the spatial density of $z>2$ sources on our slit-masks
and the possibility of biased placement of slit-masks around high-$z$
clusters.  The former can be gauged by the number of high$-z$
confirmations per slit-mask; the 36 sources of this sample are
distributed across 12 LRIS masks and 14 DEIMOS masks, with an
additional 5 LRIS masks and 3 DEIMOS masks without any high-$z$
sources; this averages to 0--2 galaxies per mask without any mask
having more than 2 sources.  Since the masks were distributed randomly
with respect to one another in each field and none of the masks were
close together, this demonstrates that these sources are indeed
randomly distributed over the surveyed area, 0.93\,deg$^{2}$ for the
whole survey.

The possibility exists that there is an intrinsic bias of the
placement of our slit-masks such that more high-$z$ sources are
observed than elsewhere.  As explained in detail in C12, masks were
placed around high-priority targets which were `red' in their {\sc
  Spire} colors (e.g. $S_{\rm 250}<S_{\rm 350}<S_{\rm 500}$ all with
$S$\simgt\,15\,mJy) and thought to be high redshift sources.  Of the 36
confirmed $z>2$ sources, seven ($\sim$\,19\%) were originally
high-priority targets.  However, an additional 44 high-priority
targets were identified at $z<2$ and 78 were unidentified.  Relative
to the number of high-priority sources targeted, we measure
5$\pm$5\%\ as identified at $z>2$, 34$\pm$5\%\ at $z<2$ and
60$\pm$4\%\ unidentified.  The same statistics for lower priority
targets are 2$\pm$1\%\ at $z>2$, 48$\pm$1\% at $z<2$, and
50$\pm$1\%\ unidentified.  Within uncertainties, the proportion of
sources identified at $z>2$ are the same between low-priority and
high-priority targets, indicating no bias or advantage in targeting
`red' sources more than any other significant {\it Herschel}-{\sc
  Spire} source.  This implies that no clustering correction on the
measured SFRD is necessary.

The lower limits to the SFRD set by {\sc Spire} sources tells us that
the early Universe potentially had a very substantial amount of star
formation in short-lived, intense $>$1000\sfr\ bursts as opposed to
slow-progressing moderate levels of star formation.  The contribution
from {\sc Spire} to the SFRD at these epochs is at least comparable to
the contribution measured from longer wavelength-selected galaxies,
like the 850\,\um-selected SMGs \citep{chapman05a,wardlow11a} or
1.2\,mm-selected MMGs \citep{roseboom12a}.  This is made more
interesting by the observation that the populations (SMG and HSG) only
overlap by 21 out of 36 galaxies (58\%).  Further work aimed at
confirming redshifts of $z>$2 {\sc Spire} sources, particularly those
without radio or 24\,\um\ counterparts, is needed to constrain these
lower limits into real measurements so the importance of
\simgt10$^{13}$\lsun\ activity in the first few Gyr of the Universe is
understood.

\section{Conclusions}\label{sec:conclusions}

The identification of submillimeter galaxies at early epochs in the
Universe's history is the key to understanding the limits of star
formation and galaxy evolution on short timescales.  This paper has
presented new observations of 36 {\it Herschel}-{\sc Spire} selected
starburst galaxies between $2<z<5$, taken from a large Keck
spectroscopic survey of 1594 {\sc Spire}-selected galaxies covering
0.93\,deg$^2$. 

We present the following conclusions:
\begin{itemize}

\item Our sample of 36\,{\it Herschel}-selected galaxies (HSGs)
  constitute some of the brightest, most extreme infrared starburst
  galaxies in the Universe.  Spanning $2<z<5$, our sample has a mean
  luminosity $\langle L_{\rm
    IR}\rangle=1.8\times10^{13}$\lsun\ (SFR\,$\approx$\,3100\sfr).

\item These $2<z<5$ HSGs have a well characterized selection across 6
  legacy fields and 0.93\,deg$^{2}$; galaxies must be $>$3$\sigma$
  significant in one of the three {\it Herschel}-{\sc Spire} bands and
  also be detected in deep 24\,\um\ and/or 1.4\,GHz survey coverage.
  Although it misses 24\um\ or 1.4\,GHz high-$z$ dropouts, the
  selection is identical to low-$z$ HSG selection.  Sources at $z>2$
  comprise 5\%\ of all galaxies selected via this method; although
  inefficient for finding high-$z$ infrared galaxies, the selection is
  easily reproducible and well suited for volume density estimates.

\item Our sample show a wide range of rest-frame ultraviolet spectral
  features: some galaxies classifiable as quasars, some as LBGs, and
  most as starbursts with a wide range of dust extinctions/reddening.
  The heterogeneous nature of their spectra provide additional
  evidence that the infrared-luminous stage might exist during
  a period when the host galaxy is rapidly evolving.

\item The radio-detected subset of our sample (23/36) follow the
  FIR/radio correlation consistent with moderate evolution in $q_{\rm
    IR}$ from previous work \citep{ivison10b}.

\item We construct composite rest-frame ultraviolet spectra and
  rest-frame infrared spectral energy distributions to assess some
  aggregate properties of HSGs.  In the rest-frame UV, we determine
  that 25\%\ of HSGs exhibit \civ emission (a signature of AGN).  In
  the infrared, HSGs exhibit a 24\,\um\ excess relative to SMGs of
  similar $L_{\rm IR}$; without mid-IR spectra, it is impossible to
  know whether this is due to enhanced PAH emission (similar to
  ``normal'' galaxies) or AGN emission.  The dust temperatures of
  radio-selected samples are warmer than those of the full sample.

\item Our spectroscopic survey is incomplete due to selection bias at
  24\,\um\ and 1.4\,GHz, as well as spectroscopic incompleteness caused
  by heavy dust obscuration in the rest-frame UV.  Therefore, we are
  able to place {\it lower limits} on the contribution of $2<z<5$ HSGs
  to the cosmic star formation rate density, which is
  $>$7$\times$10$^{-3}$\sfr\,h$^3$\,Mpc$^{-3}$ at $z\approx2.6$ and
  $>$3$\times$10$^{-3}$\sfr\,h$^3$\,Mpc$^{-3}$ at $z\approx4$,
  corresponding to $>$10\%\ and $>$20\%\ of the best-estimates of the
  total SFRD at their respective epochs.
\end{itemize}
This work highlights the importance of extremely luminous FIR-bright
galaxies to the build-up of stellar mass, particularly at early times
in the Universe's history.  Further work on constraining completeness
and the parent population of infrared-luminous galaxies at $z>2$ is
needed to understand the role that short-lived starbursts have in the
context of galaxy evolution and formation.

\section*{Acknowledgements}

We thank the anonymous referee for his/her constructive suggestions
which improved the manuscript.
CMC is generously supported by a Hubble Fellowship from Space
Telescope Science Institute, grant HST-HF-51268.01-A.
The data presented herein were obtained at the W.M. Keck Observatory,
which is operated as a scientific partnership among the California
Institute of Technology, the University of California and the National
Aeronautics and Space Administration. The Observatory was made
possible by the generous financial support of the W.M. Keck
Foundation.  The authors wish to recognize and acknowledge the very
significant cultural role and reverence that the summit of Mauna Kea
has always had within the indigenous Hawaiian community.  We are most
fortunate to have the opportunity to conduct observations from this
mountain. This work would not be possible without the hard work and
dedication of the Keck Observatory night and day staff; special thanks
to Marc Kassis, Luca Rizzi and Greg Wirth for help and advice while
observing.  The analysis pipeline used to reduce the DEIMOS data was
developed at UC Berkeley with support from NSF grant AST-0071048.

{\sc Spire} has been developed by a consortium of institutes led by
Cardiff Univ. (UK) and including: Univ. Lethbridge (Canada); NAOC
(China); CEA, LAM (France); IFSI, Univ. Padua (Italy); IAC (Spain);
Stockholm Observatory (Sweden); Imperial College London, RAL,
UCL-MSSL, UKATC, Univ. Sussex (UK); and Caltech, JPL, NHSC,
Univ. Colorado (USA).  This development has been supported by national
funding agencies: CSA (Canada); NAOC (China); CEA, CNES, CNRS
(France); ASI (Italy); MCINN (Spain); SNSB (Sweden); STFC, UKSA (UK);
and NASA (USA).

This research has made use of data from the HerMES project
(http://hermes.sussex.ac.uk/).  HerMES is a {\it Herschel} Key
Programme utilizing Guaranteed Time from the {\sc Spire} instrument
team, ESAC scientists and a mission scientist.  HerMES is described in
\citet{oliver12a}.  The {\sc Spire} data presented in this paper will
be released through the HerMES Database in Marseille, HeDaM
(http://hedam.oamp.fr/HerMES).

\clearpage
\begin{landscape}
\begin{deluxetable}{l@{ }c@{\,}l@{\,}c@{ }c@{ }c@{ }c@{ }c@{ }c@{ }c@{ }cc@{ }c@{ }c}
\tablecolumns{8}
\tablecaption{$z>$2 identified {\sc Spire}-selected Galaxies}
\tablehead{
\colhead{NAME} & \colhead{$z_{\rm spec}$}   &   \colhead{Comments} & \colhead{{\sc class}} & \colhead{$z_{\rm phot}$} & \colhead{$S_{\rm 24}$} & \colhead{$S_{\rm 100}$} & \colhead{$S_{\rm 160}$} & \colhead{$S_{\rm 250}$} & \colhead{$S_{\rm 350}$} & \colhead{$S_{\rm 500}$} & \colhead{$S_{\rm 1.4GHz}$} & \colhead{L$_{\rm IR}$} & \colhead{T$_{\rm dust}$}\\
\colhead{}     &     &                        &                      &                          & (\uJy)                 & (mJy)                   & (mJy)                   & (mJy)                   & (mJy)                   & (mJy)                   & (\uJy)                    &  (\lsun)                   & (K) \\
}
\startdata

1HERMES\,X24\,J033136.96$-$275510.9... & 3.145$^{\rm L}$ & \lya-break       & LBG & 3.44$^{d}$  & 265$\pm$11  &    $-$      &    $-$       & 15.5$\pm$3.8 & 20.8$\pm$3.8 & 24.0$\pm$4.8    & ...  & (1.0$^{+0.3}_{-0.2}$)$\times$10$^{13}$ & 40.6$\pm$3.6 \\
1HERMES\,X1.4\,J033151.94$-$275326.9... & 2.938$^{\rm L}$ & \lya;\siiv       & SB  & 2.96$^{d}$  & 397$\pm$14  &       $-$    &    $-$       & 12.9$\pm$3.9  & 21.6$\pm$3.7   & 24.2$\pm$4.2  & 54.3$\pm$12.8 & (1.2$^{+0.3}_{-0.2}$)$\times$10$^{13}$ & 43.8$\pm$3.4 \\
1HERMES\,X24\,J033319.58$-$274119.7... & 2.325$^{\rm L}$ & \lya             & SB  & 1.84$^{d}$  & 329$\pm$13  &        $-$   &    $-$       & 16.0$\pm$3.8 & 19.0$\pm$3.8 & 15.2$\pm$4.5     & ...  & (5.4$^{+1.7}_{-1.3}$)$\times$10$^{12}$ & 39.3$\pm$3.8 \\
1HERMES\,X24\,J095830.24+015633.2...   & 2.327$^{\rm L}$ & Diffuse \lya     & SB  & 1.70$^{a}$  & 418$\pm$16  & ...         & ...          & 19.2$\pm$2.2  & 23.4$\pm$3.0   & 23.0$\pm$3.3   & ...  & (5.4$^{+2.9}_{-1.9}$)$\times$10$^{12}$ & 24.6$\pm$3.7 \\
1HERMES\,X24\,J095916.08+021215.3...   & 4.454$^{\rm L}$ & \lya             & SB  & 2.97$^{a}$  & 282$\pm$14  & ...         & 10.1$\pm$3.4 & 25.8$\pm$2.2  & 24.1$\pm$2.9   & 14.7$\pm$3.2  & ...  & (1.9$^{+1.3}_{-0.7}$)$\times$10$^{13}$ & 45.9$\pm$10.7 \\
1HERMES\,X24\,J095917.28+021300.4...   & 2.101$^{\rm L}$ & \lya;\siiv;\civ;+& SB  & 1.95$^{a}$ & 472$\pm$17  & ...         & 8.4$\pm$2.7  & 14.2$\pm$2.2  & 14.7$\pm$2.7   & 7.1$\pm$3.0  & ...  & (4.4$^{+1.4}_{-1.0}$)$\times$10$^{12}$ & 51.5$\pm$5.2 \\
1HERMES\,X1.4\,J095934.08+021706.3...  & 2.926$^{\rm L}$ & \lya             & SB  & 2.94$^{a}$  & 1451$\pm$16 & ...         & 11.5$\pm$2.9 & 11.4$\pm$2.2  & 12.9$\pm$3.4   & 0.2$\pm$4.7   & 170$\pm$15 & (2.0$^{+0.4}_{-0.2}$)$\times$10$^{13}$ & 100$\pm$3.0 \\
1HERMES\,X24\,J095948.00+024140.7...   & 2.600$^{\rm L}$ & \lya;\civ        & SB+AGN & 2.27$^{a}$  & 909$\pm$16  & ...         & 13.6$\pm$4.2 & 10.5$\pm$2.2  & 10.7$\pm$2.8   & 6.4$\pm$3.0  & ...  & (5.9$^{+5.3}_{-2.7}$)$\times$10$^{12}$ & 39.4$\pm$12.2 \\
1HERMES\,X1.4\,J100008.64+022043.1...  & 2.888$^{\rm L}$ & \lya-break;\civ  & LBG & 2.18$^{a}$  & 574$\pm$14  & ...         & ...          & 16.5$\pm$2.2  & 12.3$\pm$2.9   & 9.4$\pm$3.6  & 66$\pm$12 & (7.9$^{+6.0}_{-3.4}$)$\times$10$^{12}$ & 31.4$\pm$7.2 \\
1HERMES\,X24\,J100020.16+021725.2...   & 2.105$^{\rm L}$ & \lya;\siiv;\civ  & QSO & 1.45$^{a}$  & 401$\pm$16  & ...         & ...          & 20.0$\pm$2.2  & 12.6$\pm$3.5   & 22.3$\pm$3.7 & ...  & (5.7$^{+1.3}_{-1.0}$)$\times$10$^{12}$ & 32.9$\pm$2.2 \\
1HERMES\,X1.4\,J100024.00+021210.9...  & 3.553$^{\rm L}$ & \lya;\siiv;\civ  & SB+AGN & 2.97$^{a}$  & 175$\pm$50  & 6.1$\pm$1.5 & 31.7$\pm$3.6 & 28.5$\pm$2.7  & 28.8$\pm$3.5   & 15.1$\pm$5.7 & 84$\pm$13 & (2.0$^{+0.4}_{-0.3}$)$\times$10$^{13}$ & 54.0$\pm$5.3 \\
1HERMES\,X24\,J100036.00+021127.6...   & 2.103$^{\rm L}$ & \lya;\siiv;\civ  & SB  & 1.48$^{a}$ & 158$\pm$17   & ...         & 11.2$\pm$2.9 & 13.2$\pm$2.7  & 21.2$\pm$3.6   & 4.9$\pm$11.4  & ...  & (2.7$^{+1.3}_{-0.9}$)$\times$10$^{12}$ & 29.0$\pm$3.9 \\
1HERMES\,X1.4\,J100111.52+022841.3...  & 3.975$^{\rm L}$ & \lya-break;\civ  & LBG & 2.60$^{a}$  & 201$\pm$41  & ...         & ...          & 24.5$\pm$2.2  & 32.9$\pm$4.3   & 22.8$\pm$6.8 & 59$\pm$11 & (1.8$^{+0.6}_{-0.5}$)$\times$10$^{13}$ & 46.7$\pm$7.2 \\
1HERMES\,X24\,J100133.36+023726.9...   & 2.619$^{\rm L}$ & Diffuse \lya$\dagger$& SB  & $-$99$^{a}$ & 245$\pm$15  & ...         & ...          & 10.1$\pm$2.2  & 9.1$\pm$2.9    & 0.0$\pm$9.9   & ...  & (4.4$^{+2.5}_{-1.6}$)$\times$10$^{12}$ & 51.9$\pm$8.3 \\
1HERMES\,X24\,J100146.56+024035.6...   & 2.050$^{\rm L}$ & \siiv;\civ;\heii & SB+AGN & 1.88$^{a}$  & 722$\pm$17  & 4.5$\pm$1.2 & ...          & 11.7$\pm$2.2  & 6.9$\pm$2.8    & 2.2$\pm$3.1  & ...  & (4.7$^{+1.8}_{-1.3}$)$\times$10$^{12}$ & 70.7$\pm$8.3 \\
1HERMES\,X24\,J100150.16+024017.2...   & 2.883$^{\rm L}$ & \lya             & SB  & 2.57$^{a}$  & 315$\pm$19  & ...         & 17.5$\pm$4.6 & 13.8$\pm$2.2  & 12.2$\pm$2.9   & 8.9$\pm$4.0   & ...  & (7.1$^{+6.2}_{-3.3}$)$\times$10$^{12}$ & 48.4$\pm$16.5 \\
1HERMES\,X24\,J100151.60+023909.5...   & 4.538$^{\rm L}$ & \lya$\dagger$    & SB  & 2.20$^{a}$  & 433$\pm$102 & ...         & ...          & 17.8$\pm$2.2  & 0.5$\pm$5.4    & 22.1$\pm$3.2  & ...  & (1.7$^{+2.3}_{-0.9}$)$\times$10$^{13}$ & 35.0$\pm$11.3 \\
1HERMES\,X1.4\,J104557.12+590000.4...  & 3.382$^{\rm L}$ & \lya;\civ        & SB  & 3.38$^{c}$ & 713$\pm$7   & 5.7$\pm$2.7 & 8.5$\pm$4.8  & 16.9$\pm$3.6  & 7.6$\pm$3.9   & 0.0$\pm$4.7    & 101.9$\pm$15.9  & (1.7$^{+0.5}_{-0.4}$)$\times$10$^{13}$ & 87.6$\pm$8.9 \\
1HERMES\,X1.4\,J104620.40+585933.4...  & 2.211$^{\rm L}$ & \lya;\civ        & SB+AGN & 2.12$^{c}$ & 559$\pm$7   &    $-$      &    $-$       & 13.1$\pm$3.5  & 3.4$\pm$3.9    & 0.0$\pm$4.7  & 64.5$\pm$11.8  & (5.1$^{+2.7}_{-1.8}$)$\times$10$^{12}$ & 63.1$\pm$9.7 \\
1HERMES\,X1.4\,J104636.00+585650.0...  & 2.841$^{\rm D}$ & \civ             & QSO & 1.84$^{c}$  & 166$\pm$7   & ...         & ...          & 11.3$\pm$3.5  & 5.4$\pm$4.0   & 0.0$\pm$4.9    & 19.4$\pm$3.6  & (4.1$^{+3.8}_{-2.0}$)$\times$10$^{12}$ & 48.0$\pm$10.8 \\
1HERMES\,X24\,J104642.89+585532.8...   & 3.626$^{\rm D}$ & \lya;\civ        & SB+AGN & $-$99$^{c}$ & 221$\pm$7   &    $-$      &    $-$       & 9.5$\pm$3.5   & 14.1$\pm$4.0   & 0.0$\pm$8.4  & ...  & (9.1$^{+3.7}_{-2.6}$)$\times$10$^{12}$ & 62.1$\pm$7.4 \\
1HERMES\,X1.4\,J104649.92+590039.6...  & 4.710$^{\rm L}$ & \lya$\dagger$    & SB  & 2.72$^{c}$ & 580$\pm$7   & 3.6$\pm$2.5 & ...          & 28.0$\pm$3.6  & 17.1$\pm$4.2  & 7.5$\pm$4.8   & 118.6$\pm$8.8  & (3.9$^{+0.8}_{-0.7}$)$\times$10$^{13}$ & 85.6$\pm$6.1 \\
1HERMES\,X1.4\,J104701.68+590447.6...  & 4.232$^{\rm L}$ & \lya$\dagger$    & SB  & $-$99$^{c}$ & 1000$\pm$8  &   ...       &  ...         & 26.2$\pm$3.5  & 12.0$\pm$4.2  & 4.7$\pm$4.8   & 81.8$\pm$12.2  & (4.1$^{+0.9}_{-0.7}$)$\times$10$^{13}$ & 99.9$\pm$7.2 \\
1HERMES\,X1.4\,J104707.69+585149.1...  & 4.677$^{\rm D}$ & \lya$\dagger$    & SB  & 2.74$^{c}$ & ...         & 7.1$\pm$2.7 & ...          & 14.5$\pm$3.6  & 31.5$\pm$8.5  & 0.0$\pm$10.7  & 36.0$\pm$6.3  & (2.6$^{+1.2}_{-0.8}$)$\times$10$^{13}$ & 80.2$\pm$17.8 \\
1HERMES\,X1.4\,J104709.60+590951.1...  & 2.942$^{\rm L}$ & \lya$\dagger$    & SB  & $-$99$^{c}$ & ...         & 5.9$\pm$2.6 & 16.9$\pm$4.9 & 13.0$\pm$3.5  & 0.0$\pm$5.2   & 0.0$\pm$4.8   & 99.0$\pm$11.2  & (7.8$^{+7.6}_{-3.8}$)$\times$10$^{12}$ & 61.0$\pm$21.7 \\
1HERMES\,X1.4\,J104722.56+590111.7...  & 4.521$^{\rm D}$ & \lya$\dagger$    & SB+AGN  & 2.24$^{c}$ & 444$\pm$7   & 2.8$\pm$2.5 & ...          & 19.0$\pm$3.5  & 21.6$\pm$4.1  & 17.7$\pm$5.3   & 73.7$\pm$9.4  & (2.0$^{+2.1}_{-1.0}$)$\times$10$^{13}$ & 45.2$\pm$15.6 \\
1HERMES\,X1.4\,J123536.28+623019.9...  & 3.380$^{\rm D}$ & Diffuse \lya$\dagger$ & SB  & $-$         & $-$         &      $-$    &    $-$       & 27.1$\pm$4.5  & 25.0$\pm$4.8   & 17.7$\pm$4.1  & 18.0$\pm$51.7  & (1.5$^{+1.3}_{-0.7}$)$\times$10$^{13}$ & 47.1$\pm$18.3 \\
1HERMES\,X1.4\,J123622.58+620340.3...  & 3.579$^{\rm D}$ & \lya;\civ$\dagger$ & SB  & $-$         & $-$         &     $-$     &    $-$       & 9.6$\pm$4.5   & 26.9$\pm$5.6   & 16.0$\pm$4.3 & 312.5$\pm$35.3  & (9.7$^{+16.}_{-6.1}$)$\times$10$^{12}$ & 34.7$\pm$17.1 \\
1HERMES\,X1.4\,J123732.66+621013.4...  & 4.019$^{\rm D}$ & \lya             & SB  & $-$         & $-$         &      $-$    &    $-$       & 16.4$\pm$4.5  & 9.7$\pm$4.9    & 1.1$\pm$5.3   & 36.5$\pm$8.4  & (2.9$^{+0.8}_{-0.6}$)$\times$10$^{13}$ & 100$\pm$2.0 \\
1HERMES\,X24\,J160539.72+534450.3...   & 3.546$^{\rm D}$ & \lya$\dagger$    & SB  & $-$99$^{b}$ & 251$\pm$16  &      $-$    &    $-$       & 16.2$\pm$4.3  & 16.9$\pm$4.4   & 20.4$\pm$4.6  & $-$  & (1.3$^{+0.4}_{-0.3}$)$\times$10$^{13}$ & 45.4$\pm$5.1 \\
1HERMES\,X24\,J160545.99+534544.4...  & 2.555$^{\rm D}$ & \siiv;\civ;\ciii & QSO & 2.31$^{b}$  & 670$\pm$17  &    $-$      &    $-$       & 51.5$\pm$4.7  & 41.8$\pm$4.5   & 31.2$\pm$5.2 & $-$  & (1.4$^{+0.2}_{-0.2}$)$\times$10$^{13}$ & 39.6$\pm$2.0 \\
1HERMES\,X24\,J160603.63+541245.1...   & 3.331$^{\rm D}$ & Diffuse \lya$\dagger$& SB  & $-$99$^{b}$ & 323$\pm$16  &      $-$    &    $-$       & 15.5$\pm$4.3  & 16.0$\pm$4.4   & 8.3$\pm$4.9   & $-$  & (1.0$^{+0.3}_{-0.2}$)$\times$10$^{13}$ & 60.6$\pm$6.7 \\
1HERMES\,X24\,J160639.40+533558.4...   & 3.801$^{\rm D}$ & \lya             & SB  & $-$99$^{b}$ & 188$\pm$17  &      $-$    &    $-$       & 9.0$\pm$4.5   & 16.3$\pm$4.4   & 14.6$\pm$4.8  & $-$  & (1.1$^{+0.6}_{-0.4}$)$\times$10$^{13}$ & 47.1$\pm$7.3 \\
1HERMES\,X24\,J160802.63+542638.1...  & 3.415$^{\rm D}$ & \lya             & SB+AGN & $-$99$^{b}$ & 442$\pm$18  &     $-$     &     $-$      & 32.7$\pm$4.3  & 35.9$\pm$4.6   & 37.9$\pm$4.5  & $-$  & (2.2$^{+0.3}_{-0.3}$)$\times$10$^{13}$ & 42.5$\pm$2.5 \\
1HERMES\,X24\,J160806.56+542301.6...  & 3.721$^{\rm D}$ & \lya;\siiv;\civ$\dagger$& SB+AGN & $-$99$^{b}$ & 246$\pm$16  &    $-$      &    $-$       & 9.8$\pm$4.3   & 7.4$\pm$4.6    & 16.8$\pm$4.6 & $-$  & (1.3$^{+0.7}_{-0.4}$)$\times$10$^{13}$ & 49.8$\pm$7.5 \\
1HERMES\,X24\,J161506.65+543846.9...   & 4.952$^{\rm D}$ & \lya$\dagger$    & SB  & 1.94$^{b}$  & 720$\pm$174 &      $-$    &    $-$       & 73.1$\pm$4.3  & 49.6$\pm$4.4   & 22.8$\pm$4.8  & $-$  & (1.2$^{+0.2}_{-0.2}$)$\times$10$^{14}$ & 86.2$\pm$12.2 \\
\enddata
\label{tab_fullsample}
\tablecomments{ Superscripts in the $z_{\rm spec}$ column indicate the
  instrument with which the spectroscopic redshift was obtained, LRIS
  (L) or DEIMOS (D).
{\it Herschel}-{\sc pacs} photometry come
  from PEP \citep{lutz11a} in COSMOS and from HerMES
  \citep{oliver12a} in Lockman Hole North.
Photometric redshifts come from: (a) \citet{ilbert10a}
  in COSMOS; (b) \citet{rowan-robinson08a} in ELAIS-N1; (c)
  \citet{strazzullo10a} in Lockman Hole North; and (d)
  \citet{cardamone10a} in ECDF-S.
%
The galaxy ``{\sc Class}'' column describes the rest-frame ultraviolet spectrum
as Lyman Break Galaxy (LBG), quasar (QSO), starburst (SB), or
starburst with AGN (SB+AGN).
Ellipsis in $S_{\rm 24}$, $S_{\rm 100}$, $S_{\rm 160}$, $S_{\rm 1.4}$,
or $z_{\rm phot}$ columns denotes that the source is undetected at the
corresponding wavelength or has no photometric redshift despite having
the necessary optical imaging, whereas $-$ denotes that no data
exist.  
Sources marked with a $\dagger$ in the comments are classified as
spectroscopically tentative, and excluded from the `high confidence'
SFRD estimate (see section~\ref{sec:discussion}).  }
\end{deluxetable}
\clearpage
\end{landscape}

\end{document}